\pdfoutput=1
\newcommand*{\ATLASLATEXPATH}{}
\documentclass[cernpreprint,txfonts,texlive=2017,UKenglish]{\ATLASLATEXPATH atlasdoc}

\usepackage[subfigure]{\ATLASLATEXPATH atlaspackage}
\usepackage{multirow}
\usepackage{slashed}
\usepackage{mathrsfs}
\usepackage{xspace} 
\usepackage{units}
\usepackage{rotating}
\usepackage{verbatim}
\usepackage{float}
\usepackage{hyperref}
\usepackage{cleveref}
\usepackage{arydshln}

\usepackage{color}

\usepackage{\ATLASLATEXPATH atlasbiblatex}

\usepackage{\ATLASLATEXPATH atlasphysics}

\graphicspath{{./}}

\usepackage{Paper_0lepton-defs}

\AtlasTitle{Search for squarks and gluinos in final states with jets
and missing transverse momentum using 36 fb$^{-1}$ of $\sqrt{s}$
=13 \TeV\ $pp$ collision data with the ATLAS detector}

\AtlasAbstract{%
A search for the supersymmetric partners of quarks and gluons (squarks
and gluinos) in final states containing hadronic jets and missing
transverse momentum, but no electrons or muons, is presented.
The data used in this search were recorded in 2015 and 2016 by the
ATLAS experiment in $\sqrt{s}=13~\TeV$ proton--proton collisions at
the Large Hadron Collider, corresponding to an integrated luminosity of 36.1~\ifb.
The results are interpreted in the context of various models where
squarks and gluinos are pair-produced and the neutralino is the lightest supersymmetric particle.
An exclusion limit at the 95\% confidence level on the mass of the
gluino is set at 2.03~\TeV\ for a simplified model incorporating only
a gluino and the lightest neutralino, assuming the lightest neutralino
is massless. For a simplified model involving the strong production of
mass-degenerate first- and second-generation squarks, squark masses
below 1.55~\TeV\ are excluded if the lightest neutralino is massless.
These limits substantially extend the region of supersymmetric
parameter space previously excluded by searches with the ATLAS
detector.
}

\AtlasRefCode{SUSY-2016-07}

\AtlasDate{\today}

\AtlasJournalRef{\PRD\ 97 (2018) 112001}
\AtlasDOI{10.1103/PhysRevD.97.112001}
\hypersetup{pdftitle={ATLAS Paper},pdfauthor={The ATLAS Collaboration}}

\begin{document}

\maketitle


\clearpage

\section{Introduction}
\label{sec:intro}
Supersymmetry (SUSY)
\cite{Golfand:1971iw,Volkov:1973ix,Wess:1974tw,Wess:1974jb,Ferrara:1974pu,Salam:1974ig}  
is a generalization of space-time symmetries that 
predicts new bosonic partners for the fermions and new fermionic partners for the bosons
of the Standard Model (SM). If $R$-parity is conserved~\cite{Farrar:1978xj}, 
SUSY particles, called sparticles, are produced in pairs and the lightest supersymmetric particle (LSP) is stable and represents a possible dark-matter candidate. 
The scalar partners of the left- and right-handed quarks, the squarks $\squarkL$ and $\squarkR$, mix to form two mass eigenstates  $\tilde{q}_1$ and $\tilde{q}_2$ ordered by increasing mass. Superpartners of the charged and neutral electroweak and Higgs bosons also mix, producing charginos ($\chinopm$) and neutralinos  ($\nino$). Squarks and the fermionic partners of the gluons, the gluinos ($\gluino$), could be produced in strong-interaction processes at the Large Hadron Collider (LHC)~\cite{LHC:2008}  and decay via cascades ending with the stable LSP, which escapes the detector unseen, producing substantial missing transverse momentum ($\vec{E}\mathrm{^{miss}_T}$). 

The large cross-sections predicted for the strong production of supersymmetric particles make the production of gluinos and squarks a primary target in searches for SUSY in proton--proton ($pp$) collisions at a center-of-mass energy of 13~\TeV\ at the LHC.  
Interest in these searches is motivated by the large available choice
of parameters for $R$-parity-conserving models in the Minimal Supersymmetric Standard Model (MSSM)~\cite{Fayet:1976et,Fayet:1977yc} where squarks (including anti-squarks) and gluinos can be produced in pairs ($\gluino\gluino$, $\squark \squark$, $\squark \gluino$) and can decay through $\squark \to q \ninoone$ and $\gluino \to q \bar{q} \ninoone$ to the lightest neutralino, $\ninoone$, assumed to be the LSP. 
Additional decay modes can include the production of charginos via $\squark\to q\chinopm$ (where $\squark$ and $q$ are of different flavor) and $\gluino\to qq\chinopm$, or neutralinos via $\gluino \to qq \ninotwo$. Subsequent chargino decay to $W^{\pm}\ninoone$ or neutralino decay to $Z\ninoone$ or $h\ninoone$, depending on the decay modes of $W$, $Z$ and $h$ bosons, can increase the jet multiplicity and missing transverse momentum.  
 
This paper presents two approaches to search for these sparticles in
final states containing only hadronic jets and large missing
transverse momentum. The first is an update of the
analysis~\cite{0LPaper_13TeV} (referred to as `Meff-based search' in
the following). The second is a complementary search using the
recursive jigsaw reconstruction (RJR)
technique~\cite{Buckley:2013kua,Jackson:2016mfb,Jackson:2017gcy} in
the construction of a discriminating variable set (`RJR-based
search'). By using a dedicated set of selection criteria, the
RJR-based search improves the sensitivity to supersymmetric models
with small mass splittings between the sparticles (models with
compressed spectra).  Both searches presented here adopt the same
general approach as the analysis of the 7~\TeV,  8~\TeV\  and
13~\TeV\ data collected during Run~1 and Run~2 of the LHC, described
in Ref.~\cite{0LPaper_13TeV}. The CMS Collaboration has set limits on similar models in Refs.~\cite{Khachatryan:2015vra,CMS:2015dbr,CMS:2017razor,Khachatryan:2016xvy}. 

In the searches presented here, events with reconstructed electrons or muons are rejected to avoid any overlap with a complementary ATLAS search in final states with one lepton, jets and missing transverse momentum~\cite{1leptonPaper-arxiv}, and to reduce the background from events with neutrinos ($W \rightarrow e\nu,\mu\nu$). 
The selection criteria are optimized in the $m_{\gluino}, m_{\ninoone}$ and $m_{\squark}, m_{\ninoone}$ planes, (where $m_{\gluino}$, $m_{\squark}$ and $m_{\ninoone}$ are the gluino, squark, and LSP masses, respectively) for simplified models~\cite{Alwall:2008ve,Alwall:2008ag,Alves:2011wf},
and in the $m_{\gluino}, m_{\squark}$ plane for the simplified phenomenological MSSM (pMSSM) models~\cite{Djouadi:1998di,Berger:2008cq} in which the number of MSSM parameters is reduced using existing experimental and theoretical constraints.
Although interpreted in terms of SUSY models, the results of this analysis could also constrain any model of new physics that predicts the production of jets in association with missing transverse momentum.  

The paper is organized as follows. Section \ref{sec:detector}
describes the ATLAS experiment and data samples used, and Section
\ref{sec:montecarlo} Monte Carlo (MC) simulation samples used for
background and signal modeling. Event reconstruction and
identification are presented in Section \ref{sec:objects}. The analysis strategy used by both searches is given in Section \ref{sec:strategy}. Since the RJR technique is a new approach for this search, Section \ref{sec:rjigsaw_intro} is dedicated to the description of the technique and associated variables.
Searches are performed in signal regions that are defined in Section~\ref{sec:selection}. 
Summaries of the background estimation methodology and corresponding systematic uncertainties are presented in Sections~\ref{sec:background} and~\ref{sec:systematics}, respectively. 
Results obtained using the signal regions optimized for both searches are reported in Section \ref{sec:results}. Section \ref{sec:conclusion} is devoted to conclusions.

\section{The ATLAS detector and data samples}
\label{sec:detector}
\interfootnotelinepenalty=10000
The ATLAS detector~\cite{Aad:2008zzm} is a multipurpose detector with a forward-backward symmetric cylindrical
geometry and nearly 4$\pi$ coverage in solid angle~\footnote{
ATLAS uses a right-handed coordinate system with its origin at the nominal
interaction point in the center of the detector. The positive $x$-axis is defined by the direction from the interaction point to the center
of the LHC ring, with the positive $y$-axis pointing upwards, while the beam direction defines the $z$-axis. Cylindrical coordinates $(r,\phi)$ are used in the transverse
plane, $\phi$ being the azimuthal angle around the $z$-axis. The pseudorapidity $\eta$ is
defined in terms of the polar angle $\theta$ by $\eta=-\ln\tan(\theta/2)$ and the rapidity is defined as $y = (1/2)\ln[(E+p_z)/(E-p_z)]$ where $E$ is the energy and $p_{\textrm z}$ the longitudinal
momentum of the object of interest. The transverse momentum $\pt$, the transverse energy $\ET$ and the missing transverse momentum $\met$ are defined in the $x$--$y$ plane unless stated otherwise. 
}.   
The inner detector (ID) tracking system consists of pixel and silicon microstrip detectors 
covering the pseudorapidity region $|\eta|<2.5$, surrounded by a transition radiation tracker, which improves electron identification over the region $|\eta|<2.0$.  The innermost
pixel layer, the insertable B-layer \cite{B-layerRef}, was added between Run~1 and Run~2 of the LHC, at a radius of 33 mm around a new, narrower and thinner beam pipe. 
The ID is surrounded by a thin superconducting solenoid providing an axial 2~T magnetic field and by
a fine-granularity lead/liquid-argon (LAr) electromagnetic calorimeter covering $|\eta|<3.2$.
A steel/scintillator-tile calorimeter provides hadronic coverage in
the central pseudorapidity range ($|\eta|<1.7$). 
The endcap and forward regions ($1.5<|\eta|<4.9$) are made of LAr active layers with either copper or tungsten as the absorber material for electromagnetic and hadronic measurements. 
The muon spectrometer with an air-core toroid magnet system surrounds the calorimeters.
Three layers of high-precision tracking chambers
provide coverage in the range $|\eta|<2.7$, while dedicated chambers allow triggering in the region $|\eta|<2.4$.

The ATLAS trigger system \cite{ATLASTrigger2016Paper} consists of two levels; the first level is a hardware-based system, while the second is a software-based system called the high-level trigger. The events used by the searches described in this paper were selected using a trigger logic that accepts events with a missing transverse momentum above $70 ~\GeV$ (for data collected during 2015) or above $90$--$110 ~\GeV$ (depending on data-taking period for data collected in 2016)  calculated using a vectorial sum of the jet transverse momenta. The trigger is 100\% efficient for the event selections considered in these analyses. Auxiliary data samples used to estimate the yields of background events were selected using triggers requiring at least one isolated electron ($\pt>24~\GeV$), muon ($\pt>20~\GeV$) or photon ($\pt>120~\GeV$) for data collected in 2015. For the 2016 data, the events used for the background estimation were selected using triggers requiring at least one isolated electron or muon ($\pt>26~\GeV$) or photon ($\pt>140~\GeV$). 

The data were collected by the ATLAS detector during 2015 with a peak
delivered instantaneous luminosity of $L = 5.2 \times 10^{33}~{\mathrm{cm^{-2} s^{-1}}}$, and during 2016 with a maximum of $L = 1.37\times10^{34}$~cm$^{-2}\textrm{s}^{-1}$. The mean number of $pp$
interactions per bunch crossing in the dataset was 14 in 2015 and 24
in 2016. 
Application of beam, detector and data-quality criteria resulted in a total integrated luminosity of 36.1~\ifb. The uncertainty in the integrated luminosity averaged over both years is 3.2\%.  
It is derived, following a methodology similar to that detailed in Ref.~\cite{Aad:2013ucp}, from a preliminary calibration of the luminosity scale using a pair of $x$--$y$ beam-separation scans performed in August 2015 and May 2016.

\section{Monte Carlo samples}
\label{sec:montecarlo}
A set of simulated MC event samples was used  to optimize the selections, estimate backgrounds and assess the sensitivity to specific SUSY signal models.  

Simplified models and pMSSM models are both used as SUSY signals in this paper.
Simplified models are defined by an effective Lagrangian describing the interactions of a small number of new particles, assuming one production process and one decay channel with a 100\% branching fraction. 
Signal samples are used to describe squark and gluino pair production, followed by the direct ($\squark \rightarrow q\ninoone$) or one-step ($\squark\rightarrow qW\ninoone$) decays of squarks and 
direct ($\gluino \rightarrow qq \ninoone$) or one-step ($\gluino \rightarrow qq W/Z/h \ninoone$) decays of gluinos as shown in Figure \ref{fig:feynman_directgrids}. Direct decays are those where the considered SUSY particles decay directly into SM particles and the LSP, while the one-step decays refer to the cases where the decays occur via one intermediate on-shell SUSY particle, as indicated in parentheses. 
In pMSSM models, gluino and first- and second-generation squark production are considered inclusively, followed by direct decays of squarks and gluinos,
or decays of squarks via gluinos ($\tilde{q}\rightarrow q\tilde{g}$) and decays of gluinos via squarks ($\tilde{g}\rightarrow q\tilde{q}$)
if kinematically possible.
All other supersymmetric particles, including the squarks of the third generation, have their masses set such that the particles are effectively decoupled.
These samples were generated with up to two (simplified models) or one (pMSSM models) extra partons in the matrix 
element using the MG5\_aMC@NLO~2.2.2 or 2.3.3 event generator~\cite{Alwall:2014hca} interfaced to  \pythia~8.186~\cite{Sjostrand:2014zea}. 
The CKKW-L merging scheme~\cite{Lonnblad:2011xx} was applied with a scale parameter that was set to a quarter of the mass of the gluino for $\gluino\gluino$ production or of the squark for $\squark\squark$ production in simplified models.
In pMSSM models, a quarter of the smaller of the gluino and squark masses was used for CKKW-L merging scale. 
 The A14~\cite{A14tune} set of tuned parameters (tune) was used for initial/final-state radiation (ISR/FSR) and underlying-event parameters together with the NNPDF2.3LO~\cite{Ball:2012cx} parton distribution function (PDF) set. The signal cross-sections were calculated at next-to-leading order
(NLO) in the strong coupling constant, adding the resummation of soft
gluon emission at next-to-leading-logarithmic accuracy
(NLO+NLL)~\cite{Beenakker:1996ch,Kulesza:2008jb,Kulesza:2009kq,Beenakker:2009ha,Beenakker:2011fu}. The
nominal squark and gluino cross-sections were taken from an envelope
of predictions using different PDF sets and factorization and
renormalization scales, as described in
Ref.~\cite{Borschensky:2014cia}, considering only first- and
second-generation squarks ($\tilde{u}$, $\tilde{d}$, $\tilde{s}$,
$\tilde{c}$). Eightfold degeneracy of first- and second-generation squarks is assumed for the simplified models with direct decays of squarks and
pMSSM models while fourfold degeneracy is assumed for the simplified
models with one-step decays of squarks. {\color{black} In the case of gluino pair
(squark pair) production in simplified models, cross-sections were evaluated
assuming arbitrarily high masses of 450 $\TeV$ for the first- and second-generation squarks (gluinos) in order to decouple them.} The free parameters are $m_{\ninoone}$ and $m_{\squark}$ ($m_{\gluino}$) for squark pair (gluino pair) production in simplified models,
while both $m_{\squark}$ and $m_{\gluino}$ are varied in pMSSM models with $m_{\ninoone}$ fixed.

\begin{figure}[t]
\begin{center}
\subfigure[]{\includegraphics[width=0.24\textwidth]{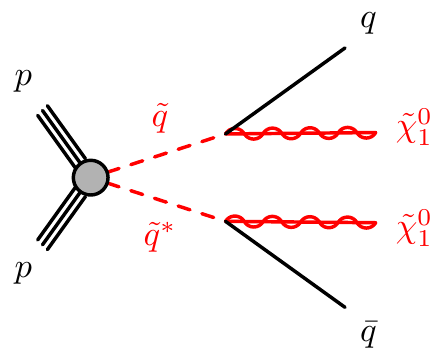}}\hspace{0.05\textwidth}
\subfigure[]{\includegraphics[width=0.24\textwidth]{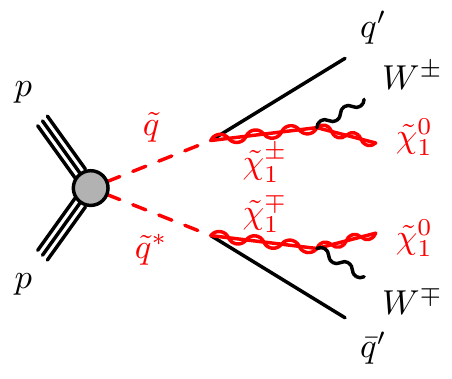}}\hspace{0.05\textwidth}
\subfigure[]{\includegraphics[width=0.24\textwidth]{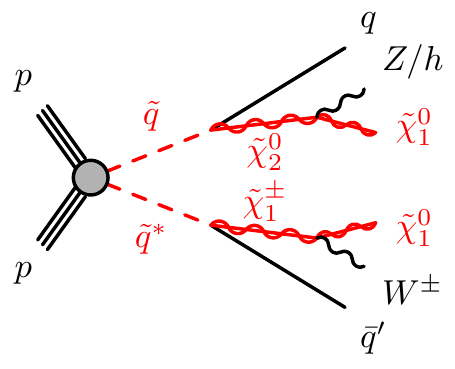}}\hspace{0.05\textwidth} \\
\subfigure[]{\includegraphics[width=0.24\textwidth]{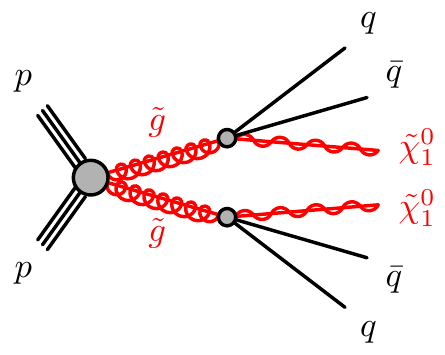}}
\subfigure[]{\includegraphics[width=0.24\textwidth]{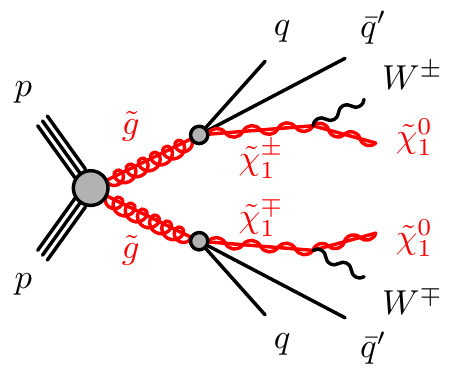}}
\subfigure[]{\includegraphics[width=0.24\textwidth]{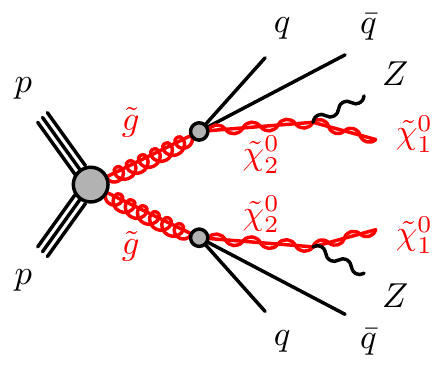}}
\subfigure[]{\includegraphics[width=0.24\textwidth]{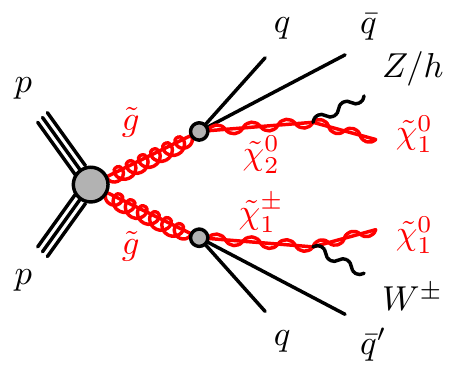}}

\caption{The decay topologies of (a,b,c) squark pair production and (d, e, f, g) gluino pair production in the simplified models with (a) direct or (b,c) one-step decays of squarks and (d) direct or (e, f, g) one-step decays of gluinos. }
\label{fig:feynman_directgrids}
\end{center}                  
\end{figure}

In the simulation of the production of $W$ or $Z/\gamma^{*}$ bosons in association with jets~\cite{ATL-PHYS-PUB-2016-003} using the \sherpa~2.2.1 event generator~\cite{Gleisberg:2008ta}, the matrix elements were calculated for up to two partons at NLO and up to two additional partons at leading order (LO) using the \textsc{Comix} \cite{comix} and \textsc{Open Loops} \cite{openloops} matrix-element generators, and merged with the \sherpa\ parton shower \cite{sherpashower} using the ME+PS@NLO prescription \cite{mepsnlo}. Simulated events containing a photon in association with jets were generated requiring a photon transverse momentum above 35~\GeV. For these events, matrix elements were calculated at LO with up to three or four partons depending on the $\pt$ of the photon, and merged with the \sherpa\ parton shower using the ME+PS@LO prescription~\cite{Hoeche:2009rj}. The $W/Z$ + jets events were normalized using their NNLO cross-sections~\cite{Catani:2009sm} while for the $\gamma$+jets process the LO cross-section, taken directly from the \sherpa\ MC event generator, was multiplied by a correction factor as described in Section~\ref{sec:background}. 

For the generation of $t\bar{t}$ and single-top processes in the $Wt$ and $s$-channel~\cite{ATL-PHYS-PUB-2016-004}, the \textsc{Powheg-Box} v2 \cite{powheg-box} generator was used, while electroweak (EW) $t$-channel single-top events was modeled using \textsc{Powheg-Box} v1. This latter generator uses the four-flavor scheme for the NLO matrix-element calculations together with the fixed four-flavor PDF set CT10f4~\cite{CT10pdf}. For each of these processes, the decay of the top quark was simulated using {\textsc MadSpin}~\cite{10a} preserving all spin correlations, while for all processes the parton shower, fragmentation, and the underlying event were generated using \pythia~6.428 \cite{pythia6} with the CTEQ6L1 \cite{Pumplin:2002vw} PDF set and the corresponding {\textsc Perugia 2012} tune (P2012) \cite{perugia}. The top quark mass was set to 172.5~\GeV. 
The $h_{\textrm{damp}}$ parameter, which controls the $\pt$ of the first
additional emission beyond the Born configuration, was set to the mass
of the top quark in the $t\bar{t}$ process. The main effect of this
parameter is to regulate the high-$\pt$ emission against which the
$t\bar{t}$ system recoils \cite{ATL-PHYS-PUB-2016-004}. The $t\bar{t}$ events were normalized using cross-sections calculated at NNLO+NNLL~\cite{Czakon:2013goa,Czakon:2011xx} accuracy, while $s$- and $t$-channel single-top events were normalized using the NLO cross-sections \cite{Aliev:2010zk,Kant:2014oha}, and the $Wt$-channel  single-top events were normalized using the NNLO+NNLL cross-sections~\cite{Kidonakis:2010ux,Kidonakis:2011wy}. Production of a top quark in association with a $Z$ boson is generated with the MG5\_aMC@NLO~2.2.1 generator at LO with CTEQ6L1 PDF set.

For the generation of $t\bar{t}$ + EW processes ($t\bar{t} + W/Z/WW$)~\cite{ATL-PHYS-PUB-2016-005}, the MG5\_aMC@NLO~2.2.3 generator at LO interfaced to the \pythia~8.186 parton-shower model was used, with up to two ($t\bar{t}+W$, $t\bar{t}+Z(\to \nu\nu/qq)$), one ($t\bar{t}+Z(\to \ell\ell)$) or no ($t\bar{t}+WW$) extra partons included in the matrix element. The events were normalized using their respective NLO cross-sections~\cite{Lazopoulos:2008de,Campbell:2012dh} and the top quark mass was set to 172.5~\GeV.

Diboson processes ($WW$, $WZ$, $ZZ$)~\cite{ATL-PHYS-PUB-2016-002} were simulated using the  \sherpa~2.1.1 generator. For processes with four charged leptons (4$\ell$), three charged leptons and a neutrino (3$\ell$+1$\nu$) or two charged leptons and two neutrinos (2$\ell$+2$\nu$), the matrix elements contain all diagrams with four electroweak couplings, and were calculated for up to one (4$\ell$, 2$\ell$+2$\nu$) or no partons (3$\ell$+1$\nu$) at NLO.
For processes in which one of the bosons decays hadronically and the other leptonically, matrix elements were calculated for up to one ($ZZ$) or no ($WW$, $WZ$) additional partons at NLO. All diboson samples also simulated up to three additional partons at LO using the \textsc{Comix} and \textsc{OpenLoops} matrix-element generators, and were merged with the \sherpa\ parton shower using the ME+PS@NLO prescription. 

A summary of the SUSY signals and the SM background processes together with the MC event generators, cross-section calculation orders in $\alpha_{\textrm s}$, PDFs, parton shower and tunes used is given in Table~\ref{tab:montecarlo}.

\begin{table}[t]
\scriptsize
\begin{center}

\caption{The SUSY signals and the SM background MC simulation samples used in this paper. The generators, the order in $\alpha_{\textrm s}$ of cross-section calculations used for yield normalization, PDF sets, parton showers and tunes used for the underlying event are shown. }
\begin{tabular}{| l l c c c c |}
\hline
Physics process & Generator& Cross-section & PDF set & Parton shower & Tune \\
&& normalization & & & \\
\hline
SUSY processes & MG5\_aMC@NLO~2.2.2--2.3.3 & NLO+NLL & NNPDF2.3LO & \pythia~8.186 & A14 \\
$W(\rightarrow \ell\nu)$ + jets              & \sherpa~2.2.1        & NNLO  &  NNPDF3.0NNLO   &  \sherpa\     & \sherpa~default \\
$Z/\gamma^{*}(\rightarrow \ell \bar \ell)$ + jets & \sherpa~2.2.1         & NNLO  &  NNPDF3.0NNLO   & \sherpa\      & \sherpa~default\\
$\gamma $ + jets & \sherpa~2.1.1         & LO  &    CT10  & \sherpa\   & \sherpa~default\\

$t\bar{t}$              & {\textsc Powheg-Box}~v2   & NNLO+NNLL                   &  CT10 &  \pythia~6.428  &\textsc{Perugia2012} \\
Single top ($Wt$-channel) & {\textsc Powheg-Box} v2  &  NNLO+NNLL  &  CT10 &  \pythia~6.428   & \textsc{Perugia2012}\\ 
Single top ($s$-channel)           & {\textsc Powheg-Box} v2  & NLO  &  CT10 &  \pythia~6.428   & \textsc{Perugia2012}\\
Single top ($t$-channel)           & {\textsc Powheg-Box} v1  & NLO  &  CT10f4 &  \pythia~6.428   & \textsc{Perugia2012}\\
Single top ($Zt$-channel) & MG5\_aMC@NLO~2.2.1 & LO & CTEQ6L1 & \pythia~6.428 & \textsc{Perugia2012} \\
$t\bar{t}+W/Z/WW$       &  MG5\_aMC@NLO~2.2.3  & NLO  & NNPDF2.3LO & \pythia~8.186 & A14    \\
$WW$, $WZ$, $ZZ$    &  \sherpa~2.1.1       & NLO  &  CT10 & \sherpa\   & \sherpa~default \\
\hline
\end{tabular}
\label{tab:montecarlo}
\end{center}
\end{table}

For all SM background samples the response of the detector to particles was modeled with a full ATLAS detector simulation \cite{:2010wqa} based on \textsc{Geant4} \cite{Agostinelli:2002hh}. 
Signal samples were prepared using a fast simulation based on a parameterization of the performance of the ATLAS electromagnetic and hadronic calorimeters \cite{ATLAS:2010bfa} and on \textsc{Geant4} elsewhere. The {\textsc EvtGen}~v1.2.0 program~\cite{evtgen} was used to describe the properties of the $b$- and $c$-hadron decays in the signal samples, and the background samples except those produced with \sherpa~\cite{Gleisberg:2008ta}.

All simulated events were overlaid with multiple $pp$ collisions simulated with \pythia~8.186 using the A2 tune~\cite{A14tune} and the MSTW2008LO parton distribution functions~\cite{Martin:2009iq}. 
The MC samples were generated with a variable number of additional $pp$ interactions (pileup), and were reweighted to match the distribution of the mean number of interactions observed in data.

\section{Event reconstruction and identification}
\label{sec:objects}

The reconstructed primary vertex of the event is required to be consistent with the luminous region and to have at least two associated tracks with $\pt > 400$~\MeV. When more than one such vertex is found, the vertex with the largest  $\sum \pt^2$ of the associated tracks is chosen.

Jet candidates are reconstructed using the anti-$k_{t}$ jet clustering algorithm~\cite{Cacciari:2008gp,Cacciari:2011ma} with a jet 
radius parameter of $0.4$ starting from clusters of calorimeter cells~\cite{Topocluster_ATLAS}. The jets are corrected for energy from pileup using the method described in Ref.~\cite{Cacciari:2007fd}: a contribution equal to the product of the jet area and the median energy density of the event is subtracted from the jet energy \cite{ATLAS-CONF-2013-083}.
Further corrections, referred to as the jet energy scale corrections, are derived from MC simulation and data, and are used to calibrate the average energies of jets to the scale of their constituent particles \cite{JetCalibRunTwo}. 
Only corrected jet candidates with $\ourpt > 20~\GeV$ and $|\eta|<2.8$  are retained. 
An algorithm based on boosted decision trees, `MV2c10'~\cite{btag_paper,ATL-PHYS-PUB-2016-012}, is used to identify jets containing a $b$-hadron ($b$-jets), with an operating point corresponding to an efficiency of 77\%, and  rejection factors of $134$ for light-quark jets and $6$ for charm jets~\cite{ATL-PHYS-PUB-2016-012} for reconstructed jets with $\ourpt > 20~\GeV$ and $|\eta|<2.5$ in simulated $t\bar{t}$ events. Candidate $b$-jets are required to have $\ourpt > 50~\GeV$ and $|\eta|<2.5$. 
Events with jets originating from detector noise and non-collision
background are rejected if the jets fail to satisfy the `LooseBad'
quality criteria, or if at least one of the two leading jets with $\pt
>100 ~\GeV$ fails to satisfy the `TightBad' quality criteria, both
described in Ref.~\cite{Aad:2013zwa}. The application of these requirements reduces the data sample by less than 1\%.
In order to reduce the number of jets coming from pileup, a
significant fraction of the tracks associated with each jet must have
an origin compatible with the primary vertex. This is enforced by
using the jet vertex tagger (JVT) output using the momentum fraction
of tracks~\cite{JVT_CONF}. The requirement JVT $> 0.59$ is only applied to jets with $\pt < 60 ~\GeV$ and $|\eta| < 2.4$.

Two different classes of reconstructed lepton candidates (electrons or muons) are used in the analyses presented here. When selecting samples for the search, events containing a `baseline' electron or muon are rejected. The selections applied to identify baseline leptons are designed to maximize the efficiency with which $W$+jets and top quark background events are rejected. When selecting `control region' samples for the purpose of estimating residual $W$+jets and top quark backgrounds, additional requirements are applied to leptons to ensure greater purity of these backgrounds. These leptons are referred to as `high-purity' leptons below and form a subset of the baseline leptons.

Baseline muon candidates are formed by combining information from the muon spectrometer and inner detector as described in Ref.~\cite{MuonPerfRun2} and are required to have $\ourpt > 7 ~\GeV$ and $|\eta| <2.7$.  High-purity muon candidates must additionally have $\pt > 27~\GeV$ and $|\eta|<2.4$, the significance of the transverse impact parameter with respect to the primary vertex $|d_0^{\mathrm{PV}}|/\sigma(d_0^{\mathrm{PV}}) <$ 3, and the longitudinal impact parameter with respect to the primary vertex  $|z_0^{\mathrm{PV}} \mathrm{sin}(\theta)|<$ 0.5~mm. Furthermore, high-purity candidates must satisfy the `GradientLoose' isolation requirements described in Ref.~\cite{MuonPerfRun2}, which rely on tracking-based and calorimeter-based variables and implement a set of $\eta$- and $\pt$-dependent criteria.  

Baseline electron candidates are reconstructed from an isolated electromagnetic calorimeter energy deposit matched to an ID track and are required to have $\ourpt > 7~\GeV$, $|\eta| < 2.47$, and to satisfy `Loose' likelihood-based identification criteria described in Ref.~\cite{Aaboud:2016vfy}.  
High-purity electron candidates additionally must satisfy `Tight' selection criteria described in Ref.~\cite{Aaboud:2016vfy}, and the leading electron must have $\pt>27 ~\GeV$. They are also required to have $|d_0^{\mathrm{PV}}|/\sigma(d_0^{\mathrm{PV}}) <$ 5, $|z_0^{\mathrm{PV}} \mathrm{sin}(\theta)|<$ 0.5~mm, and to satisfy isolation requirements similar to those applied to high-purity muons~\cite{Aaboud:2016vfy}.

After the selections described above, ambiguities between candidate jets with $|\eta|<2.8$ and leptons are resolved as follows: first, any such jet candidate that is not tagged as $b$-jet, lying within a distance $\Delta
R\equiv\sqrt{(\Delta\eta)^2+(\Delta\phi)^2}=0.2$ of a baseline electron is discarded. If a jet candidate is $b$-tagged it is interpreted as a jet and the overlapping electron is ignored. 
Additionally, if a baseline electron (muon) and a jet passing the JVT
selection described above are found within $0.2 \leq \Delta R < 0.4$
($<$ min(0.4, 0.04 + 10 \GeV/$p_{\textrm T}^{\mu}$)), it is interpreted as a jet and the nearby electron (muon) candidate is discarded. 
Finally, if a baseline muon and jet are found within $\Delta R < 0.2$, it is treated as a muon and the overlapping jet is ignored, unless the jet satisfies $N_\text{trk} < 3$, where $N_\text{trk}$ refers to the number of tracks with $\pT>500~\MeV$ that are associated with the jet, in which case the muon is ignored. 
This criterion rejects jets consistent with final-state radiation or hard bremsstrahlung.

Additional ambiguities between electrons and muons in a jet, originating from the decays of hadrons, are resolved to avoid double counting and/or remove non-isolated leptons: the electron is discarded if a baseline electron and a baseline muon share the same ID track.

Reconstructed photons are used in the missing transverse momentum
reconstruction as well as in the control region used to constrain the
$Z$+jets background, as explained in
Section~\ref{sec:background}. These latter photon candidates are
required to satisfy $\ourpt > 150~\GeV$ and $|\eta| < 2.37$,  photon
shower shape and electron rejection criteria, and to be
isolated~\cite{ATLAS:photon_Run1}. {\color{black} The reduced $\eta$ range for photons
is chosen to avoid a region of coarse granularity at high $\eta$ where
photon and $\pi^{0}$ separation worsens.} Ambiguities between candidate jets and photons (when used in the event selection) are resolved by discarding any jet candidates lying within $\Delta R$ = 0.4 of a photon candidate. 
Additional selections to remove ambiguities between electrons or muons and photons are applied such that a photon is discarded if it is within $\Delta R$ = 0.4 of a baseline electron or muon. 

The measurement of the missing transverse momentum vector
$\vec{E}\mathrm{^{miss}_T}$ (and its magnitude $\ourmagptmiss$) is based on
the calibrated transverse momenta of all electron, muon, jet
candidates, photons and all tracks originating from the primary vertex and not associated with such objects~\cite{MET2015}.

Initial jet-finding is extended using an approach called jet reclustering~\cite{Ben:2015jhep}. This allows the use of larger-radius-jet algorithms while maintaining
the calibrations  and systematic uncertainties associated with the input jets. Jets with a radius parameter 0.4 described above surviving the resolution of ambiguities and having $\pt>25~\GeV$ are used as input to
an anti-$k_{t}$ algorithm with a jet radius parameter 1.0. A grooming scheme
called ``reclustered jet trimming'' is applied to remove any small-radius jet constituent $j$ of a large-radius reclustered jet $J$ if $p_{\textrm T}^{j}<f_{\textrm{cut}}\times p_{\textrm T}^{J}$ where the parameter $f_{\textrm{cut}}$ is set to be 0.05.

Corrections derived from data control samples are applied to account for differences between data and simulation for the lepton and photon trigger and reconstruction efficiencies, the lepton momentum/energy scale and resolution, and for the efficiency and mis-tag rate of the $b$-tagging algorithm.

\section{Analysis strategy and background prediction}
\label{sec:strategy}
This section summarizes the common analysis strategy and statistical techniques that are employed in the searches presented in this paper. 

To search for a possible signal, selection criteria are defined to
enhance the expected signal yield relative to the SM
backgrounds. Signal regions (SRs) are defined using the MC simulation
of SUSY signals and the SM background processes. They are optimized to
maximize the expected discovery sensitivity for each model considered. 
To estimate the SM backgrounds in an accurate and robust fashion, control regions (CRs) are defined for each of the signal regions. 
They are chosen to be orthogonal to the SR selections in order to provide independent data samples enriched in particular backgrounds, and are used to normalize the background MC simulation. The CR selections are optimized to have negligible SUSY signal contamination for the models near the previously excluded 
boundary~\cite{0LPaper_13TeV}, while minimizing the systematic uncertainties arising from the extrapolation of the CR event yields to estimate backgrounds in the SR. 
Cross-checks of the background estimates are performed with data in several validation regions (VRs) selected with requirements such that these regions do not overlap with the CR and SR selections, and also have a low expected signal contamination. 

In order to ensure sensitivity to the variety of squark and gluino
production signals targeted in this search, a collection of inclusive
SRs is considered. Each of the SR selection requirements is optimized
to exploit expected differences in masses, kinematics, and jet
multiplicities, and each represents its own counting experiment. Two
different approaches are used in defining these SRs, with Meff-based
and RJR-based selection criteria described in
Sections~\ref{MeffbasedSR} and~\ref{RJigsawbasedSR},
respectively. These two approaches are complementary because of
differences in selected event populations and the strategy for
balancing the signal-to-background ratio against systematic uncertainties. A discussion of differences in these approaches is provided in Section~\ref{MeffRJRComparison}.

To extract the final results, three different classes of likelihood fits are employed: background-only, model-independent and model-dependent fits  \cite{HFpaper}. 
A background-only fit is used to estimate the background yields in each SR. The fit is performed using the observed event yields in the CRs associated with the SR as the only constraints, but not the yields in the SR itself. It is assumed that signal events from physics beyond the Standard Model (BSM) do not contribute to these CR yields. 
The scale factors represent the normalization of background components
relative to MC predictions ($\mu(W\textrm{+jets})$, $\mu(Z\textrm{+jets})$,
$\mu(\textrm{Top})$), and are simultaneously determined in the fit to all the CRs associated with a SR. 
The expected background in the SR is based on the yields predicted by simulation for $W/Z$+jets and background processes containing top quarks, corrected by the scale factors derived from the fit. In the case of multi-jet background, the estimate is based on the data-driven method described in Section~\ref{sec:background}. The systematic and MC statistical uncertainties in the expected values are included in the fit as nuisance parameters that are constrained by Gaussian distributions with widths corresponding to the sizes of the uncertainties considered and by Poisson distributions, respectively.  
The background-only fit is also used to estimate the background event yields in the VRs. 

A model-independent fit is used to quantify the level of agreement
between background predictions and observed yields and to set upper
limits on the number of BSM signal events in each SR. This fit
proceeds in the same way as the background-only fit, where yields in
the CRs are used to constrain the predictions of backgrounds in each
SR, while the SR yield is also used in the likelihood with an
additional nuisance parameter describing potential signal
contributions. The observed and expected upper limits at 95\%
confidence level (CL) on the number of events from BSM phenomena for
each signal region ($S_{\textrm{obs}}^{95}$ and $S_{\textrm{exp}}^{95}$) are
derived using the CL$_{\textrm s}$ prescription \cite{Read:2002hq}, neglecting any possible signal contamination in the CRs. These limits, when normalized by the integrated luminosity of the data sample, may be interpreted as upper limits on the visible cross-section of BSM physics ($\langle\epsilon\sigma\rangle_{\textrm{obs}}^{95}$), where the visible cross-section is defined as the product of production cross-section, acceptance and efficiency. The model-independent fit is also used to compute the one-sided $p$-value ($p_0$) of the background-only hypothesis, which quantifies the statistical significance of an excess.

Finally, a model-dependent fit is used to set exclusion limits on the signal cross-sections for specific SUSY models. Such a fit proceeds in the same way as the model-independent fit, except that both the signal yield in the signal region and the signal contamination in the CRs are taken into account. Correlations between signal and background systematic uncertainties are taken into account where appropriate. 
Signal-yield systematic uncertainties due to detector effects and the theoretical uncertainties in the signal acceptance are included in the fit.

\section{The Recursive Jigsaw Reconstruction technique}
\label{sec:rjigsaw_intro}

The RJR technique~\cite{Buckley:2013kua,Jackson:2016mfb,Jackson:2017gcy} is a method for defining kinematic variables event by event. While it is straightforward to fully describe an event's underlying kinematic features when all objects are fully reconstructed, events involving invisible weakly interacting particles present a challenge, as the loss of information from escaping particles constrains the kinematic variable construction to take place in the lab frame instead of the more physically natural frames of the hypothesized decays. The RJR method partially mitigates this loss of information by determining approximations of the rest frames of intermediate particle states in each event. This reconstructed view of the event gives rise to a natural basis of kinematic observables, calculated by evaluating the momenta and energy of different objects in these reference frames.

\begin{figure}[t]
\begin{center}
\includegraphics[width=0.99\textwidth]{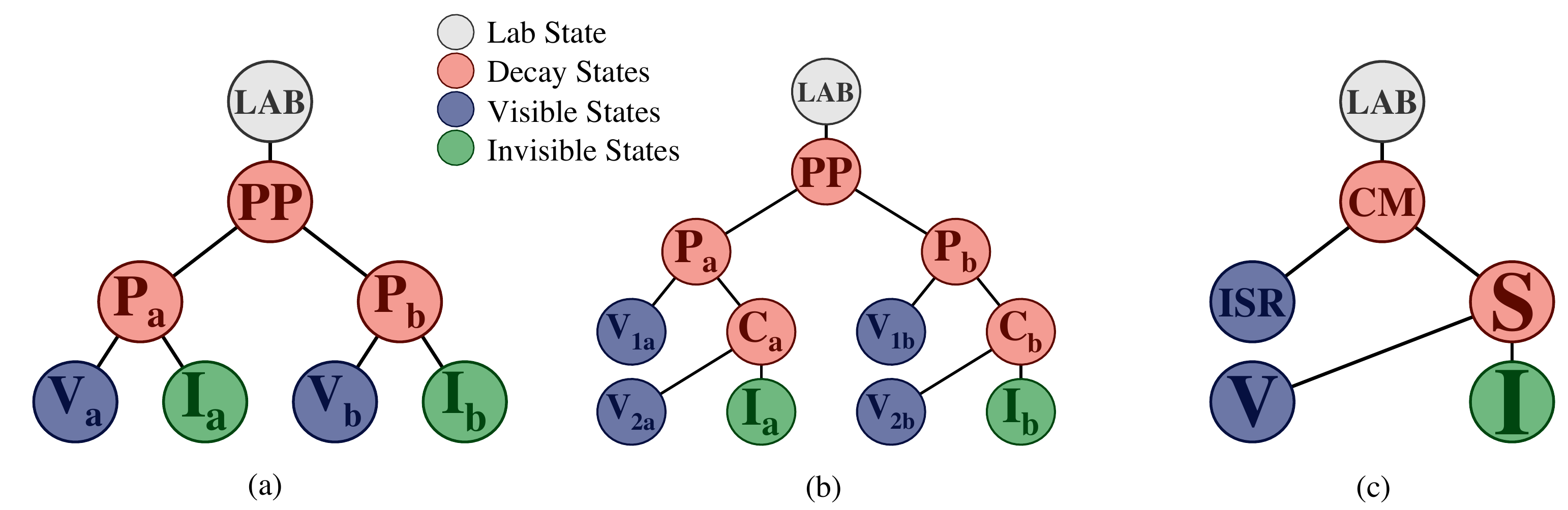}
\end{center}
\caption{\label{fig:InclusiveTree} (a) Inclusive strong sparticle production decay tree. Two sparticles ($P_{\textrm a}$ and $P_{\textrm b}$) are non-resonantly pair-produced with each decaying to one or more visible particles ($V_{\textrm a}$ and $V_{\textrm b}$) that are reconstructed in the detector, and two systems of invisible particles ($I_{\textrm a}$ and $I_{\textrm b}$) whose four-momenta are only partially constrained. (b) An additional level of decays can be added when requiring more than two visible objects. This tree is particularly useful for the search for gluino pair production described in the text. The di-sparticle production frame is denoted $PP$. Intermediate decay states are labelled $C$. (c) Strong sparticle production with ISR decay tree for use with small mass splitting spectra. CM refers to the center-of-mass of the whole reaction.
A signal sparticle system $S$ decays into visible particles ($V$) and a system of invisible particles ($I$) that recoil from a jet radiation system ISR. 
}
\end{figure}

All jets with $p_{T} > 50 ~\GeV$ and $|\eta| < 2.8$ and the missing transverse momentum are used as input to the RJR algorithm. Motivated by searches for strong production of sparticles in $R$-parity-conserving models, a decay tree, shown in Figure~\ref{fig:InclusiveTree}(a), is used in the analysis of events. Each event is evaluated as if two sparticles (the intermediate states $P_{\textrm a}$ and $P_{\textrm b}$) were produced and then decayed to the particles observed in the detector (the collections $V_{\textrm a}$ and $V_{\textrm b}$). The benchmark signal models probed in this search give rise to signal events with at least two weakly interacting particles associated with two systems of invisible particles ($I_{\textrm a}$ and $I_{\textrm b}$), the respective children of the initially produced sparticles. 

This decay tree includes several kinematic and combinatoric unknowns. In the final state with no leptons, the objects observed in the detector are exclusively jets and it is necessary to decide how to partition these jets into the two groups $V_{a}$ and $V_{b}$ in order to calculate the observables associated with the decay tree. In this analysis, the grouping that minimizes the masses of the four-momentum sum of group constituents is chosen. 

More explicitly, the collection of reconstructed jet four-momenta, $V \equiv \left\{ p_{i} \right\}$ and their four-momentum sum $p_{V}$ are considered. Each of the four-momenta is evaluated in the rest frame of $p_{V}$ ($V$ frame) and different partitions of these jets $V_{i} = \{ p_{1},\cdots, p_{N_{i}} \}$ are considered such that $V_{a} \bigcap V_{b} = 0$ and $V_{a} \bigcup V_{b} = V$. For each partition, 
the sum of four-momenta $p_{V_{i}} = \sum_{j=1}^{N_{i}}p_{j}$ is calculated and the combination that maximizes the sum of momentum of the two groups, $|\vec{p}_{V_{a}}|+|\vec{p}_{V_{b}}|$, is chosen. The axis that this partition implicitly defines in the $V$ rest frame is equivalent to the thrust axis of the jets, and the masses $M_{V_{i}} = \sqrt{p_{V_{i}}^2}$ are simultaneously minimized. 

{\color {black}When the decay tree shown in Figure~\ref{fig:InclusiveTree}(b) is used to analyze events, each of
the groups $V_{a}$ and $V_{b}$ are further subdivided, with each group
undergoing exactly the same partitioning algorithm (based on selecting
the combination maximizing the scalar sum of the momentum of the two
partitions), resulting in a finer partition with subgroups $V_{1a/2a}$ and $V_{1b/2b}$.}
Similarly, the same algorithm is used to decide which jets are
assigned to the groups $V$ and ISR when analyzing events according to
the decay tree shown in Figure~\ref{fig:InclusiveTree}(c), where
the \met, represented as $I$, is treated as an additional, massless
jet in the partitioning algorithm. {\color{black} The reconstruction code for the
algorithm can be found in Ref.~\cite{0lepHepdata}.}

The remaining unknowns in the event are associated with the two collections of weakly interacting particles: their masses, longitudinal momenta and information about how the two groups contribute to the $\vec{E}\mathrm{^{miss}_T}$.
The RJR algorithm determines these unknowns through subsequent minimizations of the intermediate particle masses appearing in the decay tree. 
In each of these newly constructed rest frames, all relevant momenta are defined and can be used to construct any variable -- multi-object invariant masses, angles between objects, etc. The primary energy-scale-sensitive observables used in the search presented here are a suite of variables denoted by $H$. These $H$ variables denote \textit{hemispheres}, with the $H$ suggesting similarities with  $H_{\textrm T}$, the scalar sum of visible transverse momenta. 
However, in contrast to $H_{\textrm T}$, these $H$ variables are constructed using different combinations of objects' momenta, including contributions from the invisible four-momenta, and are not necessarily evaluated in the lab frame, nor only in the transverse plane.

The $H$ variables are labeled with a superscript $F$ and two subscripts $n$ and $m$, $H_{n,m}^F$. The $F$ represents the rest frame in which the momenta are evaluated. In this analysis, this may be the lab frame, the proxy frame for the sparticle--sparticle frame $PP$, or the proxy frame for an individual sparticle's rest frame $P$. The subscripts $n$ and $m$ represent the number of visible and invisible momentum vectors considered, respectively. This means, given the number of visible momentum vectors in the frame, these are summed until only $n$ distinct vectors remain. The choice for which vectors are summed is made by finding jets with smallest mutual four vector dot products, using the minimization procedure described above. The same is done for the invisible system so that only $m$ distinct vectors remain. For events with fewer than $n$ visible objects, the sum only runs over the available vectors. The additional subscript ``T'' can denote a transverse version of the variable, where the transverse plane is defined in a frame $F$ as follows: The Lorentz transformation relating $F$ to the lab frame is decomposed into a boost along the beam axis, followed by a subsequent transverse boost. The transverse plane is defined to be normal to the longitudinal boost. In practice, this is similar to the plane transverse to the beam-line. 

The variables that are used to define the signal and control regions are listed below. As few requirements are placed on dimensionful variables as possible, in order to increase the generality of the signal regions' sensitivity. Additional discrimination is achieved through a minimal set of dimensionless variable requirements with selections imposed on unit-less quantities exploiting common mass-independent features of the signals considered. 

To select signal events in models with squark pair production, the following variables are used:

\begin{itemize}
\item $H_{1,1}^{~PP}$: scale variable as described above. Measures the momentum of missing particles in the $PP$ frame and behaves similarly to $\ourmagptmiss$.
\item $H_{\textrm T~2,1}^{~PP}$: scale variable as described above. Behaves similarly to effective mass, $\meff$ (defined as the scalar sum of the transverse momenta of the two leading jets and $\met$) for squark pair production signals with two-jet final states.
\item $H_{1,1}^{~PP}/H_{2,1}^{~PP}$: provides additional information in testing the balance of the two scale variables, where in the denominator the $H_{2,1}^{~PP}$ is no longer solely transverse. This provides excellent discrimination against unbalanced events where the large scale is dominated by a particular object \pt\ or by high $\ourmagptmiss$.  
\item $p_{PP,~z}^{\textrm{lab}}/(p_{PP,~z}^{\textrm{lab}} + H_{\textrm T~2,1}^{~PP})$: compares the $z$-momentum of all the objects associated with the $PP$ system in the lab frame ($p_{PP,~z}^{\textrm{lab}}$) to the overall transverse scale variable considered. This variable tests for significant boost in the $z$ direction.
\item $p_{\textrm{T~j2}}^{~PP}/H_{\textrm T~2,1}^{~PP}$: the ratio of the \pt\ of the second leading jet, evaluated in the $PP$ frame ($p_{\textrm{T~j2}}^{~PP}$) to the transverse scale variable, with small values generally more background-like.
\end{itemize}

For signal topologies with higher jet multiplicities, there is the option to exploit the internal structure of the hemispheres by using a decay tree with an additional decay. For gluino pair production, the tree shown in Figure~\ref{fig:InclusiveTree}(b) can be used and the variables used by this search are:

\begin{itemize}
\item $H_{1,1}^{~PP}$: described above.
\item $H_{\textrm T~4,1}^{~PP}$: analogous to the transverse scale variable described above but more appropriate for four-jet final states expected from gluino pair production.
\item $H_{1,1}^{~PP}/H_{4,1}^{~PP}$: analogous to $H_{1,1}^{~PP}/H_{2,1}^{~PP}$ for the squark search.
\item $H_{\textrm T~4,1}^{~PP}/H_{4,1}^{~PP}$: a measure of the fraction of the momentum that lies in the transverse plane.
\item $p_{PP,~z}^{\textrm{lab}}/(p_{PP,~z}^{\textrm{lab}} + H_{\textrm T~4,1}^{~PP})$: analogous to $p_{PP,~z}^{\textrm{lab}}/(p_{PP,~z}^{\textrm{lab}} + H_{\textrm T~2,1}^{~PP})$ above.
\item $\min_{i}$ ($p_{{\textrm{T~j2}} i}^{~PP}/H_{{\textrm T~2,1} i}^{~PP}$): represents the fraction of a hemisphere's overall scale due to the second-highest-\pt\ jet (in the $PP$ frame) compared to the overall scale, independently for each hemisphere. The smaller of the values in the two hemispheres is used, corresponding to the index $i$. 
\item $\max_{i}$ ($H_{1,0}^{~P_{i}}/H_{2,0}^{~P_{i}}$): testing balance of solely the jets momentum in a given hemisphere's approximate sparticle rest frame ($P_{i}$, index $i$ indicating each hemisphere) provides additional discrimination against a small but otherwise signal-like subset of background events with a vector boson and associated jets.
\end{itemize}

In order to reject events where the $\met$ results from mismeasurements of jets, the $\met$ is attributed to one or more jets using a transverse clustering scheme. The transverse components of reconstructed jet four vectors and the \met, treated as massless, are organized into a binary decay tree by choosing associations through the recursive minimization of subgroup masses at each decay step using the previously described algorithm. The jet(s) appearing in the decay step where the \met appears alone are those that have the smallest inner product with the system of invisible particles in the event, and their mutual transverse momentum is compared with the \met using the ratio $R_{\textrm{QCD}}$:
\begin{equation}
R_{\textrm{QCD}} = \frac{ {\textrm{max}}( \vec{p}_{\mathrm{T}}^{\textrm{~jets}} \cdot \vec{E}_{\textrm T}^{\textrm{~miss}},0) }{ (\met)^{2} + {\textrm{max}}(\vec{p}_{\mathrm{T}}^{\textrm{~jets}} \cdot \vec{E}_{\textrm T}^{\textrm{~miss}},0)}~,
\end{equation}
where $\vec{p}_{\mathrm{T}}^{\textrm{~jets}}$ is the transverse
momentum of the \met-associated jet(s) or {\color{black} system of jets} in the lab frame. Alternatively, the magnitude and direction of these jets can be compared with the \met by considering the ``decay angle'' of the jet(s)/\met system, $\cos{ (\phi_{\mathrm{j},~\met})}$, defined using the transverse jet(s) and \met four vectors of the binary decay tree. These quantities are combined into a discriminant $\Delta_{\textrm{QCD}}$, defined as
\begin{equation}
{\color{black} \Delta_{\textrm{QCD}} = \frac{1 + \cos(\phi_{\mathrm{j},~\met}) - 2R_{\textrm{QCD}} }{1 + \cos(\phi_{\mathrm{j},~\met}) + 2R_{\textrm{QCD}} }~.}
\end{equation}
This observable is used to quantify the likelihood that mismeasurements of these jets were responsible for the $\met$.  Multi-jet events with severe jet mismeasurements tend to have $\Delta_{\textrm{QCD}}$ values in the interval $[-1,0]$ while events with $\met$ from weakly interacting particles are more likely to have values in the interval $[0,1]$. 

In addition to trying to resolve the entirety of the signal event, it can be useful for sparticle spectra with smaller mass splittings and lower intrinsic $\ourmagptmiss$ to instead select events with a partially resolved sparticle system recoiling from a high-\pt\ jet from initial-state radiation (ISR). To target such topologies, a separate tree for compressed spectra is shown in Figure~\ref{fig:InclusiveTree}(c). This tree is somewhat simpler and attempts to identify visible ($V$) and invisible ($I$) systems that are the result of an intermediate state corresponding to the system of sparticles and their decay products ($S$). As the \met is used to choose which jets are identified as ISR, a transverse view of the reconstructed event is used which ignores the longitudinal momentum of the jets. The reference frames appearing in the decay tree shown in Figure~\ref{fig:InclusiveTree}(c), such as the estimate of the center-of-mass frame (CM), are then approximations in this transverse projection. This tree yields a slightly different set of variables:

\begin{itemize}
\item $p_{\mathrm{T}~S}^{\textrm{~CM}}$: the magnitude of the vector-summed transverse momenta of all $S$-associated jets ($|\vec{p}_{\mathrm{T}~S}^{\textrm{~CM}}|$) and \met evaluated in the CM frame.
\item $R_{\textrm{ISR}} \equiv \vec{p}_{I}^{\textrm{~CM}}\cdot \hat{p}_{\mathrm{T}~S}^{\textrm{~CM}}/p_{\mathrm{T}~S}^{\textrm{~CM}}$: serves as an estimate of $m_{\tilde{\chi}}/m_{\tilde{g}/\tilde{q}}$. This is the fraction of the momentum of the $S$ system that is carried by its invisible system $I$, with momentum $\vec{p}_{I}^{\textrm{~CM}}$ in the CM frame. As $p_{\mathrm{T}~S}^{\textrm{~CM}}$ grows it becomes increasingly hard for backgrounds to possess a large value in this ratio -- a feature exhibited by compressed signals.
\item $M_{\mathrm{T}~S}$: the transverse mass of the $S$ system.
\item $N_{\textrm{jet}}^{V}$: number of jets assigned to the visible system ($V$) and not associated with the ISR system.
\item $\Delta\phi_{{\textrm{ISR}},~I}$: the azimuthal opening angle between the ISR system and the invisible system in the CM frame.
\end{itemize}

\section{Event selection and signal regions definitions}
\label{sec:selection}

Following the event reconstruction described in Section~\ref{sec:objects}, in both searches documented here, events are discarded if a baseline electron or muon with $\pt>7~\GeV$ remains, or if they contain a jet failing to satisfy quality selection criteria designed to suppress detector noise and non-collision backgrounds (described in Section~\ref{sec:objects}). 
Events are rejected if no jets with $\pt >50~\GeV$ are found. The remaining events are then analyzed in two complementary searches, both of which require the presence of jets and significant missing transverse momentum. 
The selections in the two searches are designed to be generic enough to ensure sensitivity in a broad set of models with jets and \met{} in the final state.

In order to maximize the sensitivity in the $m_{\gluino},m_{\squark}$ plane, a variety of signal regions are defined. 
Squarks typically generate at least one jet in their decays, for instance through $\squark \to q \ninoone$, while gluinos typically generate at least two jets, for instance through $\gluino\to q \bar{q} \ninoone$. Processes contributing to $\squark\squark$ and $\gluino\gluino$ final states therefore lead to events containing at least two or four jets, respectively. Decays of heavy SUSY and SM particles produced in longer $\squark$ and $\gluino$ decay cascades (such as those involving chargino production with subsequent decays e.g. $\chinoonepm\to qq'\ninoone$) tend to further increase the jet multiplicity in the final state. To target different scenarios, signal regions with different jet multiplicity requirements
(in the case of Meff-based search) or different decay trees (in the case of RJR-based search) are assumed. The optimized signal regions used in both searches are summarized in the following.

\subsection{The jets+\met{} Meff-based search}
\label{MeffbasedSR}

Due to the high mass scale expected for the SUSY models considered in this study, the `effective mass', $\meff$~\cite{Hinchliffe:1996iu}, is a powerful discriminant between the signal and SM backgrounds. 
When selecting events with at least $N_{\textrm  j}$ jets, $\meff(N_{\textrm  j})$ is defined to be the scalar sum of the transverse momenta of the leading $N_{\textrm  j}$ jets and \met{}. Requirements placed on $\meff(N_{\textrm j})$ and \met{} form the basis of the Meff-based search by strongly suppressing the multi-jet background where jet energy mismeasurement generates missing transverse momentum.
 The final signal selection uses a requirement on $\meff({\textrm{incl.}})$, which sums over all jets with $\ourpt>50 ~\GeV$ and \met{} to suppress SM backgrounds, which tend to have low jet multiplicity.

Twenty-four inclusive SRs characterized by increasing the minimum jet multiplicity, from two to six, are defined in Table~\ref{tab:srdefs}: eight regions target models characterized 
by the squark pair production with the direct decay of squarks, seven regions target models with gluino pair production followed by the direct decay of gluinos and nine regions 
target squark pair or gluino pair production followed by the one-step decay of squarks/gluinos via an intermediate chargino or neutralino. 
Signal regions requiring the same jet multiplicity are distinguished
by increasing the threshold of the \meff({\textrm{incl.}}) and
$\met/\meff(N_{\textrm  j})$ or $\MET/\sqrt{H_{\textrm  T}}$ requirements. This ensures the sensitivity to a range of sparticle
masses for each decay mode. All signal regions corresponding to the
Meff-based approach are labeled with the prefix `Meff'. {\color{black}For SR's with a low number of hard jets, $\MET/\sqrt{H_{\textrm T}}$ is found to be more discriminant than $\met/\meff(N_{\textrm j})$.} 

{\color {black}In each region, different requirements are applied for
jet momenta and pseudorapidities. These thresholds are defined to reduce the SM background while keeping
high efficiency for targeted signal events. Signal regions with high
\meff({\textrm{incl.}}) thresholds are optimized for large mass differences,
leading to hard jets in the central region of the detector. For the SRs Meff-2j-2100, Meff-3j-1300 (and Meff-5j-1700) that are optimized for
small mass differences between $\squark$ ($\gluino$) and $\ninoone$,
a very high $\ourpt$ threshold is applied to the leading jet in order to
explicitly tag a jet originating from initial-state radiation, which results in asymmetric $\ourpt$ requirements on the leading jet and the other jets.}

{\color{black}Two signal regions, Meff-2jB-1600/2400, optimized for one-step decay models are designed to improve the sensitivity to models with the cascade squark decay via $\chinopm$ to $qW\ninoone$ 
(Figure~\ref{fig:feynman_directgrids}(b)) or gluino decay via $\chinopm$ ($\ninotwo$) to $qqW\ninoone$ (or $qqZ\ninoone$) (Figures~\ref{fig:feynman_directgrids}(e) and~\ref{fig:feynman_directgrids}(f)), 
in cases where the $\chinopm$ ($\ninotwo$) is nearly degenerate in mass with the squarks or the gluino. These signal regions place additional requirements on the mass of the 
large-radius jets to select the candidate hadronically decaying $W$ or
$Z$  bosons that, due to the small mass difference between the parent
SUSY particles and intermediate chargino or neutralino, can have
significant transverse momentum and appear as a single high-mass jet.}
{\color{black} The signal regions Meff-5j-2000/2600 target similar
models and have similar $\MET/\sqrt{H_{\textrm  T}}$
and $\meff({\textrm{incl.}})$ selections to the 2jB signal regions, filling the coverage gaps between the 2jB SRs
and the other non-boosted SRs.} In the other regions with at least four jets in the final state, jets from signal processes are distributed isotropically. 
Additional suppression of background processes is based on the aplanarity variable, which is defined as $A = 3/2 \lambda_3$, where $\lambda_3$ is the smallest eigenvalue of the normalized momentum tensor of the jets~\cite{Chen:2011aa}.

{\color{black}To reduce the background from multi-jet processes, requirements are
placed on two variables: $\ourdeltaphifull$ and
$\ourmagptmiss/\meff(N_{\textrm  j})$. The former is defined to be the
smallest azimuthal separation between $\vec{E}\mathrm{^{\textrm{miss}}_T}$ and the momentum vector of any of the
reconstructed jets with $\ourpt>50~\GeV$. The exact requirements, which
depend on the jet multiplicity in each SR, are summarized in Table~\ref{tab:srdefs},
where the criteria for all the Meff-based signal regions can also be
found.}

\begin{table}[H]
\fontsize{9}{10}\selectfont
     \begin{center}
      \begin{tabular}{|l|c|c|c|c|c|c|c|c|}
      \hline
       Targeted signal   & \multicolumn{8}{c|}{$\tilde{q}\tilde{q}$, $\tilde{q} \rightarrow q \tilde{\chi}_{1}^{0}$} \\
       \hline \hline
        \multirow{2}{*}{Requirement} &\multicolumn{8}{c|}{Signal Region  [\textbf{Meff-}]} \\
\cline{2-9} & \textbf{2j-1200} & \textbf{2j-1600} & \textbf{2j-2000} & \textbf{2j-2400} & \textbf{2j-2800} & \textbf{2j-3600} &  \textbf{2j-2100} & \textbf{3j-1300} \\ \hline 
            \met [\GeV] $>$&  \multicolumn{8}{c|}{ 250 }\\ \hline
   $\pt(j_1)$ [\GeV] $>$& 250 & 300 & \multicolumn{4}{c|}{350} & 600 & 700 \\ \hline
   $\pt(j_2)$ [\GeV] $>$& 250 & 300 & \multicolumn{4}{c|}{350} & \multicolumn{2}{c|}{50} \\ \hline
   $\pt(j_3)$ [\GeV] $>$&  \multicolumn{7}{c|}{--} & 50 \\ \hline
   $|\eta(j_{\textrm{1,2}})|<$ & 0.8 & \multicolumn{4}{c|}{1.2} & \multicolumn{3}{c|}{\color{black} 2.8} \\ \hline 
   $\ourdeltaphishort(\textrm{jet}_{1,2,(3)},\ourvecptmiss)_\mathrm{min}>$ & \multicolumn{6}{c|}{0.8} & \multicolumn{2}{c|}{ 0.4} \\ \hline
   $\ourdeltaphishort(\textrm{jet}_{i>3},\ourvecptmiss)_\mathrm{min}>$ & \multicolumn{6}{c|}{0.4} & \multicolumn{2}{c|}{0.2} \\ \hline
   $\met/\sqrt{H_{\textrm  T}}$ [\GeV$^{1/2}$] $>$ & 14 & \multicolumn{5}{c|}{18} & 26 & 16 \\ \hline
   $ \meff({\textrm{incl.}})$ [\GeV] $>$ & 1200  & 1600 & 2000 & 2400 & 2800 & 3600 & 2100 & 1300 \\ \hline
    \end{tabular}
    
    \vspace{0.01\textheight}
    \begin{tabular}{|l|c|c|c|c|c|c|c|}
      \hline
           Targeted signal  &  \multicolumn{7}{c|}{ $\tilde{g}\tilde{g}$, $\tilde{g} \rightarrow q\bar{q} \tilde{\chi}_{1}^{0}$ } \\
            \hline \hline
            \multirow{2}{*}{Requirement} &\multicolumn{7}{c|}{Signal Region  [\textbf{Meff-}]} \\
            \cline{2-8}   &\textbf{4j-1000} &  \textbf{4j-1400} & \textbf{4j-1800} & \textbf{4j-2200} & \textbf{4j-2600} & \textbf{4j-3000} & \textbf{5j-1700} \\ \hline
            \met [\GeV] $>$  &\multicolumn{7}{c|}{ 250 }\\ \hline
            $\pt(j_1)$ [\GeV] $>$ &\multicolumn{6}{c|}{ 200 } & 700 \\ \hline
            $\pt(j_4)$ [\GeV] $>$ & \multicolumn{4}{c|}{100} & \multicolumn{2}{c|}{150} & 50 \\ \hline
            $\pt(j_5)$ [\GeV] $>$&\multicolumn{6}{c|}{--}   & 50 \\ \hline
             $|\eta(j_{1,2,3,4})|<$ & 1.2 & \multicolumn{5}{c|}{2.0} & {\color{black} 2.8} \\ \hline 
            $\ourdeltaphishort(\textrm{jet}_{\textrm 1,2,(3)},\ourvecptmiss)_\mathrm{min}>$ &\multicolumn{7}{c|}{0.4} \\ \hline
            $\ourdeltaphishort(\textrm{jet}_{i>3},\ourvecptmiss)_\mathrm{min}>$  &\multicolumn{6}{c|}{0.4} & 0.2\\ \hline
            $\met/\meff(N_{\textrm j})>$  & 0.3 & \multicolumn{3}{c|}{0.25} & \multicolumn{2}{c|}{0.2} & 0.3\\ \hline
            Aplanarity $>$  & \multicolumn{6}{c|}{0.04} & --\\ \hline
            $ \meff({\textrm{incl.}})$ [\GeV] $> $  & 1000 & 1400 & 1800 & 2200 & 2600 & 3000 & 1700 \\\hline
    \end{tabular}

    \vspace{0.01\textheight}
    \begin{tabular}{|l |c|c|c|c|c|c|c |}
      \hline
            Targeted signal & \multicolumn{7}{c|}{ $\tilde{g}\tilde{g}$, $\tilde{g} \rightarrow q\bar{q}W \tilde{\chi}_{1}^{0}$ and $\tilde{q}\tilde{q}$, $\tilde{q} \rightarrow qW\tilde{\chi}_{1}^{0}$ } \\ \hline \hline
            \multirow{2}{*}{Requirement} &\multicolumn{7}{c |}{Signal Region  [\textbf{Meff-}]} \\ 
            \cline{2-8}   & \textbf{5j-1600} & \textbf{5j-2000} & \textbf{5j-2600} & \textbf{6j-1200} & \textbf{6j-1800} &  \textbf{6j-2200} & \textbf{6j-2600} \\ \hline
            \met [\GeV] $>$&\multicolumn{7}{c |}{ 250 }  \\ \hline
            $\pt(j_1)$ [\GeV] $>$&\multicolumn{7}{c |}{ 200 }  \\ \hline
            $\pt(j_5)$ [\GeV] $>$ & \multicolumn{4}{c|}{50} & \multicolumn{3}{c |}{100}   \\ \hline
            $\pt(j_6)$ [\GeV] $>$ & \multicolumn{3}{c|}{--} & 50 & \multicolumn{3}{c |}{100}   \\ \hline
             $|\eta(j_{\textrm{1,...,6}})|<$    & \multicolumn{3}{c|}{{\color{black} 2.8}}    & \multicolumn{2}{c|}{2.0} & \multicolumn{2}{c|}{{\color{black} 2.8}} \\ \hline 
            $\ourdeltaphishort(\textrm{jet}_{\textrm 1,2,(3)},\ourvecptmiss)_\mathrm{min}>$ & \multicolumn{2}{c |}{0.4} & 0.8 & \multicolumn{4}{c|}{0.4} \\ \hline
            $\ourdeltaphishort(\textrm{jet}_{i>3},\ourvecptmiss)_\mathrm{min}>$ & 0.2 & \multicolumn{2}{c |}{0.4} & \multicolumn{4}{c|}{0.2} \\ \hline
            $\met/\meff(N_{\textrm  j})>$ & 0.15 & \multicolumn{2}{c|}{--} & 0.25 & \multicolumn{2}{c |}{0.2} & 0.15 \\ \hline
            $\met/\sqrt{H_{\textrm  T}}$ [\GeV$^{1/2}$] $>$ & -- & 15 & 18 & \multicolumn{4}{c|}{--} \\ \hline
            Aplanarity $>$& 0.08 & \multicolumn{3}{c|}{--} & 0.04 & \multicolumn{2}{c |}{0.08}  \\ \hline
            $ \meff({\textrm{incl.}})$ [\GeV] $> $& 1600 & 2000 & 2600 & 1200 & 1800 & 2200 & 2600 \\\hline
       \end{tabular}

    \vspace{0.01\textheight}
    \begin{tabular}{|l |c|c|}
      \hline
            Targeted signal & \multicolumn{2}{c|}{ $\tilde{g}\tilde{g}$, $\tilde{g} \rightarrow q\bar{q}W \tilde{\chi}_{1}^{0}$ and $\tilde{q}\tilde{q}$, $\tilde{q} \rightarrow qW\tilde{\chi}_{1}^{0}$ } \\ \hline \hline
            \multirow{2}{*}{Requirement} &\multicolumn{2}{c |}{Signal Region  [\textbf{Meff-}]} \\ 
            \cline{2-3}   & \textbf{2jB-1600} & \textbf{2jB-2400} \\ \hline
            \met [\GeV] $>$&\multicolumn{2}{c |}{ 250 }  \\ \hline
            $\pt({\textrm{large}\mathchar`-R}\ j_1)$ [\GeV] $>$&\multicolumn{2}{c |}{ 200 }  \\ \hline
            $\pt({\textrm{large}\mathchar`-R}\ j_2)$ [\GeV] $>$ & \multicolumn{2}{c|}{200} \\ \hline
            $m({\textrm{large}\mathchar`-R}\ j_1)$ [\GeV] &  \multicolumn{2}{c |}{ [60,110] }  \\ \hline
            $m({\textrm{large}\mathchar`-R}\ j_2)$ [\GeV] &  \multicolumn{2}{c |}{ [60,110] }  \\ \hline
            $\ourdeltaphishort(\textrm{jet}_{\textrm{1,2,(3)}},\ourvecptmiss)_\mathrm{min}>$ & \multicolumn{2}{c |}{0.6} \\ \hline
            $\ourdeltaphishort(\textrm{jet}_{i>3},\ourvecptmiss)_\mathrm{min}>$ & \multicolumn{2}{c |}{0.4} \\ \hline
            $\met/\sqrt{H_{\textrm  T}}$ [\GeV$^{\textrm{1/2}}$] $>$ & \multicolumn{2}{c|}{20} \\ \hline
            $ \meff({\textrm{incl.}})$ [\GeV] $> $& 1600 & 2400 \\\hline
       \end{tabular}

\caption{\label{tab:srdefs} Selection criteria and targeted signal models from Figure~\ref{fig:feynman_directgrids} used to define signal regions in the Meff-based search, indicated by the prefix `Meff'. The first block of SRs
targets Figure~\ref{fig:feynman_directgrids}(a), the second block of
SRs targets Figure~\ref{fig:feynman_directgrids}(d). The third and
fourth blocks of SRs target Figures~\ref{fig:feynman_directgrids}(b)
and~\ref{fig:feynman_directgrids}(e). Each SR is labeled with the
inclusive jet multiplicity considered (`2j', `3j' etc.) together with
the the \meff\ requirement. The $\met/\meff(N_{\textrm  j})$ cut in any $N_{\textrm  j}$-jet channel uses a value of $\meff$ constructed from only the leading $N_{\textrm  j}$ jets ($\meff(N_{\textrm  j})$).  However, the final $\meff({\textrm{incl.}})$ selection, which is used to define the signal regions, includes all jets with $\pt>50~\GeV$. Large-radius reclustered jets are denoted by large-R $j$.
}
\end{center}
\end{table}

\subsection{The jets+\met{} RJR-based search}
\label{RJigsawbasedSR}

The procedure adopted is such that, as the mass splitting between parent sparticle and the LSP increases, the criteria applied to the scale variables are tightened, while the criteria for dimensionless variables are loosened. In searching for the squark pair production, the overall balance of the events is studied with $H_{\textrm{1,1}}^{~PP}/H_{\textrm{2,1}}^{~PP}$. The range selected in this ratio rejects those events where the missing transverse momentum dominates the scale (upper bound) and ensures the sufficient balance between the scales of visible and invisible particles (lower bound). The selection on the $p_{\textrm{T~j2}}^{\textrm ~PP}/H_{\textrm{T~2,1}}^{~PP}$ ratio serves to ensure that each of the jets contributes to the overall scale significantly. This particular ratio is a powerful criterion against imbalanced events with $W/Z$+jets, where one of the jets has a much higher momentum than the subleading jet. 

For signals of gluino pair production, the same principles are followed. Tight requirements are placed on $H_{1,1}^{~PP}/H_{4,1}^{~PP}$ and $H_{\textrm T~4,1}^{~PP}/H_{4,1}^{~PP}$  to target scenarios with more compressed spectra. A selection is applied to the ratio $p_{PP,~z}^{\textrm{~lab}} / \left( p_{PP,~z}^{\textrm{~lab}}+H_{\textrm T~4,1}^{~PP} \right)$ to test the size of the total $z$-component of momentum relative to the overall scale, requiring that it should be small. A lower bound is placed on $p_{\textrm{T~j2}}^{~PP}/H_{\textrm T~2,1}^{~PP}$. 
This provides a very strong constraint against events where the two hemispheres are well balanced but one of the jets dominates the scale variable contribution. In order to reject events where the $\met$ results from mismeasurements of jets a requirement on the variable $\Delta_{\textrm{QCD}}$ is applied, rejecting events where this is deemed likely. 

Additionally, separate SRs are defined for models with extremely compressed spectra. Following the pattern of successive SRs targeting larger mass splitting scenarios, several regions designed to be sensitive to various mass splittings utilize the ISR-boosted compressed decay tree described in Section~\ref{sec:rjigsaw_intro}. These regions target mass splittings between parent squarks and gluinos and \ninoone from roughly $25 ~\GeV$ to $200 ~\GeV$.

The selection criteria of the resulting 19 signal regions are summarized in Table~\ref{tab:RJsrdefs}. The entries for $|\eta_{\textrm{j1,j2}}|$ and $|\eta_{\mathrm{j}1,2,a,b}|$ correspond to upper bounds on the pseudorapidities of the leading two jets in each event and the leading two jets in each hemisphere $a,b$, respectively, while $|\eta_{\mathrm{j}V}|$ corresponds to the jets associated with the system $V$. All signal regions included in the RJR-based search have an `RJR' prefix.

\begin{table}[H]
\footnotesize
\begin{center}
\begin{tabular}{|r|c|c|c|c|c|c|c|}
\hline
Targeted signal   & \multicolumn{7}{c|}{$\tilde{q}\tilde{q}$, $\tilde{q} \rightarrow q \tilde{\chi}_{1}^{0}$} \\
\hline\hline
\multirow{2}{*}{Requirement} &\multicolumn{7}{c|}{Signal Region} \\
\cline{2-8}  & \multicolumn{2} {c|}{ \textbf{RJR-S1}}  & \multicolumn{2} {c|}{\textbf{RJR-S2}}   & \multicolumn{2} {c|}{\textbf{RJR-S3}} & \textbf{RJR-S4}  \\
\hline
$H_{\textrm 1,1}^{~PP}/H_{\textrm 2,1}^{~PP} \geq$ &   \multicolumn{2} {c|}{$ 0.55$}      & \multicolumn{2} {c|}{$ 0.5$}       &  \multicolumn{2} {c|}{$ 0.45$}   &  $-$ \\ \hline
$H_{\textrm 1,1}^{~PP}/H_{\textrm 2,1}^{~PP} \leq$ &   \multicolumn{2} {c|}{$ 0.9$}      & \multicolumn{2} {c|}{$ 0.95$}       &  \multicolumn{2} {c|}{$ 0.98$}   & $-$  \\ \hline
$p_{\textrm{T~j2}}^{~PP}/H_{\textrm T~2,1}^{~PP} \geq $ &   \multicolumn{2} {c|}{$ 0.16$}      & \multicolumn{2} {c|}{$ 0.14$}    &   \multicolumn{2} {c|}{$ 0.13$}  & $ 0.13$ \\ \hline
$|\eta_{\textrm{j1,j2}}| \leq$      & \multicolumn{2} {c|}{$\ 0.8$} & \multicolumn{2} {c|}{$\ 1.1$} & \multicolumn{2} {c|}{$ 1.4$} & 2.8 \\ \hline
$\Delta_{\textrm{QCD}} \geq$ &  \multicolumn{2} {c|}{$ 0.1$} & \multicolumn{2} {c|}{$ 0.05$} & \multicolumn{2} {c|}{$ 0.025$} & 0 \\  \hline
$p_{PP,~\mathrm{T}}^{\textrm{~lab}} / \left( p_{PP,~\mathrm{T}}^{\textrm{~lab}}+H_{\textrm T~2,1}^{~PP}\right) \leq$  &  \multicolumn{7} {c|}{$ 0.08$} \\ \hline
\hline
&  \textbf{RJR-S1a}  & \textbf{RJR-S1b} & \textbf{RJR-S2a} & \textbf{RJR-S2b} & \textbf{RJR-S3a} & \textbf{RJR-S3b} & \textbf{RJR-S4} \\ 
\hline
$H_{\textrm T~2,1}^{~PP}$ [\GeV] $>$ &   1000  & 1200  & 1400 & 1600 & 1800 & 2100 & 2400 \\
\hline
$H_{\textrm 1,1}^{~PP}$ [\GeV] $>$ &   800  & 1000  & 1200 & 1400 & 1700 & 1900 & 2100   \\
\hline
\end{tabular}

\vspace*{0.01\textheight}

\begin{tabular}{|r|c|c|c|c|c|c|c|}
\hline
Targeted signal   & \multicolumn{7}{c|}{ $\tilde{g}\tilde{g}$, $\tilde{g} \rightarrow q\bar{q} \tilde{\chi}_{1}^{0}$} \\
\hline \hline
\multirow{2}{*}{Requirement} &\multicolumn{7}{c|}{Signal Region} \\
\cline{2-8}  & \multicolumn{2} {c|}{\textbf{RJR-G1}}  & \multicolumn{2} {c|}{\textbf{RJR-G2}}   & \multicolumn{2} {c|}{\textbf{RJR-G3}} & \textbf{RJR-G4} \\
\hline 
$H_{\textrm 1,1}^{~PP}/H_{\textrm 4,1}^{~PP} \geq$ &   \multicolumn{2} {c|}{$ 0.45$}      & \multicolumn{2} {c|}{$ 0.3$}       &  \multicolumn{2} {c|}{$ 0.2$}  & $-$    \\ \hline
$H_{\textrm T~4,1}^{~PP}/H_{\textrm 4,1}^{~PP} \geq$ &   \multicolumn{2} {c|}{$ 0.7$}      & \multicolumn{2} {c|}{$ 0.7$}       &  \multicolumn{2} {c|}{$ 0.65$}  & $ 0.65$    \\ \hline
min~$\left( p_{{\textrm{T~j2}}i}^{~PP}/H_{{\textrm T~2,1}i}^{~PP} \right) \geq$ &   \multicolumn{2} {c|}{$ 0.12$}      & \multicolumn{2} {c|}{$ 0.1$}    &   \multicolumn{2} {c|}{$ 0.08$}  & $ 0.07$ \\ \hline
max~$\left( H_{\textrm 1,~0}^{~P_i}/H_{\textrm 2,~0}^{~P_i} \right) \leq$ &   \multicolumn{2} {c|}{$ 0.96$}      & \multicolumn{2} {c|}{$ 0.97$}    &   \multicolumn{2} {c|}{$ 0.98$} & $ 0.98$  \\  \hline
$|\eta_{\mathrm{j}1,2,a,b}| \leq$      & \multicolumn{2} {c|}{$\ 1.4$} & \multicolumn{2} {c|}{$\ 2.0$} & \multicolumn{2} {c|}{$ 2.4$} & 2.8 \\ \hline
$\Delta_{\textrm{QCD}} \geq$ &  \multicolumn{2} {c|}{$ 0.05$} & \multicolumn{2} {c|}{$ 0.025$} & \multicolumn{2} {c|}{$ 0$} & 0 \\  \hline
$p_{PP,~z}^{\textrm{~lab}} / \left( p_{PP,~z}^{\textrm{~lab}}+H_{\textrm T~4,1}^{~PP}\right) \leq$  &   \multicolumn{2} {c|}{$ 0.5$}      & \multicolumn{2} {c|}{$ 0.55$}    &   \multicolumn{2} {c|}{$ 0.6$}  & $0.65$ \\ \hline
$p_{PP,~\mathrm{T}}^{\textrm{~lab}} / \left( p_{PP,~\mathrm{T}}^{\textrm{~lab}}+H_{\textrm T~4,1}^{~PP}\right) \leq$  &  \multicolumn{7} {c|}{$ 0.08$} \\ \hline
\hline 
&  \textbf{RJR-G1a}  & \textbf{RJR-G1b} & \textbf{RJR-G2a} & \textbf{RJR-G2b} & \textbf{RJR-G3a} & \textbf{RJR-G3b} & \textbf{RJR-G4} \\
\hline
$H_{\textrm T~4,1}^{~PP}$ [\GeV] $>$ &   1200  & 1400 & 1600  & 2000 &  2400 & 2800 & 3000\\
\hline
$H_{\textrm 1,1}^{~PP}$ [\GeV] $>$ & \multicolumn{2} {c|}{700}  & \multicolumn{2} {c|}{800}  &  \multicolumn{2} {c|}{900} & 1000  \\
\hline
\end{tabular}

\vspace*{0.01\textheight}

\begin{tabular}{|r|c|c|c|c|c|}
\hline
       Targeted signal   & \multicolumn{5}{c|}{ compressed spectra in $\tilde{q}\tilde{q}$ ($\tilde{q} \rightarrow q \tilde{\chi}_{1}^{0}$); $\tilde{g}\tilde{g}$ ($\tilde{g} \rightarrow q\bar{q} \tilde{\chi}_{1}^{0}$) } \\
       \hline \hline
      \multirow{2}{*}{Requirement} &\multicolumn{5}{c|}{Signal Region} \\
 \cline{2-6}  &  \textbf{RJR-C1}  &  \textbf{RJR-C2}   &  \textbf{RJR-C3}   &  \textbf{RJR-C4} &  \textbf{RJR-C5} \\
\hline 
$R_{\textrm{ISR}} \geq $ & $ 0.95$ & $ 0.9$   & $ 0.8$   &  $ 0.7$  &  $ 0.7$   \\ \hline
$p_{\mathrm{T}~S}^{\textrm{~CM}}$ [\GeV]  $\ge$ & $ 1000$   & $ 1000$  & $ 800$  & $ 700$  & $ 700$  \\ \hline
$ \Delta\phi_{\mathrm{ISR},~I}/\pi \geq$ &   $ 0.95$  &  $ 0.97$ &  $ 0.98$  &  $ 0.95$ &  $ 0.95$ \\ \hline
$\ourdeltaphishort(\textrm{jet}_{1,2},\ourvecptmiss)_\mathrm{min}>$  &   $-$  &  $-$ &  $-$  &  0.4 &  0.4 \\ \hline
$M_{\mathrm{T}~S}$ [\GeV] $\geq$ &  $-$     & $ 100$  &  $ 200$   &  $ 450$ &  $ 450$   \\ \hline
$N_{\textrm{jet}}^{~V} \geq$ & $ 1$  & $ 1$  &  $ 2$  &  $ 2$  & $ 3$   \\ \hline
$|\eta_{\mathrm{j}V}| \leq$ & $ 2.8$  & $ 1.2$  &  $ 1.4$  &  $ 1.4$  & $ 1.4$   \\ \hline

\end{tabular}

\caption{Selection criteria and targeted signal model from Figure~\ref{fig:feynman_directgrids} used to define signal regions in the RJR-based search, indicated by the prefix `RJR'. Each SR is labeled with the targeted SUSY particle or the targeted region of parameter space, such that `S', `G' and `C' denote search regions for squark pairs, gluino pairs, or compressed spectra, respectively. 
\label{tab:RJsrdefs}}
\end{center}
\end{table}

\subsection{Meff-based and RJR-based signal region comparison}
\label{MeffRJRComparison}

Even though the selection requirements that define the Meff-based and RJR-based SRs use different sets of kinematic observables, the regions are not necessarily orthogonal. The fraction of events common to different regions, for both the SM backgrounds and the SUSY signals, reflects the complementarity of using these two approaches. For models with large \squark/\gluino masses, the signal efficiency is prioritized due to low production cross-sections. In these cases, stringent requirements on the similarly behaving $\meff$ and $H_{\textrm T~2,1}^{~PP}$/$H_{\textrm T~4,1}^{~PP}$ variables result in a larger overlap between the Meff-based and RJR-based signal regions. Conversely, signal regions designed for increasingly compressed mass spectra have looser $\meff$ and $H_{\textrm T~2,1}^{~PP}$/$H_{\textrm T~4,1}^{~PP}$, and backgrounds must be suppressed with other, complementary, kinematic requirements. As these additional kinematic observables can be quite different between Meff-based and RJR-based approaches, the orthogonality of these respective SRs increases with decreasing sparticle mass splittings. 

\begin{figure}[t]
\begin{center}
\vspace*{-0.01\textheight}
\includegraphics[width=0.7\textwidth]{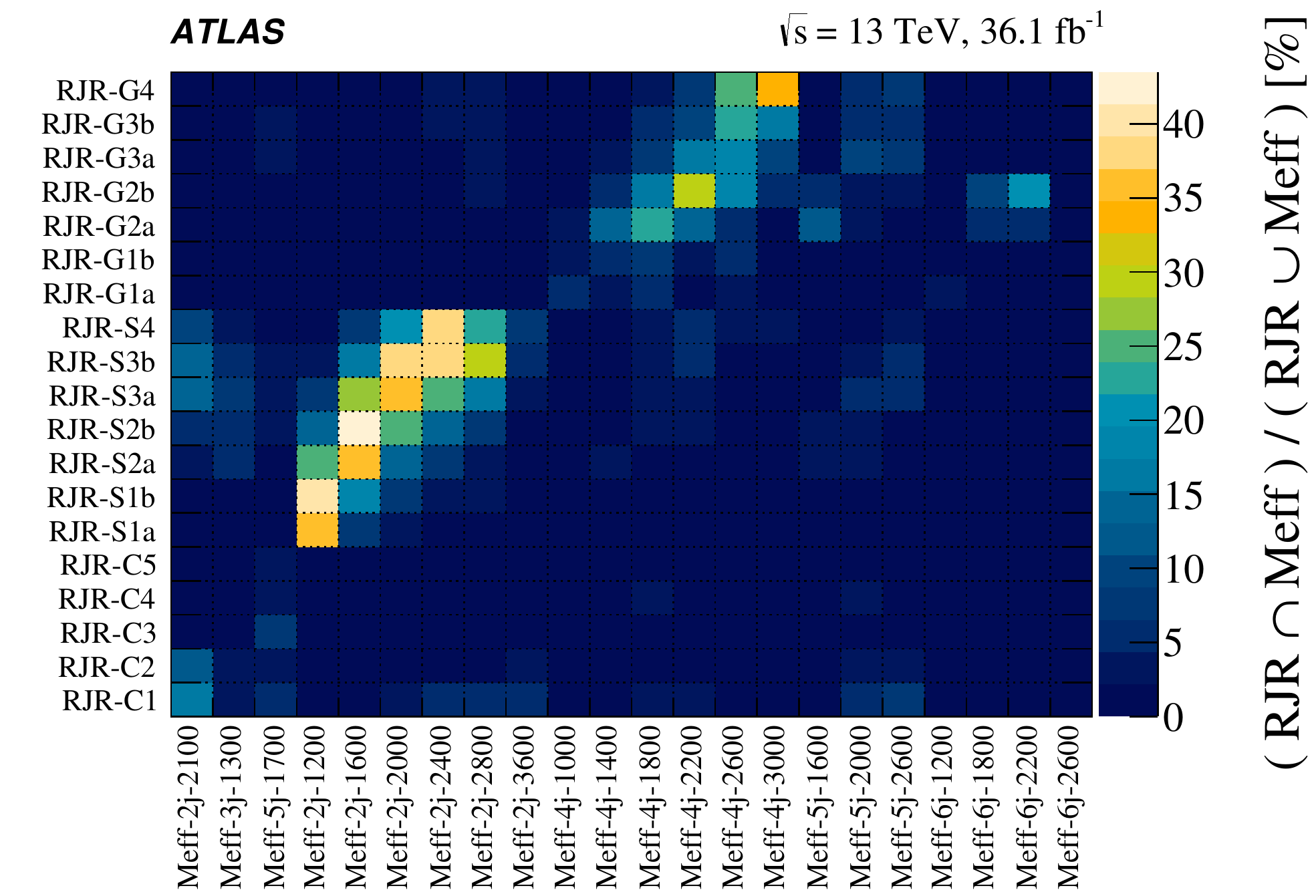}
\end{center}
\vspace*{-0.03\textheight}\caption{\label{fig:Overlap2D_data} Fractional overlap of data events selected in Meff-based and RJR-based SRs. Meff-based SRs are listed along the $x$-axis with RJR-based regions on the $y$-axis. The intersection events falling in each pair of regions, normalized by the union, is shown on the $z$-axis. The Meff-based boosted boson SRs (Meff-2jB-1600,Meff-2jB-2400) are not included as they have negligible overlap with other regions due to their unique requirements.
}
\end{figure}

This behavior can be observed in Figure~\ref{fig:Overlap2D_data}, which shows the fractional overlap of selected events in data between the Meff-based and RJR-based SRs. Each of the axes listing the various SRs are organized in the same order, with SRs targeting compressed mass spectra in the lower left of the figure, followed by squark regions with increasing sparticle masses, and then gluinos with increasing mass. This ordering results in a diagonal pattern of larger overlap, as SRs targeting the same signals are more similar. The SRs searching for evidence of squark production (RJR-Sx and Meff-2j-x) have fractions of overlapping events between 25\% and 45\%, while those targeting gluino production (RJR-Gx and Meff-4j-x) have smaller intersections, ranging from a few percent to 35\%. This decrease in overlap for gluino SRs follows from increasing differences between the selections used in the Meff-based and RJR-based approaches. While observables such as $\met/\meff(N_{\textrm j})$ and aplanarity are sensitive to global event properties, the RJR-based analysis for gluinos attempts to decompose the event into two hemispheres representing each gluino. Kinematic variables used in the definitions of SRs are calculated from each hemisphere independently, providing complementarity to those describing the total event. Using this additional information in the RJR-based selections leads to generally tighter SRs, adding increased sensitivity for intermediate mass splittings. 

Similar trends in event overlaps between SRs are expected for signal contributions, as shown in Figures~\ref{fig:Overlap2D_signal}(a) and~\ref{fig:Overlap2D_signal}(b) where a simulated squark signal with $m_{\squark}=1.5~\TeV$ and massless $\ninoone$,
and a gluino signal with $m_{\gluino} = 2 ~\TeV$ and massless \ninoone are used as examples. In these cases, the SRs targeting squarks and gluinos share a large fraction of their events, with the RJR-S4 and Meff-2j-2800 regions best suited to this squark signal having 45\% of selected events in common and the analogous gluino SRs (RJR-G4 and Meff-4j-3000) having an overlap of 40\%.
In the case of a squark signal, the largest overlap of 65\% is seen with the RJR-S2a and Meff-2j-1600, with smaller overlap between tighter SRs favored for this signal point.

The RJR-Cx SRs targeting signals with the most compressed  mass
spectra ($0 < m_{\squark/\gluino}-m_{\ninoone} \lsim 200$ \GeV) are
the most dissimilar from their Meff-based analogs. They attempt to
explicitly identify the strong initial-state radiation system that
provides the escaping \ninoone pair the \met needed to satisfy trigger
and selection requirements and use kinematic requirements based on
this interpretation of the event. The Meff-based SRs designed for
these signals (Meff-2j-2100/3j-1300/5j-1700) exploit this
compressed-mass-spectra event topology by requiring large
$\MET/\sqrt{H_{\textrm  T}}$ {\color{black} or large $\met/\meff(N_{\textrm  j})$} and a hard leading jet corresponding to the ISR system, and the modest $\meff$ requirements result in SRs with relatively large expected background yields and low systematic uncertainties. The RJR-Cx SRs take a more restrictive approach, using observables designed specifically for this ISR event topology, with the corresponding SRs having much lower event yields, higher signal-to-background ratios, but larger uncertainties. This results in much smaller event overlap for both signal and background, as seen in Figures~\ref{fig:Overlap2D_signal}(c) and~\ref{fig:Overlap2D_signal}(d) for an example simulated squark signal with $m_{\squark} = 700 ~\GeV$ and $m_{\ninoone} = 600 ~\GeV$, and a gluino signal with $m_{\gluino} = 1 ~\TeV$ and $m_{\ninoone} = 800 ~\GeV$. For these signals, the overlap between the best-suited SRs (RJR-C3 and Meff-3j-1300 for the squark signal, RJR-C5 and Meff-5j-1700 for the gluino) is only about $10\%$.  On the other hand, $65\%$ ($35\%$) of the signal events in RJR-C3 (RJR-C4) are also selected in Meff-3j-1300 (Meff-5j-1700). The more stringent selection strategy employed in the RJR-Cx regions leads to increased sensitivity for compressed mass spectra for each of the signal variants considered in this analysis.

\begin{figure}[t]
\begin{center}
\vspace*{-0.01\textheight}
\subfigure[]{\includegraphics[width=0.48\textwidth]{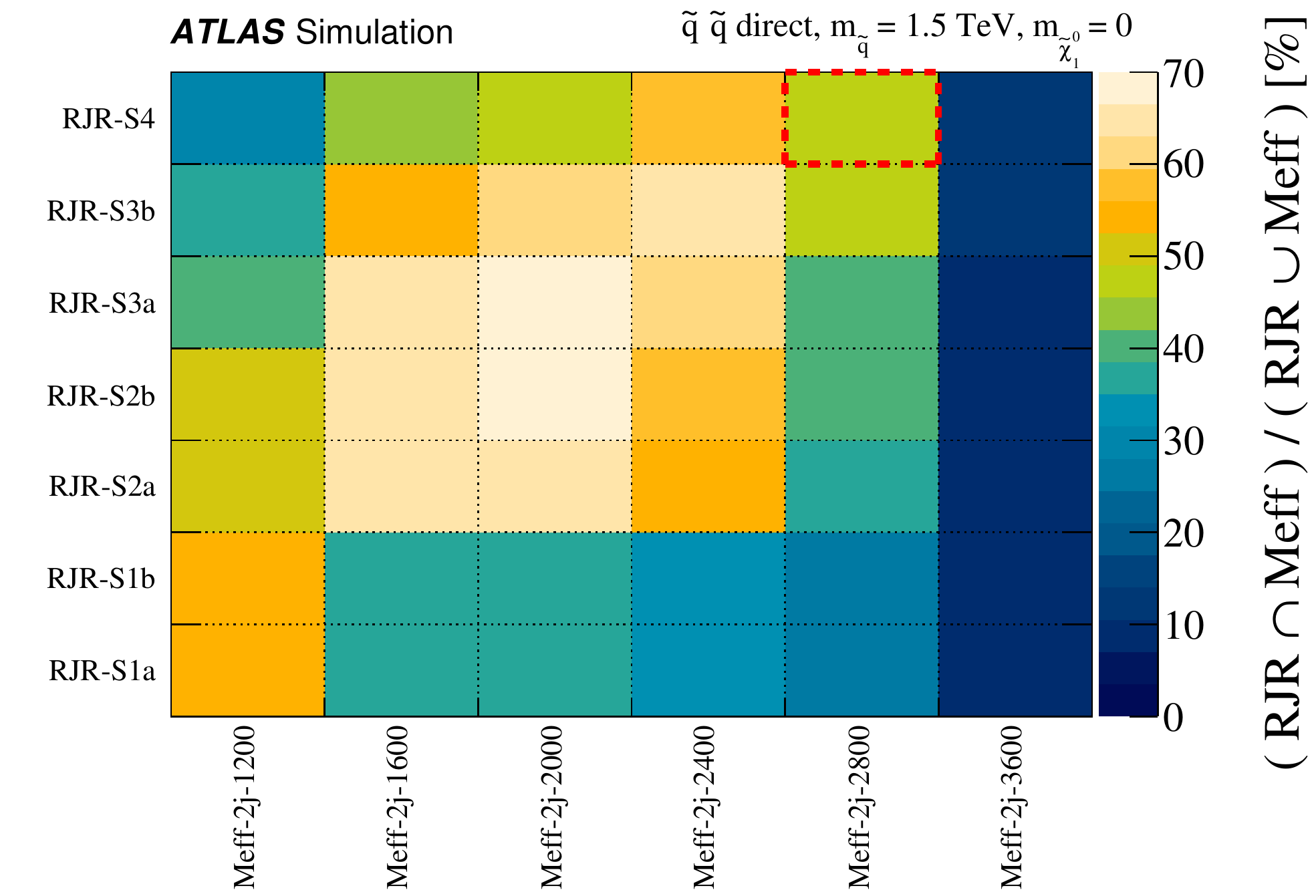}} 
\subfigure[]{\includegraphics[width=0.48\textwidth]{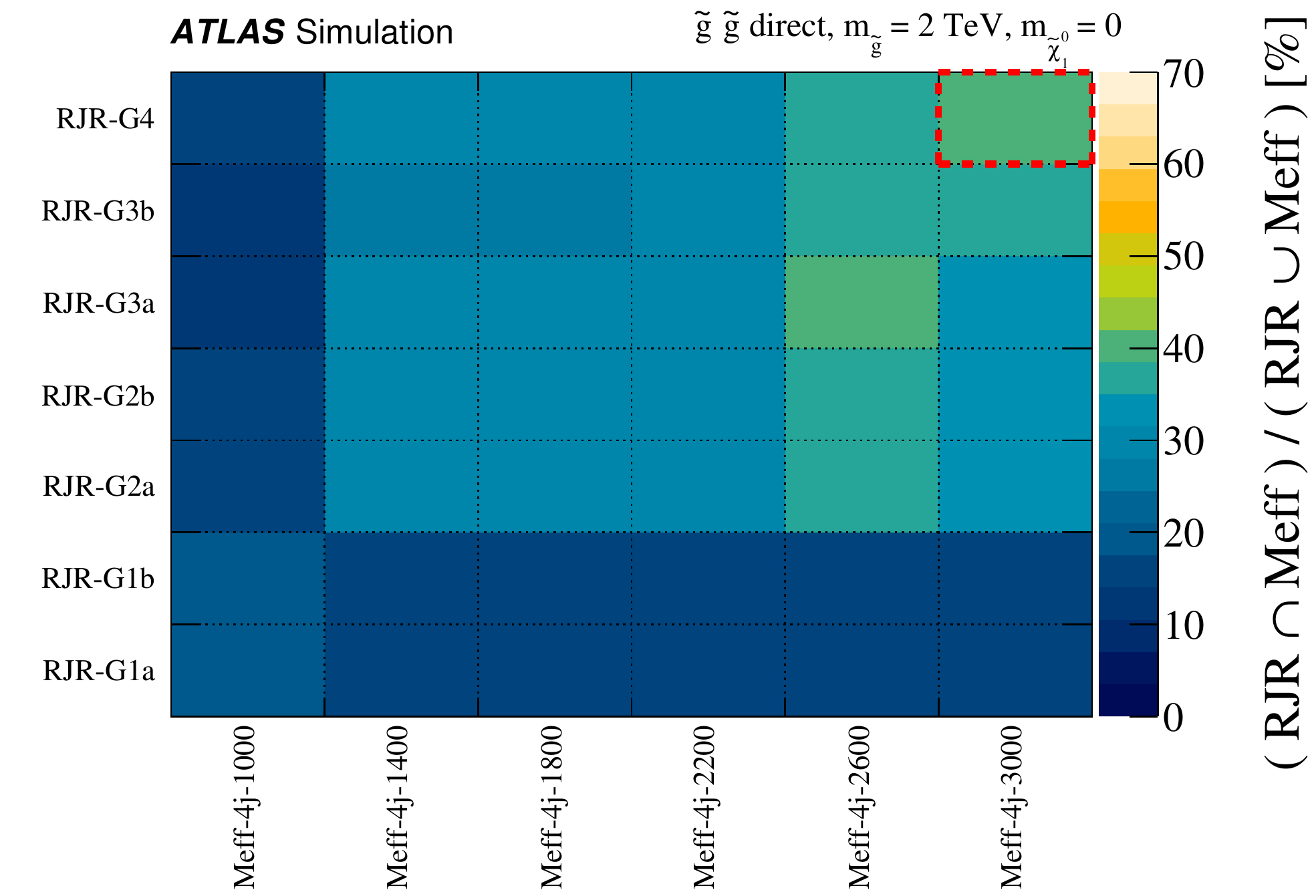}} 
\subfigure[]{\includegraphics[width=0.48\textwidth]{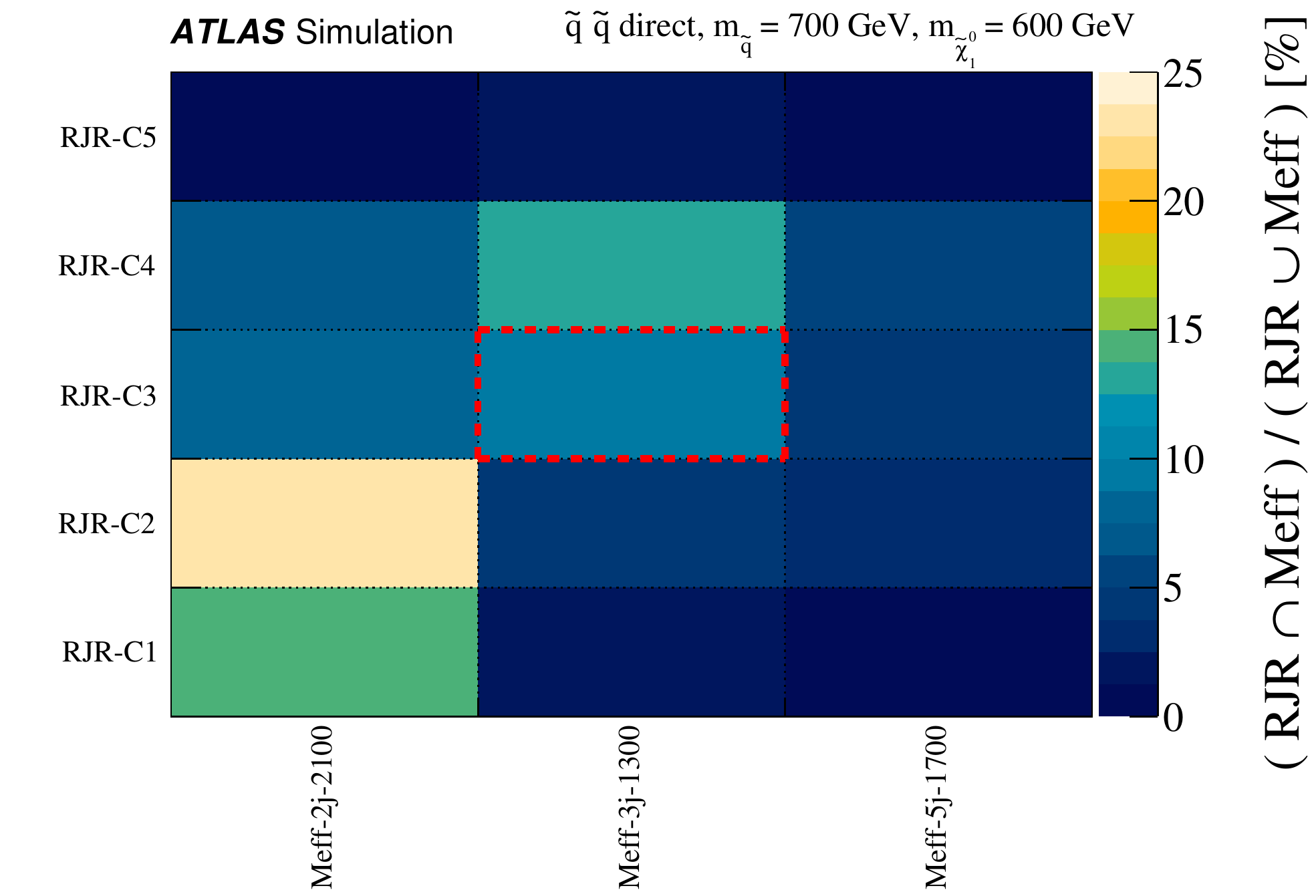}} 
\subfigure[]{\includegraphics[width=0.48\textwidth]{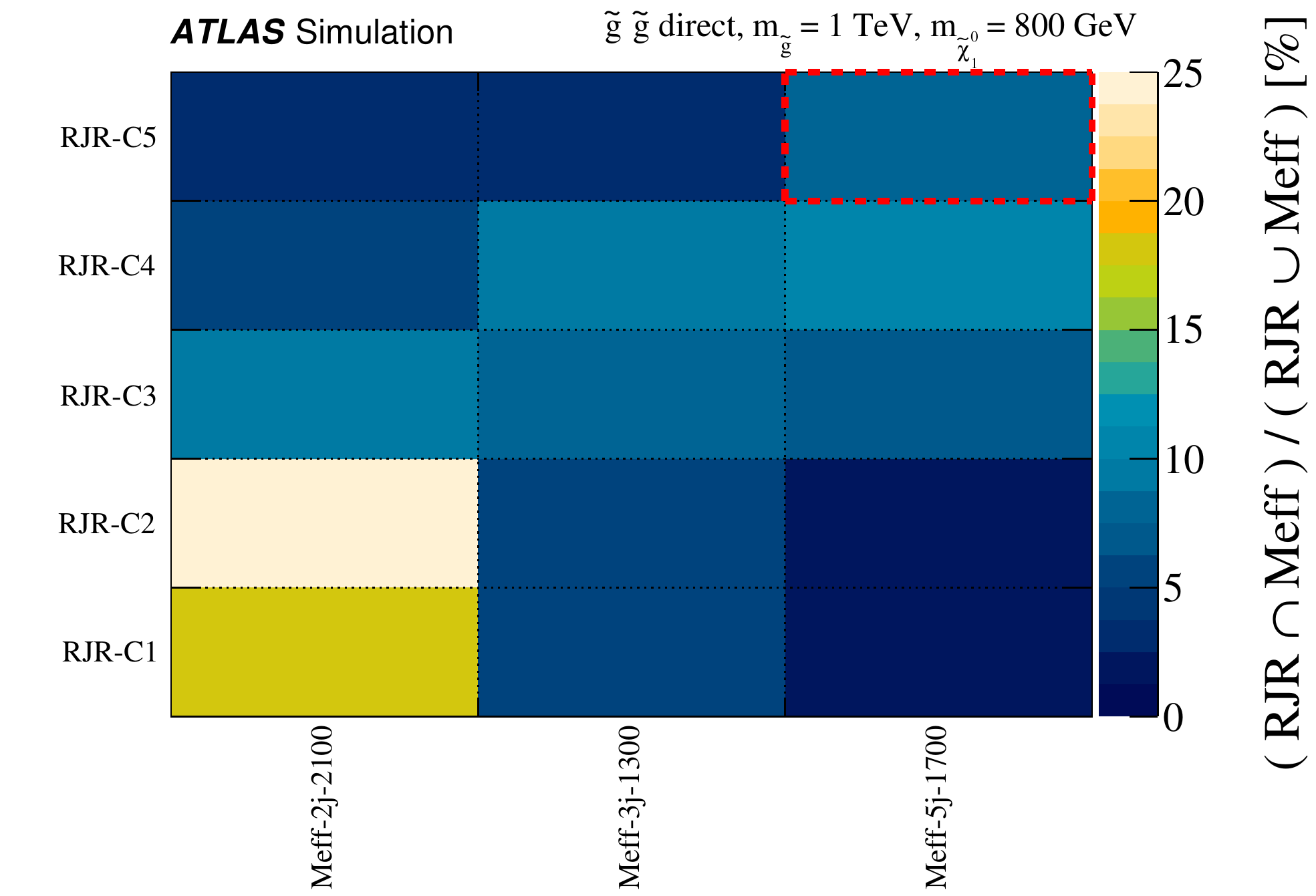}} 
\end{center}
\vspace*{-0.03\textheight}\caption{\label{fig:Overlap2D_signal} Fractional overlap of simulated squark and gluino pair events selected in Meff-based and RJR-based SRs. For these signals each squark (gluino) decays to one (two) quarks and a \ninoone. Figures correspond to simulated signals with (a) $m_{\squark} = 1.5 ~\TeV$, $m_{\ninoone} = 0$, (b) $m_{\gluino} = 2 ~\TeV$, $m_{\ninoone} = 0$, (c) $m_{\squark} = 700 ~\GeV$, $m_{\ninoone} = 600 ~\GeV$, and  (d) $m_{\gluino} = 1 ~\TeV$, $m_{\ninoone} = 800 ~\GeV$. These selected signal points are near the limit of expected sensitivity for these SRs. Meff-based SRs are listed along the $x$-axis with RJR-based regions on the $y$-axis. The intersection events falling in each pair of regions, normalized by the union, is shown on the $z$-axis. The Meff- and RJR-based SRs best suited to each signal, respectively, are indicated by dashed red boxes.
}
\end{figure}

\section{Background estimation}
\label{sec:background}

Standard Model background processes contribute to the event counts in
the signal regions. The largest backgrounds in both searches presented here are: $Z+$jets, $W+$jets, top quark
pair, single top quark, diboson and multi-jet production. Non-collision backgrounds are negligible. 

Generally, the largest background results from an irreducible component of $Z$+jets events in which $Z\to\nu\bar\nu$ decays generate large $\ourmagptmiss$. Similarly, most of the $W$+jets background is composed of $W\to \tau\nu$ events in which the $\tau$-lepton decays to hadrons, with additional contributions from $W\to e\nu, \mu\nu$ events in which no baseline electron or muon is reconstructed, with $\ourmagptmiss$ due to neutrinos. Top quark pair production, followed by semileptonic decays, in particular $t \bar t \to b \bar b \tau \nu q q' $ (with the $\tau$-lepton decaying to hadrons), as well as single-top-quark events,
can also generate large $\ourmagptmiss$ and satisfy the jet
and lepton-veto requirements. Each of these primary backgrounds is estimated using dedicated control regions, as described in the following section, while diboson production is estimated with MC simulated data normalized using NLO cross-section predictions, as described in Section~\ref{sec:montecarlo}. 

The multi-jet background in the signal regions is due to missing transverse momentum from misreconstruction of jet energies in the calorimeters,
jets misidentified as electrons, jets lost due to the JVT requirement, 
as well as neutrinos from semileptonic decays of heavy-flavor hadrons. After applying the requirements based on $\ourdeltaphifull$ and $\ourmagptmiss/\meff(N_{j})$ in the Meff-based search, or $\Delta_{\textrm{QCD}}$,  $p_{\textrm{T~j2}}^{~PP}/H_{\textrm T~2,1}^{~PP}$ and $\ourdeltaphishort(\textrm{jet},\ourvecptmiss)_\mathrm{min}$ in the RJR-based search, as indicated in Tables~\ref{tab:srdefs} and \ref{tab:RJsrdefs}, the remaining multi-jet background is negligible.

\subsection{Control regions}

In order to estimate the expected background yields, control regions are defined for each of the signal regions in four different final states. 
In the Meff-based search, each SR has its own set of four CRs, while in the RJR-based search, a common set of CRs is used for all SRs in every targeted signal category (RJR-S, RJR-G or RJR-C). The CR selections are optimized to maintain adequate statistical precision while minimizing the systematic uncertainties arising from the extrapolation of the CR event yield to estimate the background in the SR. The latter is addressed through the fact that the jet $\pt$ thresholds and $\meff$(incl.) selections in the CRs are the same as used for the SR in the Meff-based search. In the RJR-based search, requirements on discriminating variables are chosen to match those used in the SRs as closely as possible. The basic CR definitions in both searches are listed in Table~\ref{tab:crdefs}.

\begin{table}[H]
\scriptsize
\begin{center}\renewcommand\arraystretch{1.2}
\begin{tabular}{| l | c | c | c | c |}
\hline
CR                                & SR background                       &  CR process                            & CR selection                                                              & CR selection   \\ 
                                     &                                                 &                                                 & (Meff-based)                                                              &  (RJR-based)  \\ \hline \hline
Meff/RJR-CR$\gamma$ & $Z(\to\nu\bar\nu)$+jets            & $\gamma$+jets                       & Isolated photon                                                                                 &  Isolated photon      \\ \hline
Meff/RJR-CRQ               & Multi-jet                                    & Multi-jet                                   & SR with reversed requirements on                                                   &  $\Delta_{\textrm{QCD}} < 0$     \\ 
                                      &                                               &                                                    &    (i) $\ourdeltaphifull$     and (ii) $\met/\meff(N_{\textrm  j})$                  &  reversed requirement on    \\
                                    &                                                  &                                                   &or $\met/\sqrt{H_{\textrm  T}}$                                                                   &   $H_{\textrm 1,1}^{~PP} $ (RJR-S/G)   \\ 
                                                                        &                                                  &                                                   &                                                 &   or $R_{\textrm{ISR}} <$ 0.5 (RJR-C)   \\ \hline
Meff/RJR-CRW              & $W(\to\ell\nu)$+jets                  & $W(\to\ell\nu)$+jets                 & \multicolumn{2}{c|}{30~\GeV $<m_{\textrm  T}(\ell,\met) < 100$~\GeV, $b$-veto     }              \\ 
\hline
Meff/RJR-CRT               & $t\bar{t}$(+EW) and single top & $t\bar{t}\to b\bar{b}qq'\ell\nu$ & \multicolumn{2}{c|}{30~\GeV $<m_{\textrm  T}(\ell,\met) < 100$~\GeV, $b$-tag      }           \\ 
\hline
\end{tabular}
\caption{\label{tab:crdefs} Summary of CRs for the Meff-based and RJR-based searches. Also listed are the main targeted SR backgrounds in each case, the process used to model the background, and the main CR requirement(s) used to select this process. The transverse momenta of high-purity leptons (photons) used to select CR events must exceed 27 (150)~\GeV. The jet $\pt$ thresholds and $\meff({\textrm{incl.}})$ selections match those used in the corresponding SRs of the Meff-based search. For the RJR-based search, selections are based on the discriminating variables used in the SRs, as described in the text. }
\end{center}
\end{table}

The $\gamma$+jets region in both searches (labeled as Meff/RJR-CR$\gamma$ in Table~\ref{tab:crdefs}) is used to estimate the contribution of 
$Z(\to\nu\bar\nu)$+jets background events to each SR by selecting a sample of $\gamma$+jets events with $\pt(\gamma)>150~\GeV$ and then treating the reconstructed photon as invisible in the $\met$ calculation. For $\pt(\gamma)$ significantly larger than $m_Z$ the kinematic properties of such events strongly resemble those of $Z$+jets events \cite{Aad:2012fqa}. 
In order to reduce the theoretical uncertainties associated with the $Z/\gamma^*$+jets background predictions in SRs arising from the use of LO $\gamma$+jets cross-sections, a correction factor is applied to 
the Meff/RJR-CR$\gamma$ events as a function of the requirement on the number of jets. 
This correction factor,  $\kappa$, ranges from 1.41 to 2.26 for two to six jets, and is determined by comparing Meff-CR$\gamma$ observations with those in a highly populated auxiliary control region 
defined by selecting events with two electrons or muons for which the invariant mass lies within 25~\GeV\ of the mass of the $Z$ boson, 
satisfying $\met > 250~\GeV$, $\MET/\sqrt{H_{\textrm  T}} > 14~\GeV^{1/2}$ 
and $\meff({\textrm{incl.}}) > 1200~\GeV$ where two leptons are treated as contributing to $\met$.

The $W$ and top regions in both searches (labeled as Meff/RJR-CRW and Meff/RJR-CRT in Table~\ref{tab:crdefs}) aim to select samples rich in $W(\to \ell \nu)$+jets and semileptonic $t\bar{t}$ background events, respectively. They use events with one high-purity lepton with  $p_{\textrm  T}$ > 27 \GeV\, and differ in their number of $b$-jets 
(zero or $\geq 1$, respectively). In both searches, the requirement on
the transverse mass $m_{\textrm  T}$ formed by the $E_{\textrm  T}^{\mathrm{miss}}$ and a selected lepton is applied, as indicated in
Table~\ref{tab:crdefs}. The lepton is treated as a jet with the same
momentum to model background events in which a hadronically decaying
$\tau$-lepton is produced.
{\color{black} This estimation procedure is used to try to get a better idea of the
$W(\to \ell \nu)$+jets and $t\bar{t}$ cross
section in a restricted kinematic phase space, by normalizing the MC
to the data for the electron and muon channels, respectively.}
{\color{black}The propagation of the number of background events from the control
region to the signal region is done purely by Monte Carlo which takes
into account the impact of all the differences in selection criteria
between the control and signal regions.}
The Meff-CRW and Meff-CRT criteria omit the SR selection requirements on $|\eta_{\textrm{jet}}|$, $\ourdeltaphifull$ and aplanarity for all SRs, while for the SRs requiring $\meff({\textrm{incl.}}) > 2200~\GeV$ the requirements on $\met/\meff(N_{\textrm  j})$ are not applied. 
This is done in order to increase the number of CR data events without significantly increasing the theoretical uncertainties associated with the CR-to-SR extrapolation in the background estimation procedure. 

The multi-jet background in both searches is estimated using a data-driven technique \cite{Aad:2012fqa}, which applies a resolution function to well-measured multi-jet events in order to estimate the impact of jet energy mismeasurement and heavy-flavor semileptonic decays on \met{} and other variables. The resolution function of jets is initially estimated from MC simulation by matching `truth' jets reconstructed from generator-level particles including muons and neutrinos to detector-level jets
with $\Delta R<0.1$ in multi-jet samples, and then is modified to agree with data in dedicated samples to
measure the resolution function.
The Meff-CRQ region uses reversed selection requirements on $\ourdeltaphifull$ and on $\met/\meff(N_{\textrm  j})$ (or $\met/\sqrt{H_{\textrm  T}}$ where appropriate) to produce samples enriched in multi-jet background events. 

In the RJR-based search, all CRs corresponding to RJR-S (RJR-G) SRs are required to satisfy $H_{\textrm 1,1}^{~PP} >$ 800 (700) \GeV\,. 
Additionally, $H_{\textrm T~2,1}^{~PP} > 1000 ~\GeV$ (for RJR-S), $H_{\textrm T~4,1}^{~PP} > 1200 ~\GeV$ (for RJR-G) and $M_{\mathrm{T}~S} > 0$ (for RJR-C) are 
required for RJR-CRW, RJR-CRT and RJR-CRQ regions. 
In RJR-CRW and RJR-CRT, the requirements on all the other variables used for the RJR-SR selections are chosen such that the loosest value in the SR category (RJR-S, RJR-G or RJR-C) indicated in Table~\ref{tab:RJsrdefs} is used. No requirement on $p_{PP,~z}^{\textrm{~lab}} / \left( p_{PP,~z}^{\textrm{~lab}}+H_{\mathrm{T}~N,1}^{~PP}\right)$ is used for the RJR-CRQ selections in all RJR-SRs, where $N=2$ or $4$. 

\begin{figure}[t]
  \begin{center}
    \subfigure[]{\raisebox{0.05\textheight}{\includegraphics[width=0.51\textwidth]{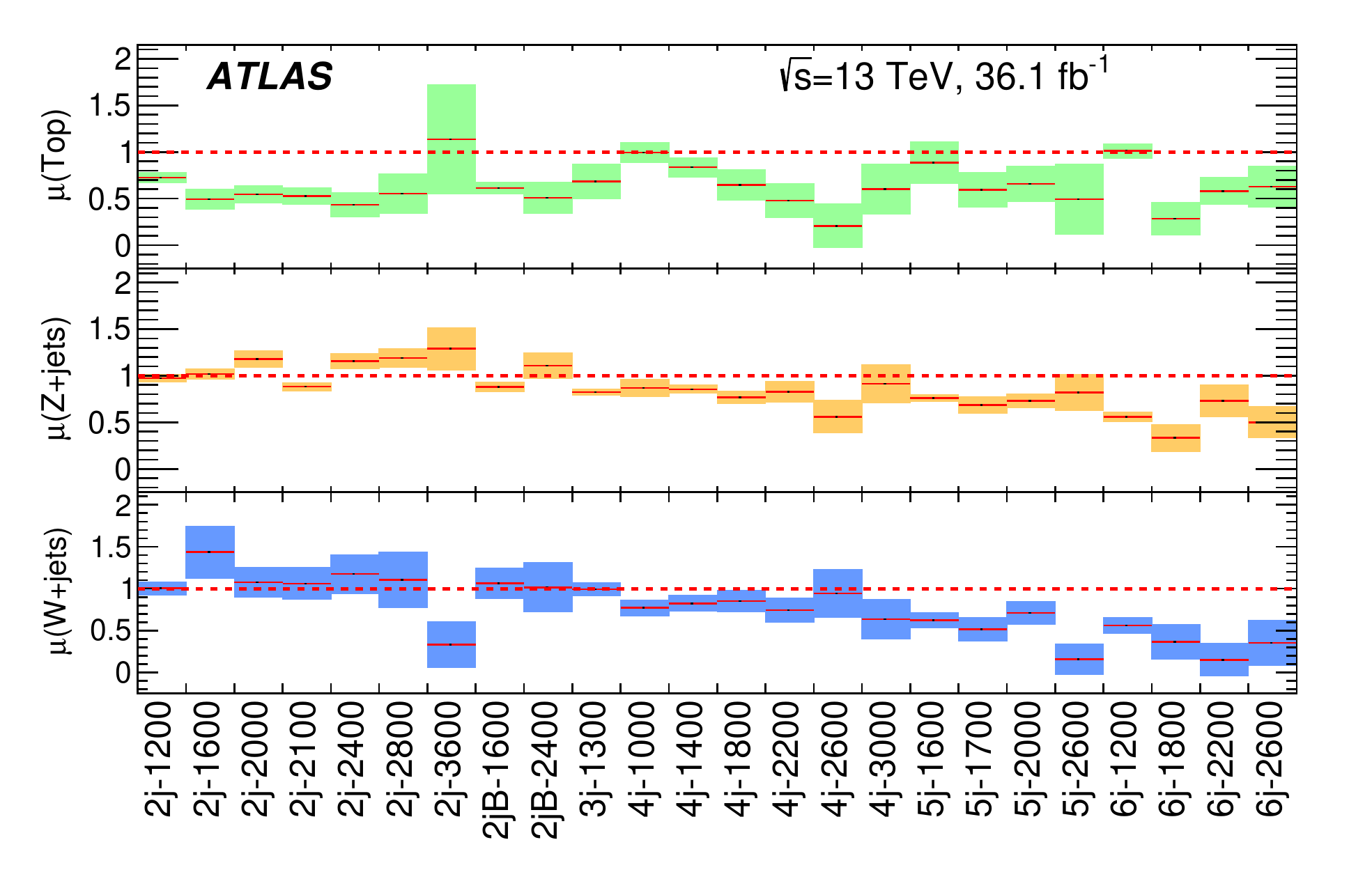}}}
    \subfigure[]{\includegraphics[width=0.48\textwidth]{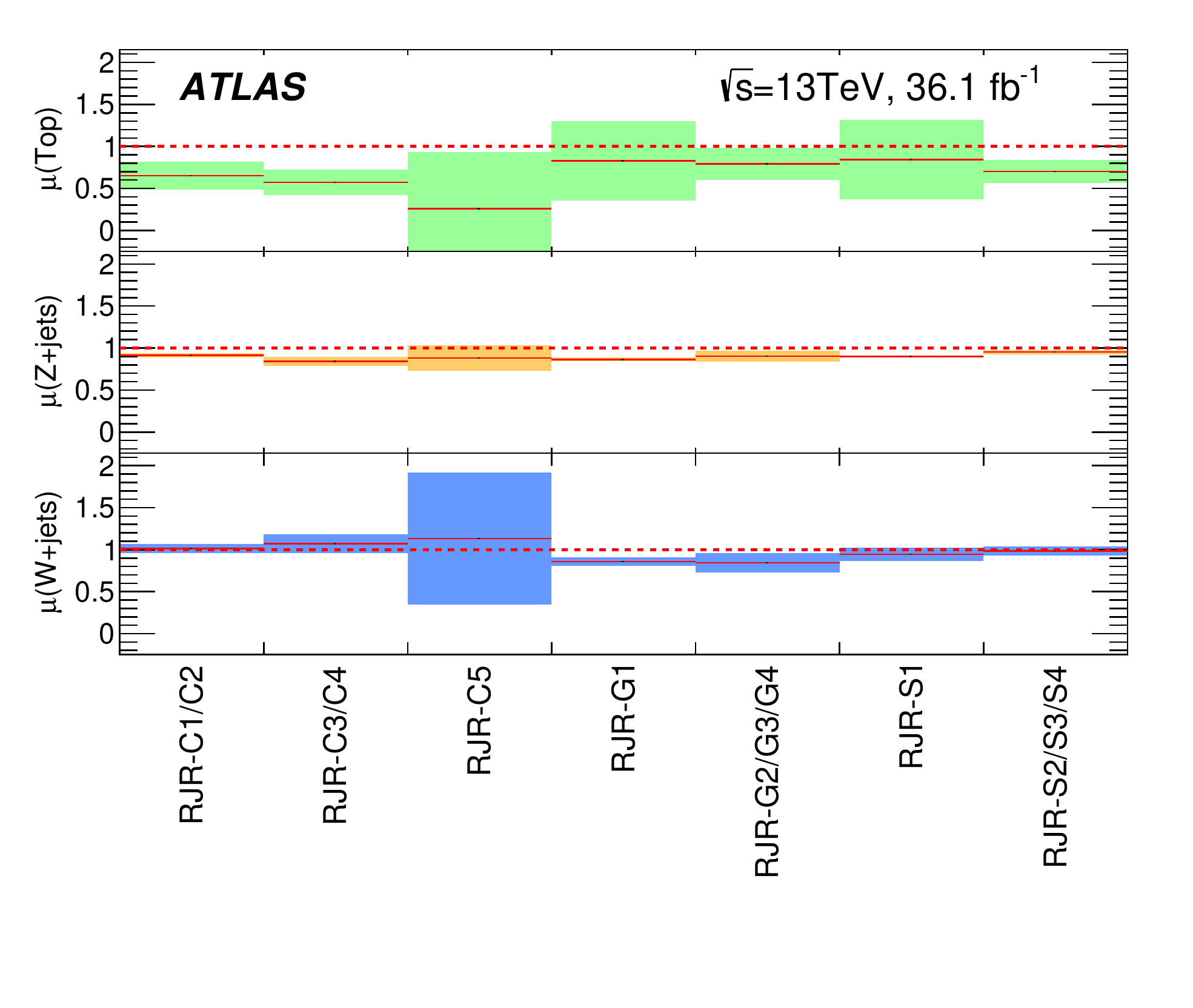}}
  \end{center}
  \vspace{-0.03\textheight}
  \caption{\label{fig:summaryMu} Fitted normalization factor per process as a function of the channel considered in the (a) Meff-based and (b) RJR-based searches. The dashed horizontal lines at 1 correspond to pure MC estimates with the vertical size of the colored regions corresponding to the total uncertainty in each background source.}
\end{figure}

The normalization factors determined from the background-only fits in each CR for each background process are shown in Figure~\ref{fig:summaryMu}.
Some trends in these factors are observed, with the normalization factors for top background becoming smaller with increasingly tight $m_{\textrm{eff}}$ requirements for the Meff-based regions. Similarly, the measured top normalization factors decrease with increasingly tight $M_{\mathrm{T}~S}$ and $N_{\textrm{jet}}^{~V}$ requirements in the RJR-based search. This behavior follows from the simulated top MC samples exhibiting generally harder kinematics than observed in data, as seen in Figures~\ref{fig:cr4j_Meff}(d) and~\ref{fig:CRC1}(d). The normalization factors for $W$+jets and $Z$+jets processes are generally stable with changing kinematic selections but with a clear indication that they become systematically smaller with increasingly strict requirements on the jet multiplicity. This is due to the MC simulation predicting jet multiplicities higher than observed in data events.

Example $ \meff({\textrm{incl.}})$ distributions in control regions associated with Meff-4j-2200 selections are shown in Figure~\ref{fig:cr4j_Meff}. Figure~\ref{fig:CRC1} shows the  $p_{\mathrm{T}~S}^{\textrm{~CM}}$ discriminating variable distributions in control regions corresponding to RJR-C1 signal region selections. 
In all CRs, the data distributions are consistent with the MC background prediction within uncertainties after normalizing the dominant process in each CR. 

\clearpage

\begin{figure}[t]
\begin{center}
\subfigure[]{\includegraphics[width=0.42\textwidth]{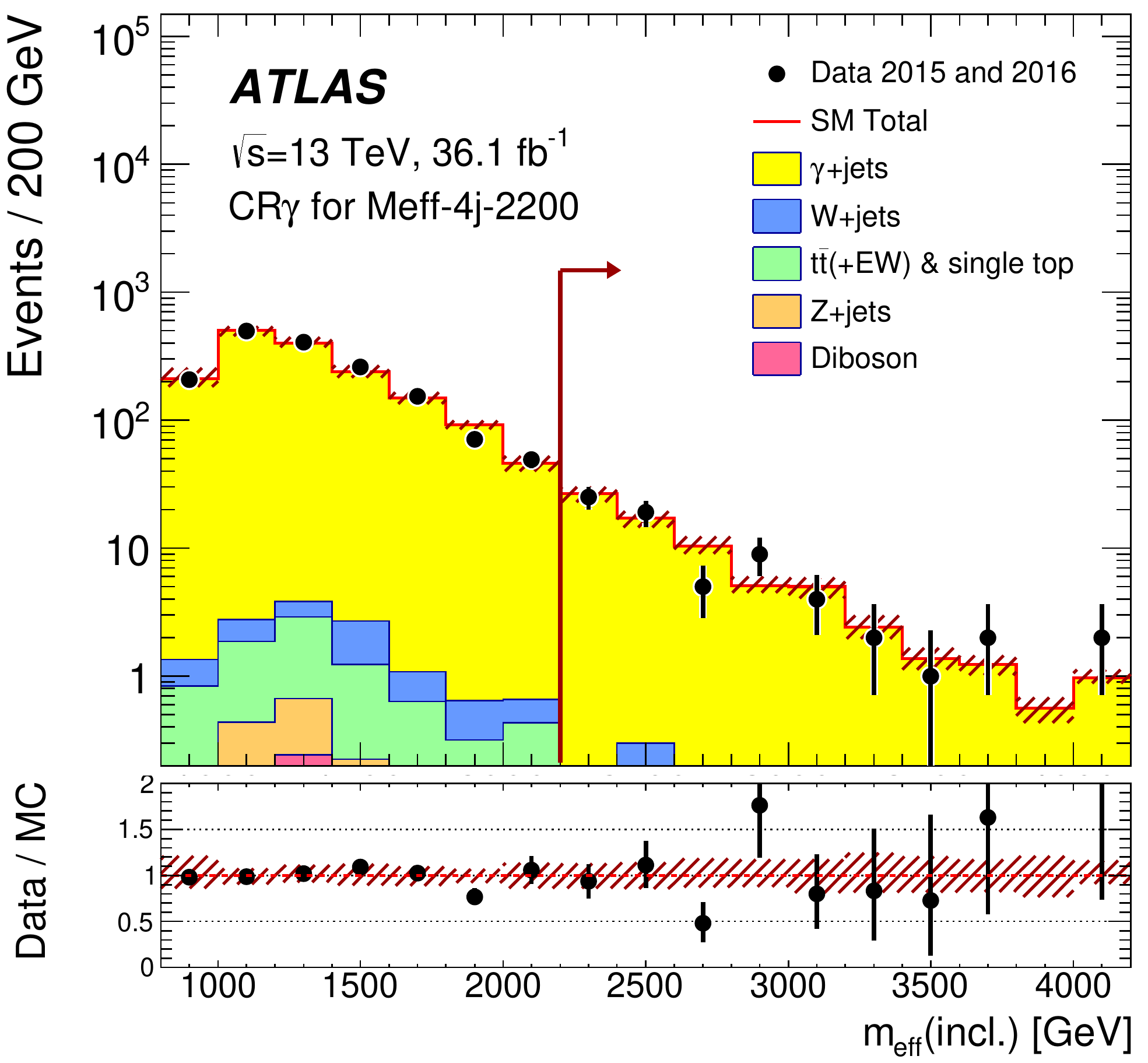}}
\subfigure[]{\includegraphics[width=0.42\textwidth]{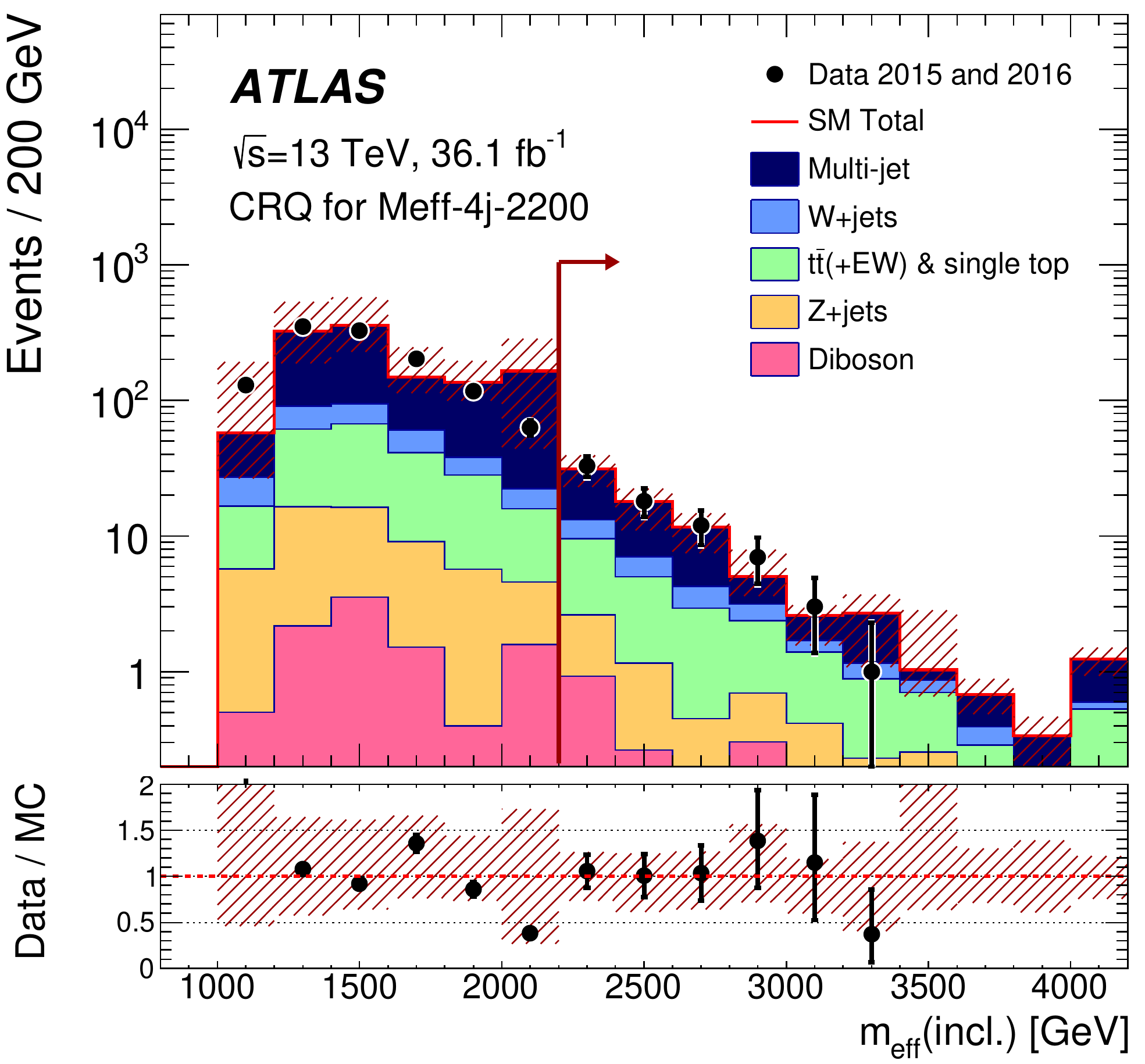}}
\subfigure[]{\includegraphics[width=0.42\textwidth]{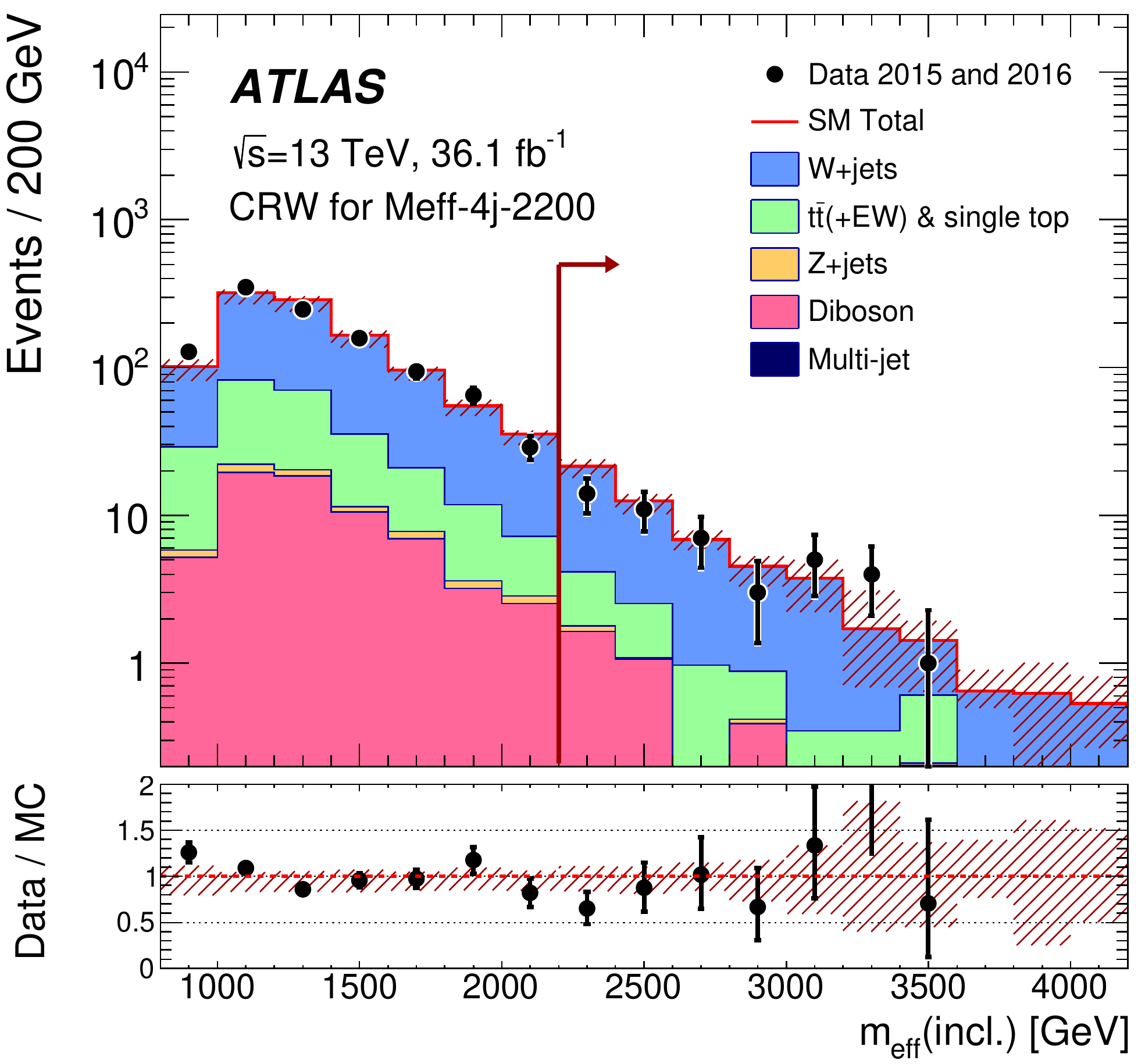}}
\subfigure[]{\includegraphics[width=0.42\textwidth]{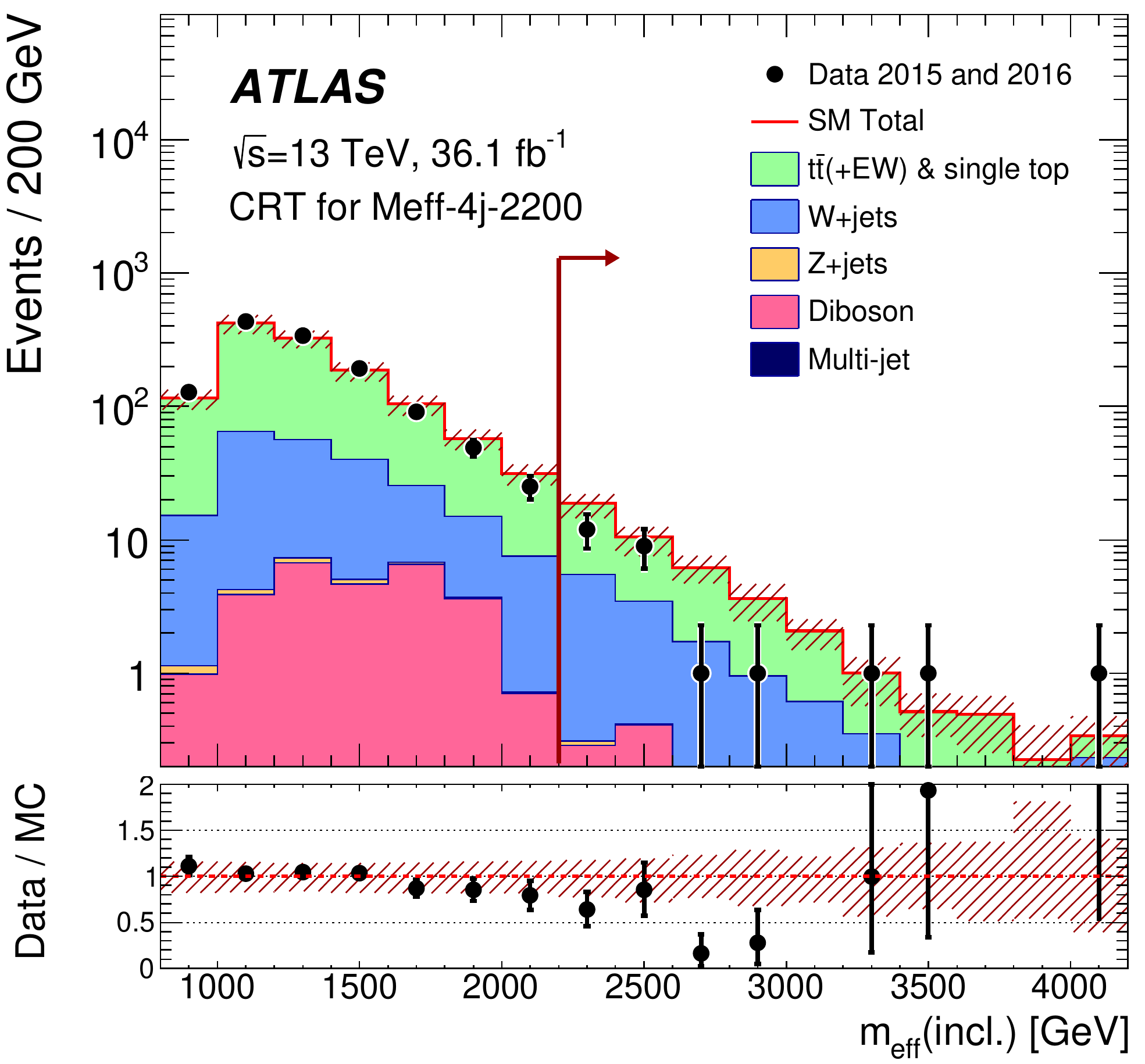}}
\end{center}
\caption{\label{fig:cr4j_Meff}Observed $\meff({\textrm{incl.}})$ distributions in control regions (a) Meff-CR$\gamma$, (b) Meff-CRQ, (c) Meff-CRW and (d) Meff-CRT after applying the Meff-4j-2200 selection requirements listed in Table~\ref{tab:srdefs}, except those on the plotted variable. No selection requirements on $\ourdeltaphifull$ are applied in Meff-CRW and Meff-CRT regions. The arrows indicate the values at which the requirements on $\meff({\textrm{incl.}})$ are applied. The histograms show the MC background predictions, normalized using cross-section times integrated luminosity and the dominant process in each CR is normalized using data.
 In the case of the $\gamma$+jets background, a $\kappa$ factor described in the text is applied. The last bin includes overflow events. The hatched (red) error bands indicate the combined experimental, MC statistical and theoretical modeling uncertainties.
}
\end{figure}

\clearpage

\begin{figure}[t]
\begin{center}
\subfigure[]{\includegraphics[width=0.42\textwidth]{{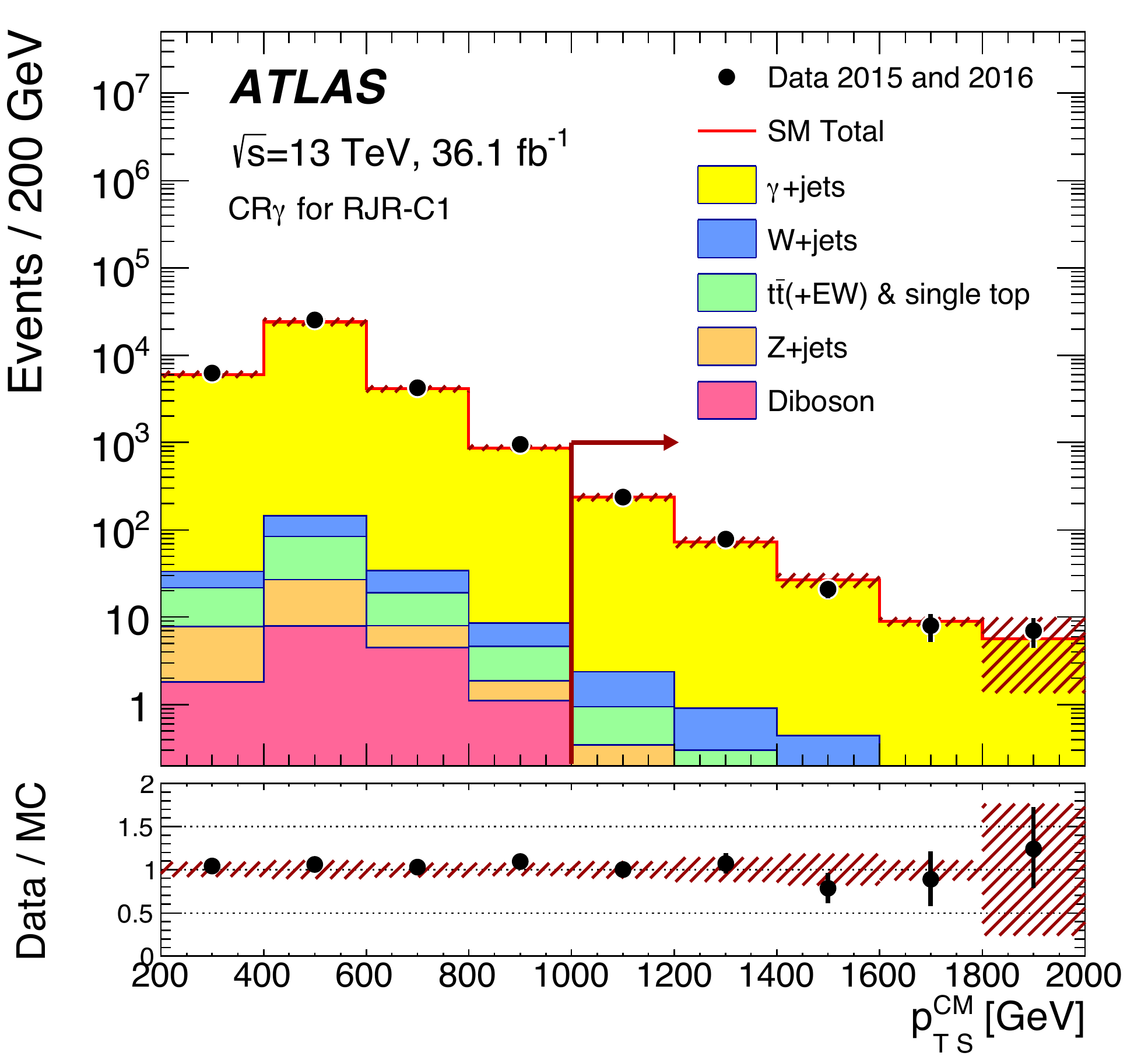}}} 
\subfigure[]{\includegraphics[width=0.42\textwidth]{{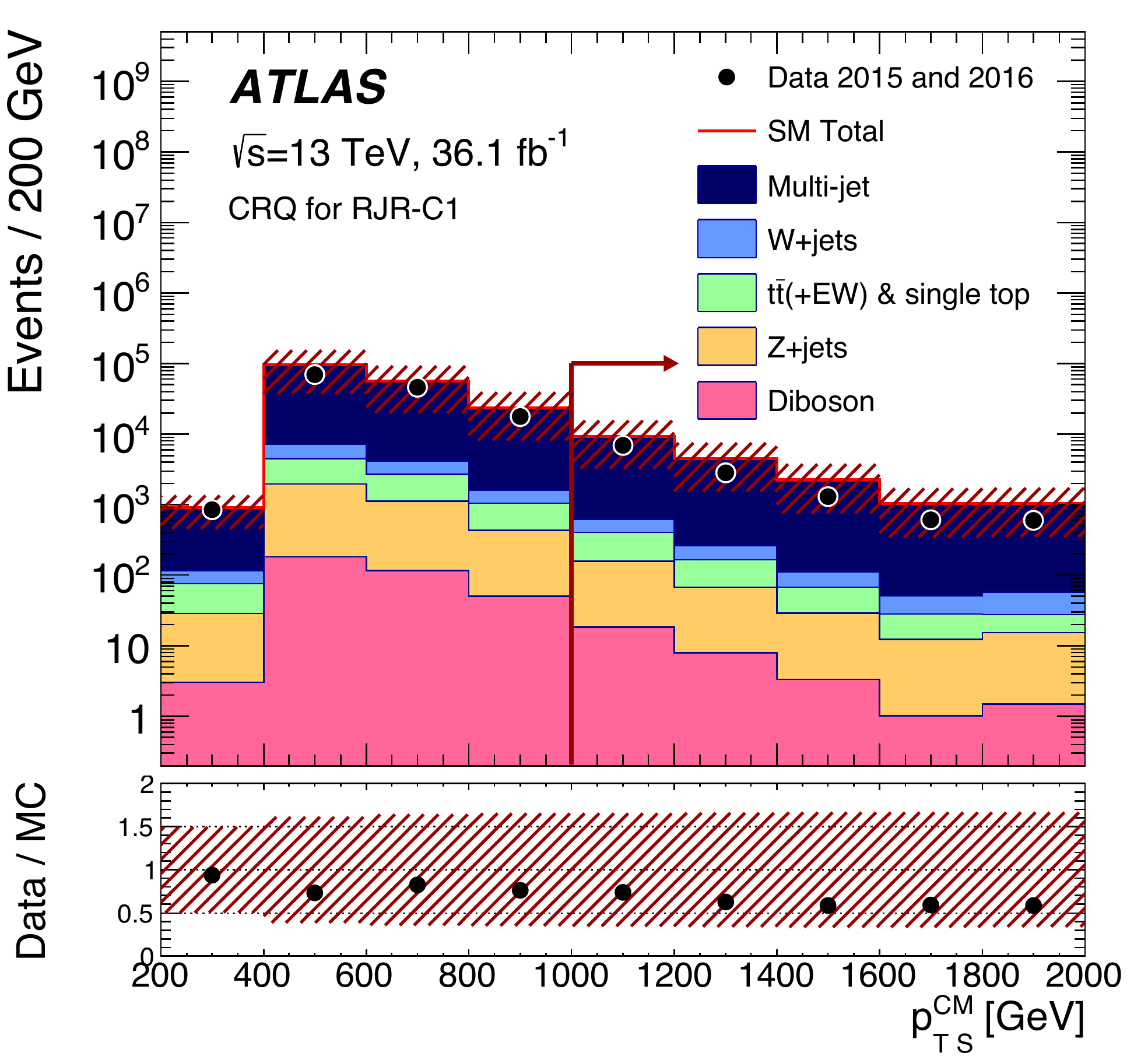}}} 
\subfigure[]{\includegraphics[width=0.42\textwidth]{{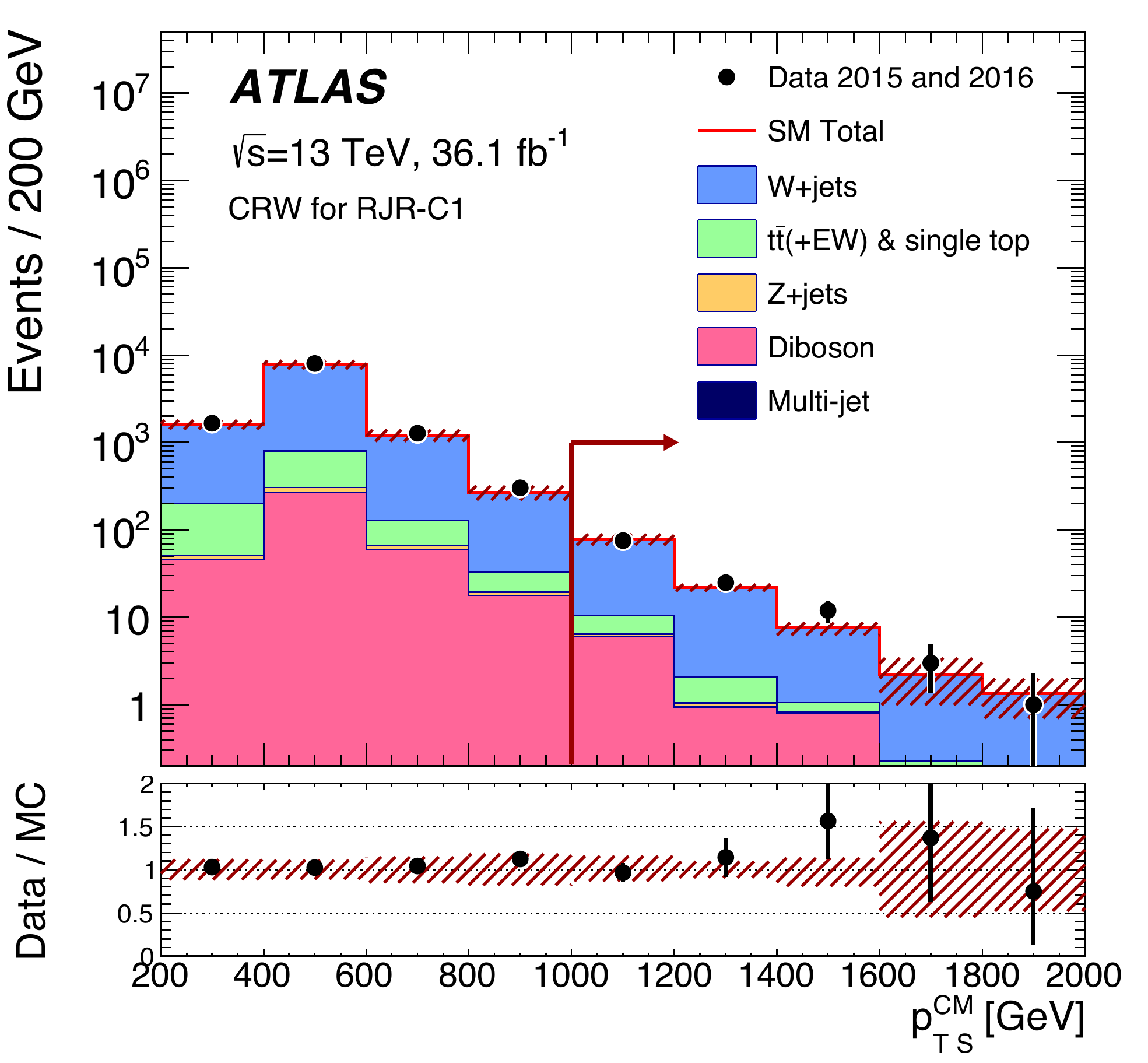}}} 
\subfigure[]{\includegraphics[width=0.42\textwidth]{{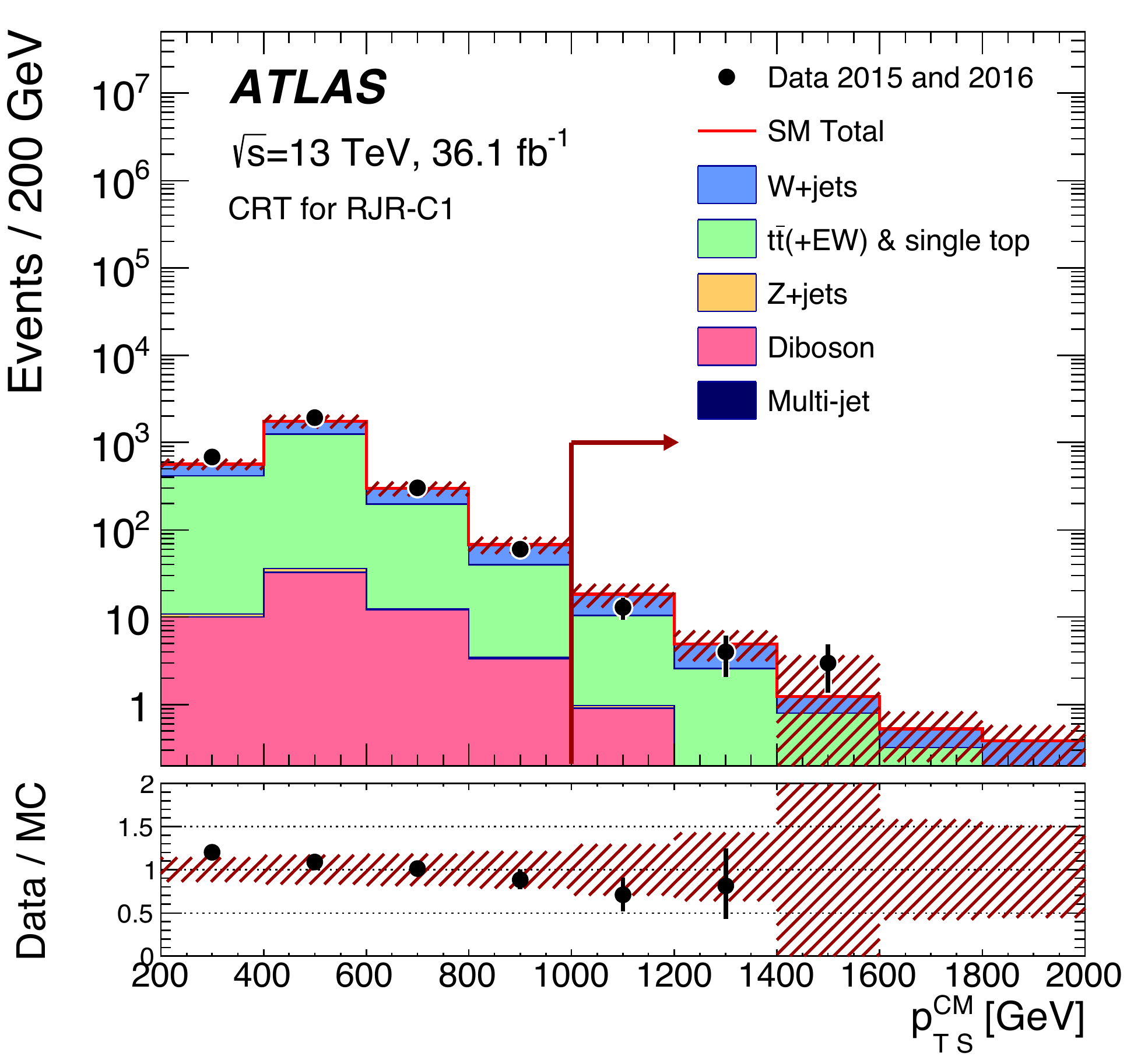}}} 
\end{center}
\caption{\label{fig:CRC1}Observed  $p_{\mathrm{T}~S}^{\textrm{~CM}}$ distribution in control regions (a) RJR-CR$\gamma$, (b) RJR-CRQ, (c) RJR-CRW and (d) RJR-CRT  after selecting events for the corresponding control regions as explained in the text for RJR-C1 region and after applying all selection requirements except those on the plotted variable. The arrows indicate the values at which the requirements are applied. The histograms show the MC background predictions, normalized using cross-section times integrated luminosity and the dominant process in each CR is normalized using data. 
In the case of $\gamma$+jets background, a $\kappa$ factor described in the text is applied. The last bin includes overflow events. The hatched (red) error bands indicate the combined experimental, MC statistical and theoretical modeling uncertainties.
}
\end{figure}

\subsection{Validation regions}

The background estimation procedure is validated by comparing the numbers of events observed in the VRs to the corresponding SM background predictions obtained from the background-only fits. Several VRs are defined in both searches, with requirements distinct from those used in the CRs and that maintain low expected signal contamination. Like the CRs, the majority of the VRs are defined in final states with leptons and photons, allowing the different expected background contributions to the SRs to be validated almost separately with high-purity selections.

The Meff/RJR-CR$\gamma$ estimates of the $Z(\to \nu\bar{\nu})$+jets background are validated using samples of $Z(\to\ell\bar\ell)$+jets events selected by requiring high-purity lepton pairs of opposite sign and identical flavor for which the dilepton invariant mass lies within 25~\GeV\ of the $Z$ boson mass (Meff/RJR-VRZ). In Meff/RJR-VRZ regions, the leptons are treated as contributing to $\met$. Additional VRs designed to validate the $Z(\to\nu\bar{\nu})$+jets estimate in the RJR-based search are also used: the VRZc region, which selects events with no leptons but inverts the $\Delta\phi_{\mathrm{ISR},~I}$ requirement of the SR selection (Table~\ref{tab:RJsrdefs}) and VRZca, which further loosens some other criteria to match the RJR-CRW and RJR-CRT regions. The VRZc regions have a purity of $Z(\to \nu\bar{\nu})$+jets of 50\%--70\%. In order to increase yields in the dilepton final state RJR-VRZ regions, two additional regions, RJR-VRZa and RJR-VRZb are constructed with $H_{\textrm{1,1}}^{~PP}$ and $H_{\textrm T~2,1}^{~PP}$ (or $H_{\textrm T~4,1}^{~PP}$ where appropriate) loosened, respectively, relative to the values used for the RJR-CRW and RJR-CRT regions.  

The Meff-CRW and Meff-CRT estimates of the $W$+jets and top quark
background are validated with the same Meff-CRW and Meff-CRT
selections, but applying the $\ourdeltaphifull$ requirement and
treating the lepton as a jet (Meff-VRW, Meff-VRT). To further validate
the extrapolation over the $\ourdeltaphifull$ and aplanarity variables
from the dedicated $W$+jets and top quark CRs to the SRs, additional
validation regions Meff-VRW$\Delta\Phi$ and Meff-VRT$\Delta\Phi$ as
well as Meff-VRWAp and Meff-VRTAp are defined by relaxing
$\ourdeltaphifull$ and aplanarity requirements, respectively, relative
to Meff-VRW and Meff-VRT.

Similarly, the RJR-CRW and RJR-CRT estimates of the $W$+jets and top quark backgrounds are validated using the same selections as for the corresponding CRs, except that the requirements on $H_{\textrm 1,1}^{~PP}$ and $M_{\mathrm{T}~S}$ (RJR-VRWa, RJR-VRTa) or $H_{\textrm T~2,1}^{~PP}$ and $H_{\textrm T~4,1}^{~PP}$ (RJR-VRWb, RJR-VRTb) are omitted. Two additional VRs that require the presence of a high-purity lepton and either veto (RJR-VRW) or require the presence of at least one $b$-jet (RJR-VRT), and require no additional SR selection criteria, are also used in the analysis. 

The Meff-CRQ estimates of the multi-jet background are validated with VRs for which the Meff-CRQ selection is applied, but with the SR $\met/\meff(N_{\textrm j})$ ($\met/\sqrt{H_{\textrm  T}}$) requirement reinstated (Meff-VRQa), or with a requirement of an intermediate value of $\ourdeltaphifull$ applied (Meff-VRQb). The RJR-VRQ regions use the same selection as the corresponding RJR-CRQ, except that the requirements on $H_{\textrm 1,1}^{~PP}$, $H_{\textrm T~2,1}^{~PP}$ (or $H_{\textrm T~4,1}^{~PP}$ where appropriate) and $M_{\mathrm{T}~S}$ are omitted depending on the region. Additional VRs with inverted $\Delta_{\textrm{QCD}}$ (RJR-VRQa), $H_{\textrm 1,1}^{~PP}$ (RJR-VRQb) for RJR-S and RJR-G signal regions, and with 0.5 $ < R_{\textrm{ISR}} < $ SR requirement (RJR-VRQc) for the RJR-C region (Table~\ref{tab:RJsrdefs}), are also used. 

\begin{figure}[t]
\begin{center}
\subfigure[]{\includegraphics[width=0.49\textwidth]{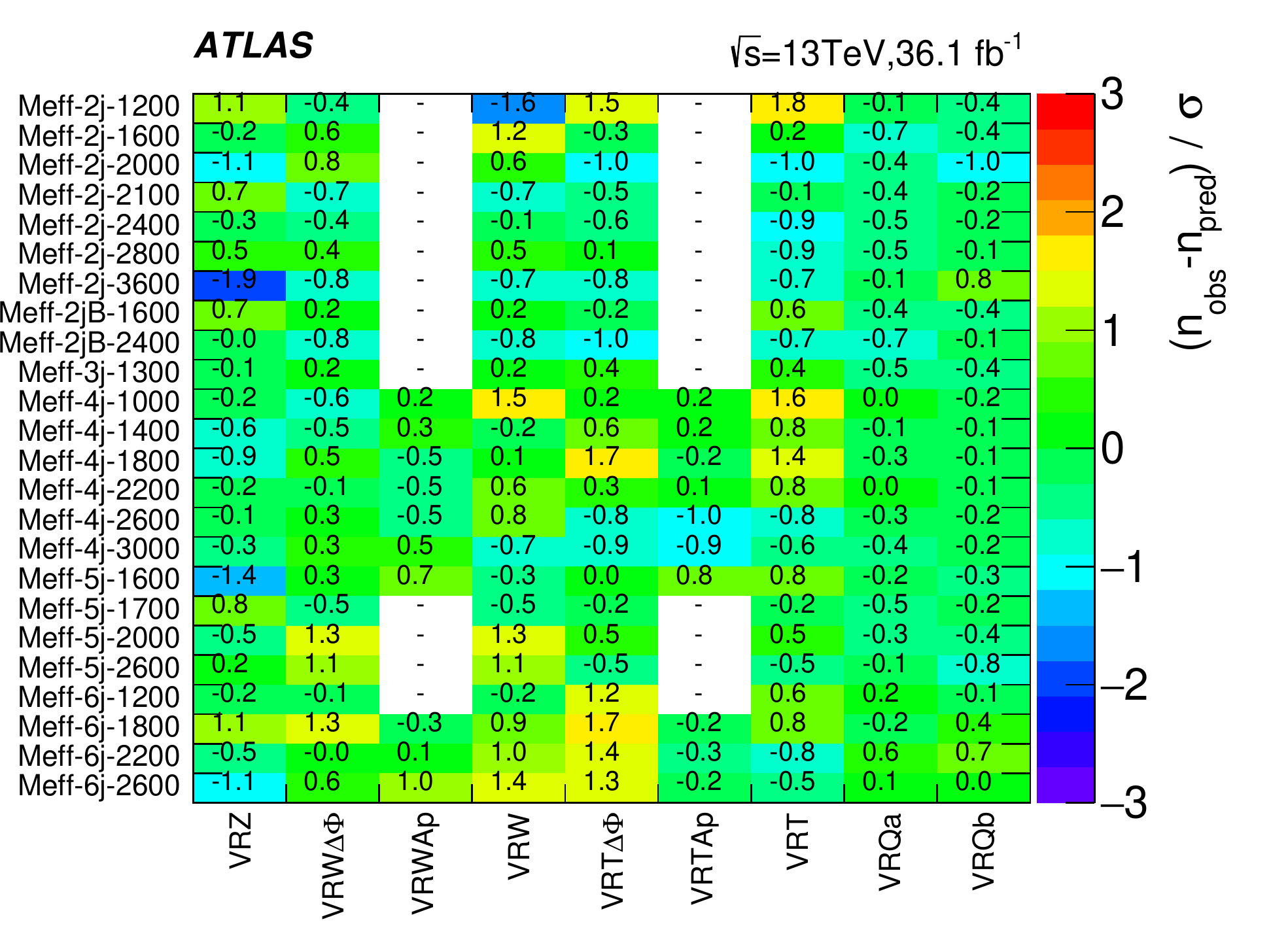}}
\subfigure[]{\includegraphics[width=0.49\textwidth]{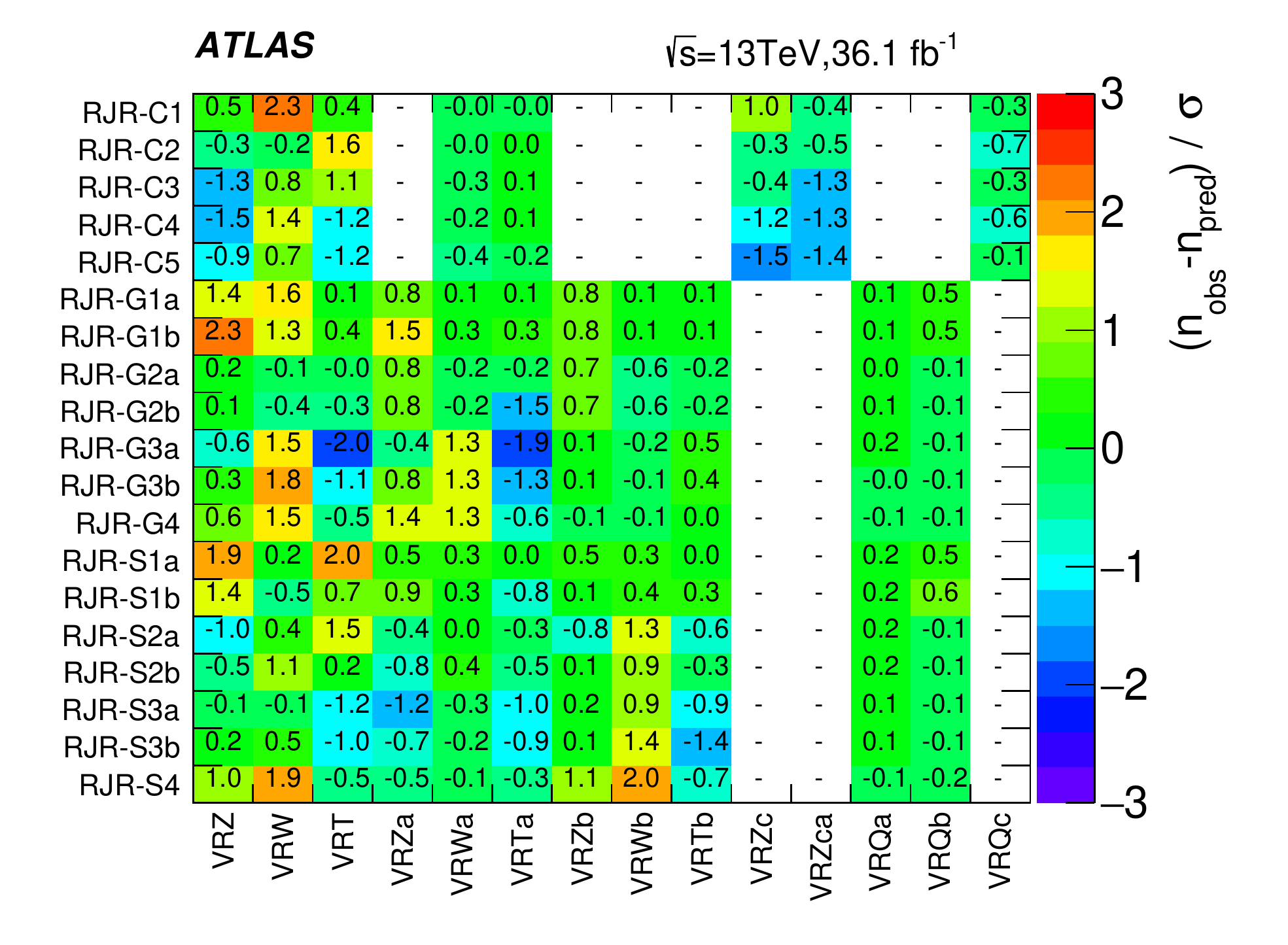}}
\end{center}
\caption{\label{fig:VRpulls} Differences between the numbers of observed events in data and the SM background predictions for each VR used in the (a) Meff-based and (b) RJR-based searches, expressed as a fraction of the total uncertainty, which combines the uncertainty in the background predictions, and the expected statistical uncertainty of the test obtained from the number of expected events. Empty boxes (indicated by a `-') are when the VR is not used for the corresponding SR selection. }
\end{figure}

The results of the validation procedure are shown in
Figure~\ref{fig:VRpulls}, where the difference in each VR between the
numbers of observed and expected events, expressed as fractions of the
one-standard deviation $(1\sigma)$ uncertainties in the latter, are
summarized. No significant systematic biases are observed for both
searches, with the largest discrepancies being $1.9\sigma$ in the Meff-VRZ associated with the SR Meff-2j-3600 out of 190 VRs and $2.3\sigma$ in RJR-VRW associated with the SR RJR-G1b out of 194 VRs.

\section{Systematic uncertainties}
\label{sec:systematics}

Systematic uncertainties in background estimates arise from the use of extrapolation factors that relate observations in the control regions to background predictions in the signal regions, and from the MC modeling of minor backgrounds. 

The overall background uncertainties, detailed in Figure~\ref{fig:BreakdownSysSRCompressed}, range from 6\% in SR Meff-2j-1200 to 67\% in SR Meff-6j-2600 and from 10\% in SRs RJR-S1a, RJR-S2a, RJR-G1a and RJR-C2 to 30\% in SR RJR-G4. 

For the backgrounds estimated with MC simulation-derived extrapolation factors, the primary common sources of systematic uncertainty are the jet energy scale (JES) calibration, jet energy resolution (JER), theoretical uncertainties, and limited event yields in the MC samples and data CRs. Correlations between uncertainties (for instance between JES or JER uncertainties in CRs and SRs) are taken into account where appropriate. 

The JES and JER uncertainties are estimated using the methods discussed in Refs.~\cite{Aad:2011he,Aad:2012hg,JetCalibRunTwo}. 
An additional uncertainty in the modeling of energy not associated with reconstructed objects, used in the calculation of $\met$ and measured with unassociated charged tracks, is also included.
The combined JES, JER and $\met$ uncertainty ranges from 1\% of the expected background in 2-jet Meff-SRs to 12\% in SR Meff-6j-2600. In the RJR-based search, the same uncertainties range from 1\% in RJR-C4 to 14\% in RJR-G4. 
Uncertainties in jet mass scale (JMS) and jet mass resolution (JMR) are additionally assigned to SR Meff-2jB-1600 and Meff-2jB-2400, which have requirements on the masses of large-radius jets.
The JMS uncertainty is estimated using the same methodology as Ref.~\cite{ATLAS-CONF-2016-035}. A 20\% uncertainty is conservatively assigned to the JMR. The combined JMS and JMR uncertainty is 3.2\% of the expected background in Meff-2j-1600 and 5.1\% in Meff-2j-2400. 

Uncertainties arising from theoretical modeling of background
processes are estimated by comparing samples produced with different
MC generators or by varying the scales. Uncertainties in $W/Z$+jets
production are estimated by increasing and decreasing the
renormalization, factorization and resummation scales by a factor of two, and by increasing and decreasing the nominal CKKW matching scale, $20~\GeV$, by $10~\GeV$ and $5~\GeV$, respectively.    
Uncertainties in the modeling of top quark pair production are
estimated by comparing samples generated with \textsc{Powheg-Box} and
MG5\_aMC@NLO, and by comparing the nominal sample with samples generated using different shower tunes.
Uncertainties associated with PDF modeling of top quark pair production are found to be negligible.  
Uncertainties in diboson production due to PDF, renormalization and factorization scale uncertainties (estimated by increasing and decreasing the scales used in the MC generators by a factor of two for all combinations and taking the envelope of them) are accounted for. The combined theoretical uncertainty ranges from 1\% in Meff-2j-1200 to 45\% in Meff-6j-2600 for Meff SRs. 
In the RJR-based search, the same uncertainties range from  8\% in RJR-S1a to 18\% in RJR-G4, with the smaller range largely due to the absence of 6-jet SRs. Uncertainties associated with the modeling of $Z$+jets production are largest in the 2-jet Meff-SRs (7\%). In the RJR-based search, these uncertainties are largest in RJR-S2b and RJR-S3b SR (8\%). The impact of lepton reconstruction uncertainties, and of the uncertainties related to the $b$-tag/$b$-veto efficiency, on the overall background uncertainty is found to be negligible for all SRs. 

\begin{figure}[t]
\begin{center}
\subfigure[]{\includegraphics[width=0.9\textwidth]{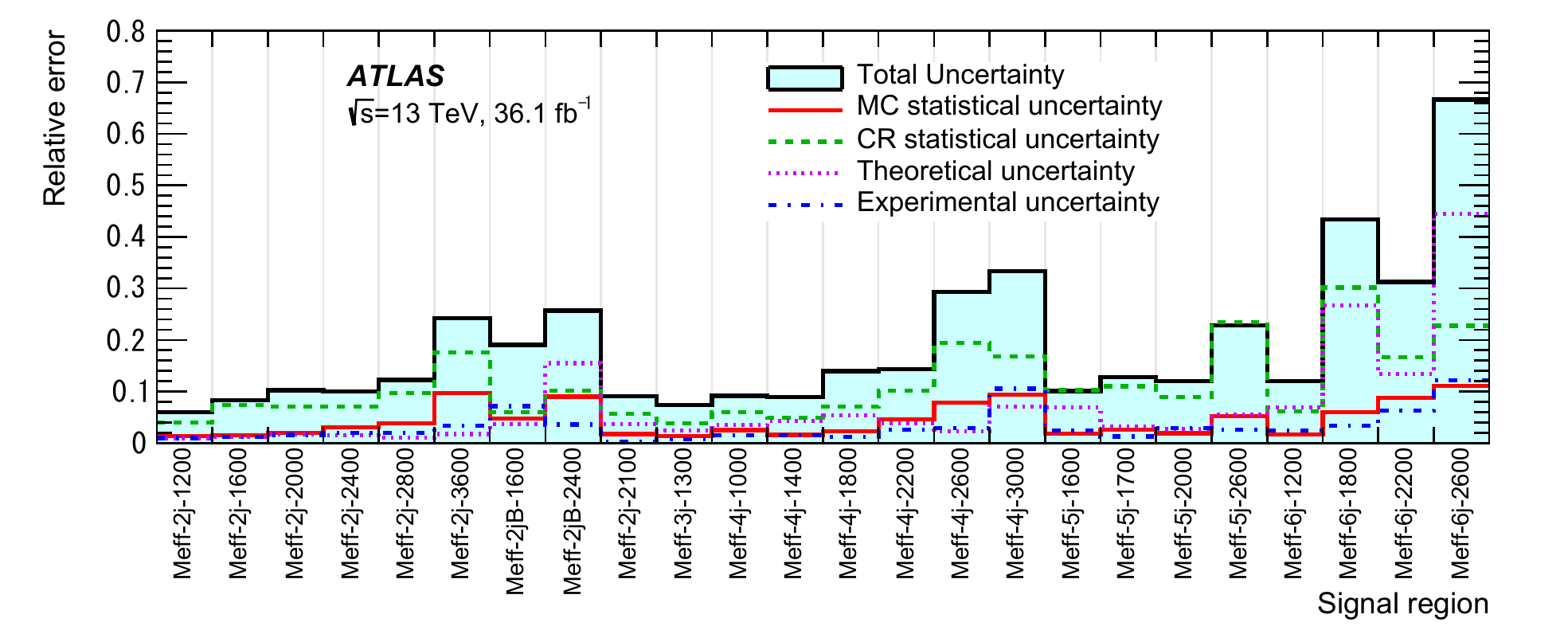}}
\subfigure[]{\includegraphics[width=0.9\textwidth]{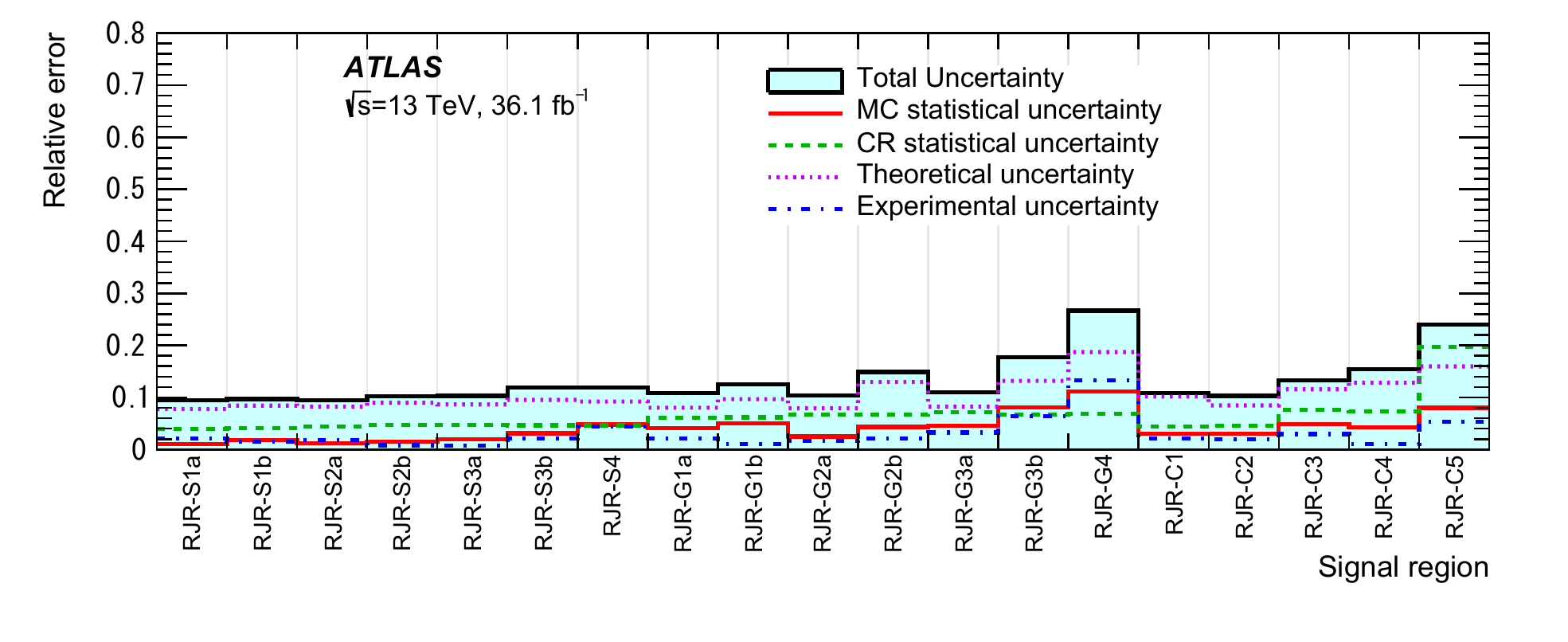}}
\end{center}
\caption{
Breakdown of the largest systematic uncertainties in the background estimates for the (a) Meff-based and (b) RJR-based searches.
The individual uncertainties can be correlated, such that the total background uncertainty is not necessarily their sum in quadrature.
\label{fig:BreakdownSysSRCompressed}}
\end{figure}

The uncertainties arising from the data-driven correction procedure applied to events selected in the CR$\gamma$ region, described in Section~\ref{sec:background}, are included in Figure~\ref{fig:BreakdownSysSRCompressed}
 under `CR statistical uncertainty'. Other uncertainties due to CR
 data sample size range from 4\% to 30\% for Meff SRs and from 4\% to
 20\% for RJR SRs. The statistical uncertainty arising from the use of
 MC samples is largest in SRs Meff-6j-2600 (11\%) and RJR-G4
 (12\%). Uncertainties related to the multi-jet background estimates
 are taken into account by applying a uniform 100\% uncertainty to the
 multi-jet yield in all SRs. In most of the SRs these uncertainties are negligible, and the maximum resulting contribution to the overall background uncertainty is less than 1\%.

Experimental uncertainties (JES, JER, JMS, JMR and $\met$) and MC statistical uncertainty in the SUSY signals are estimated in the same way as for the background
and are less than a few percent for most of the signals, except that 7\% is assigned as JMS and JMR uncertainties
in Meff-2jB-1600 and Meff-2jB-2400. The signal cross-section uncertainty is estimated  
by computing the changes when the renormalization and factorization scale, PDF and the strong coupling constant ($\alpha_{\textrm s}$) are varied.
The uncertainties in the amount of ISR and FSR in the SUSY signals are estimated by varying generator tunes in the simulation as well as scales
used in the matrix-element generator as a function of the mass difference, $\Delta m$, between gluino (or squark) and $\ninoone$.
This uncertainty reaches $20\%$ in the limit of no mass difference and is negligible for $\Delta m>200~\GeV$.

\section{Results, interpretation and limits}
\label{sec:results}

Distributions of $\meff({\textrm{incl.}})$ from the Meff-based search for selected signal regions, obtained before the final selections on this quantity (but after applying all other selections), are shown in Figure~\ref{fig:srMeff} for data and the different MC samples normalized using the theoretical cross-sections. Similarly, distributions of the final discriminating variables used in the RJR-based search, $H_{\textrm T~2,1}^{~PP}$ ($H_{\textrm T~4,1}^{~PP}$ where appropriate) in selected RJR-S and RJR-G regions, and $p_{\mathrm{T}~S}^{\textrm{~CM}}$ in selected RJR-C regions, after applying all other selection requirements except those based on the plotted variable, are shown in Figure~\ref{fig:srRJR}.
Examples of SUSY signals are also shown for illustration. These signals correspond to the processes to which each SR is primarily sensitive: $\squark\squark$ production for the lower jet-multiplicity SRs and $\gluino\gluino$ production for the higher jet-multiplicity SRs. 
In these figures, data and background distributions largely agree within uncertainties. 

The number of events observed in the data and the number of SM events expected to enter each of the signal regions, determined using the background-only fit, are shown in Tables~\ref{tab:p0_UL} and \ref{tab:p0_UL_RJ} and in Figure~\ref{fig:PlotSR}.  The pre-fit background predictions are also shown in Tables~\ref{tab:p0_UL} and \ref{tab:p0_UL_RJ} for comparison. 

The background normalizations for each SR are fit to reproduce the
event yields observed in the CRs. This is in particular seen in Figure~\ref{fig:summaryMu}, leading to agreement between data and post-fit background predictions in most of the SRs. 
The most significant observed excess in the signal regions for the
Meff-based search, with a $p$-value for the background-only hypothesis
of 0.02, corresponding to a significance of 2.0 standard deviations,
occurs in SR Meff-2j-1200 and Meff-2j-2100 (Table~\ref{tab:p0_UL}).
The most significant observed excess across the signal regions for RJR-based search, with a $p$-value for the background-only hypothesis of 0.01, corresponding to a significance of 2.5 standard deviations, occurs in SR RJR-S1a (Table~\ref{tab:p0_UL_RJ}).

\begin{figure}[t]
\begin{center}
\vspace*{-0.03\textheight}
\subfigure[]{\includegraphics[width=0.4\textwidth]{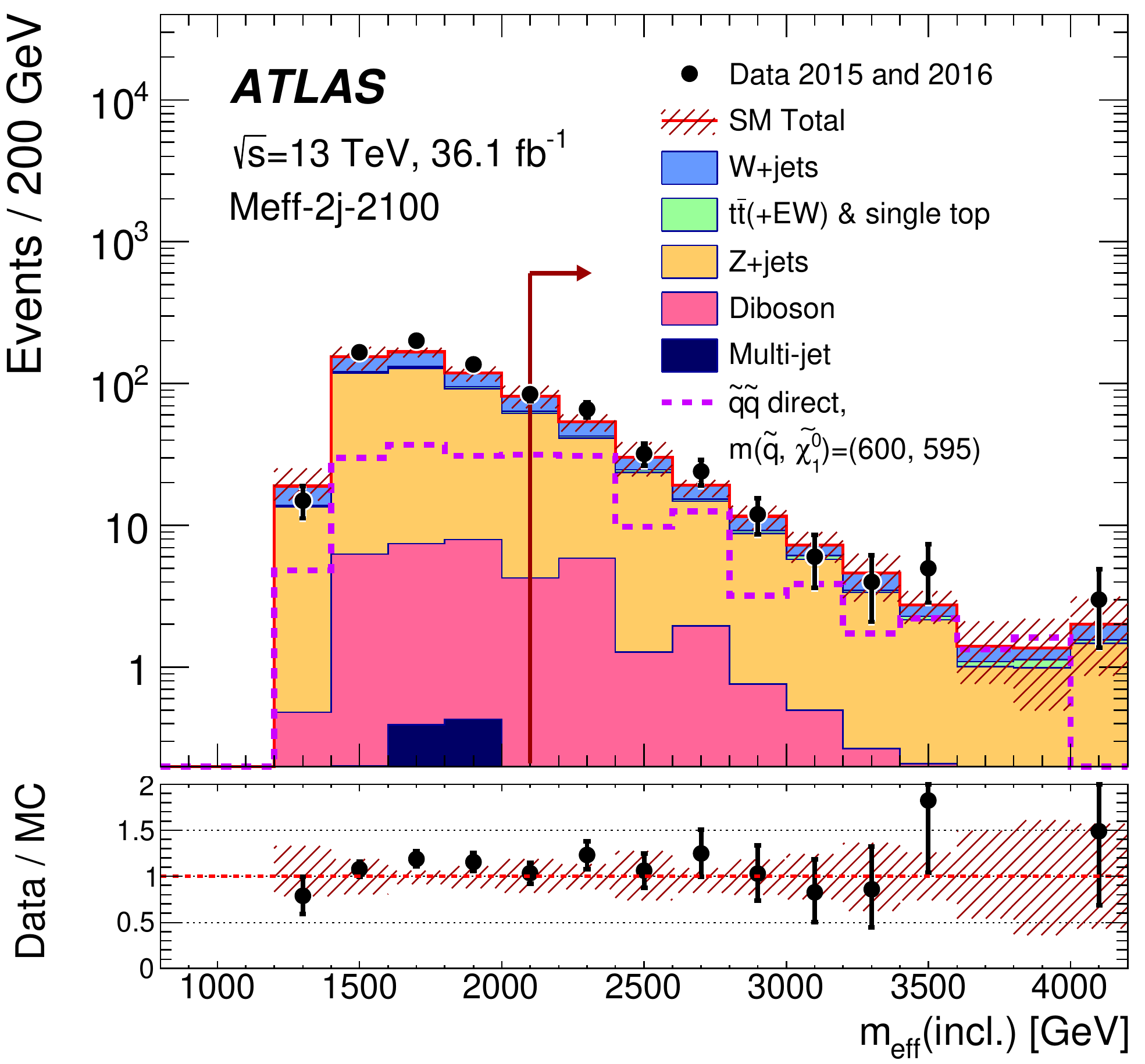}}
\subfigure[]{\includegraphics[width=0.4\textwidth]{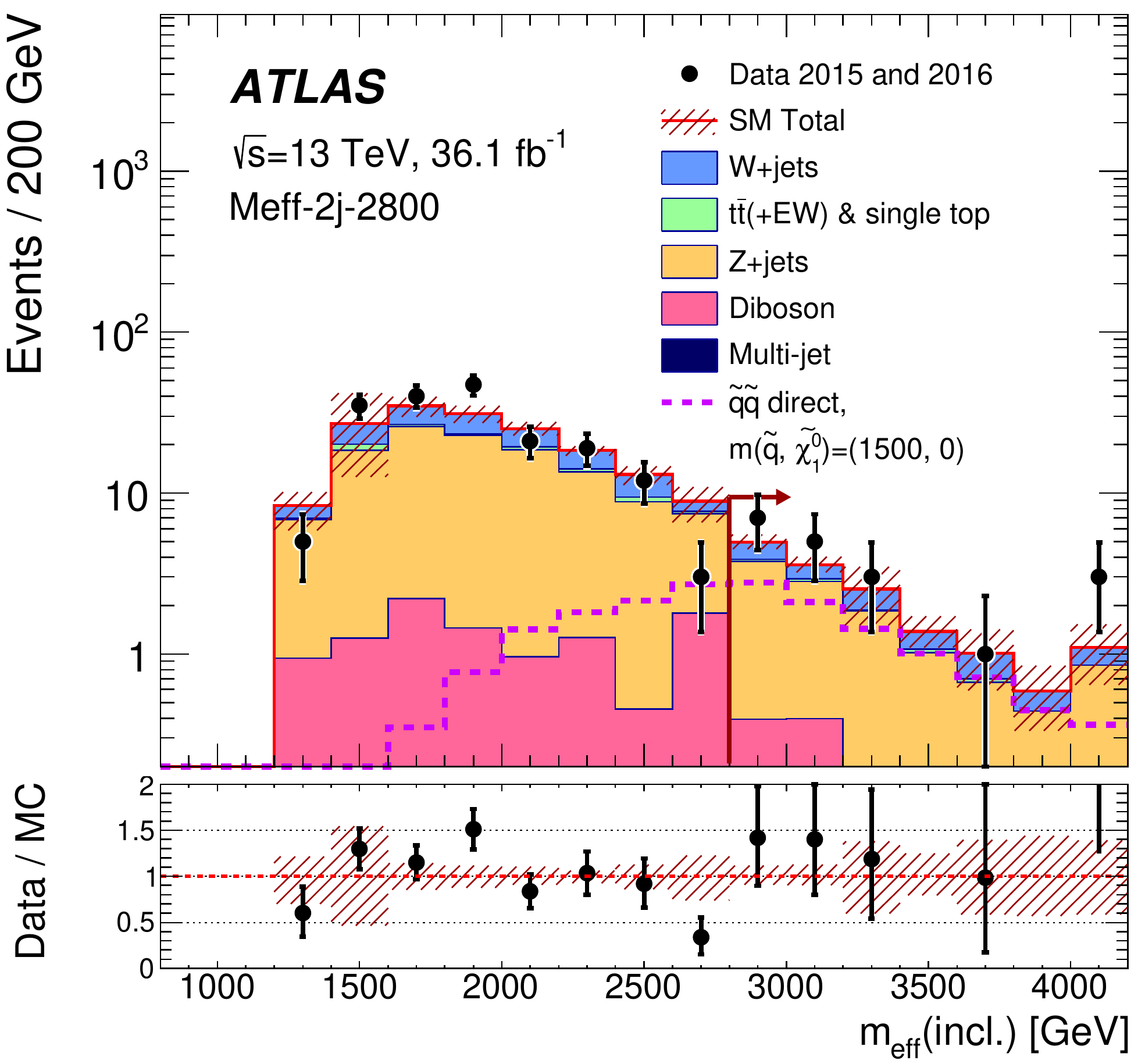}}\\
\subfigure[]{\includegraphics[width=0.4\textwidth]{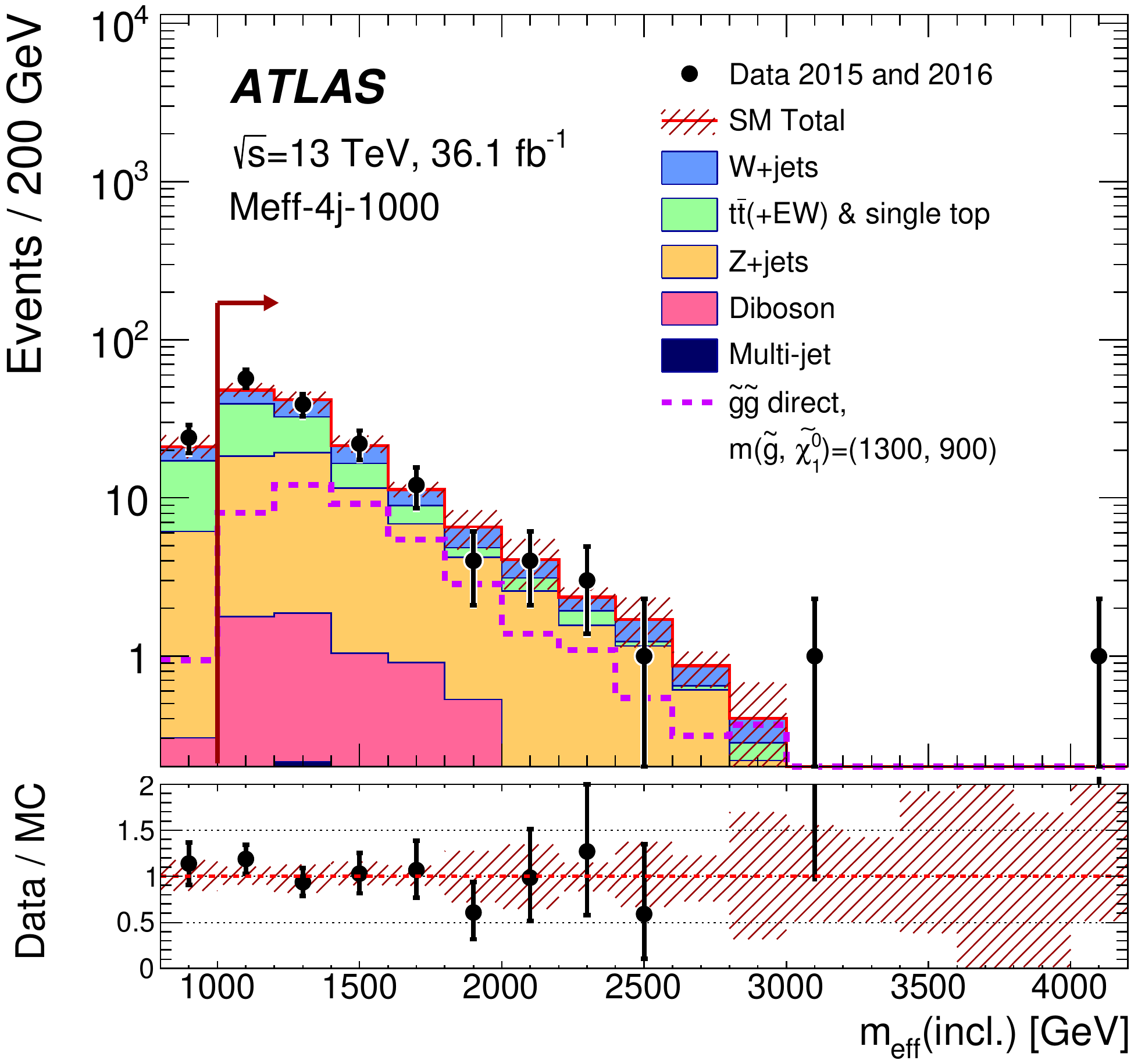}}
\subfigure[]{\includegraphics[width=0.4\textwidth]{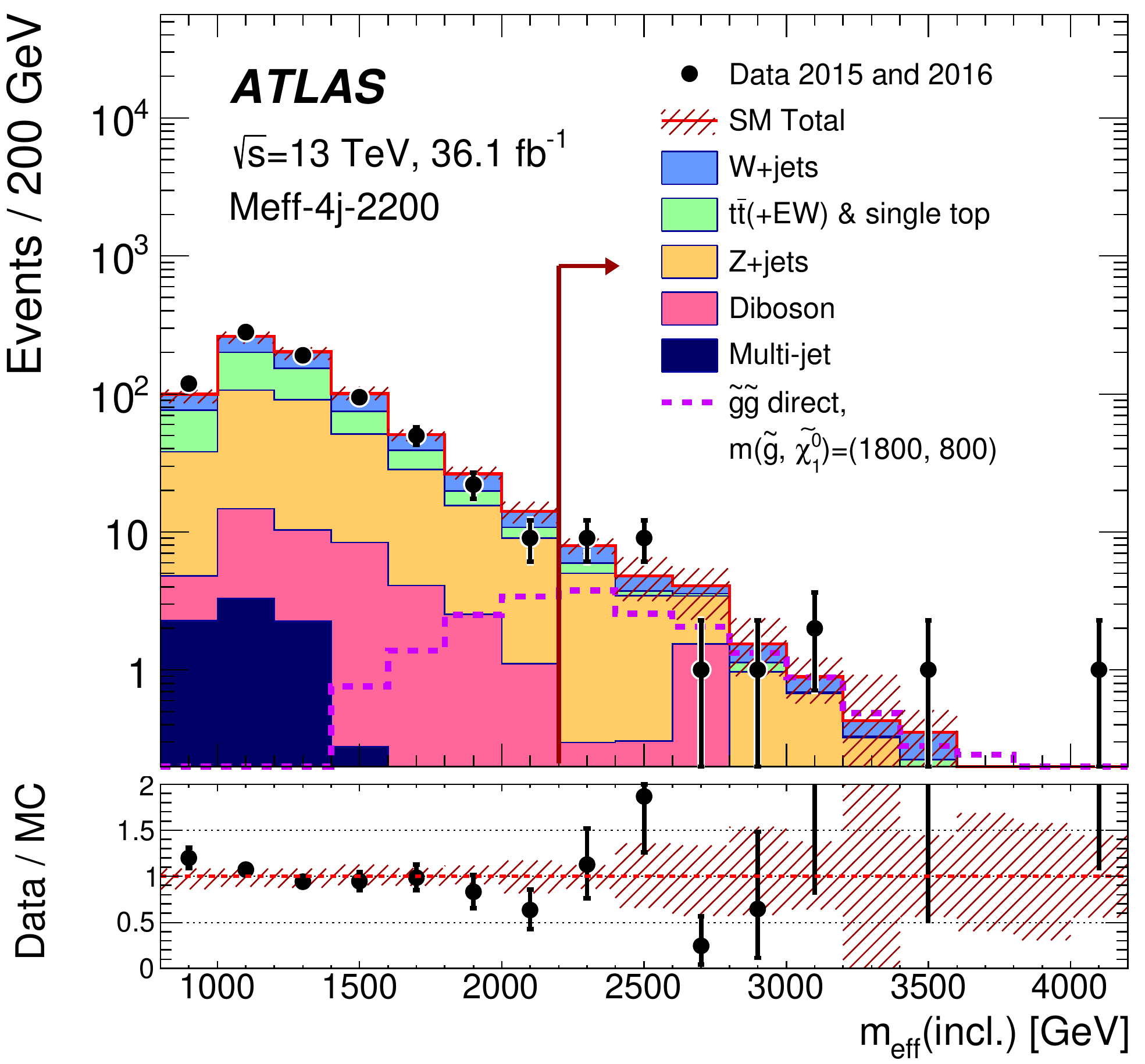}}\\
\subfigure[]{\includegraphics[width=0.4\textwidth]{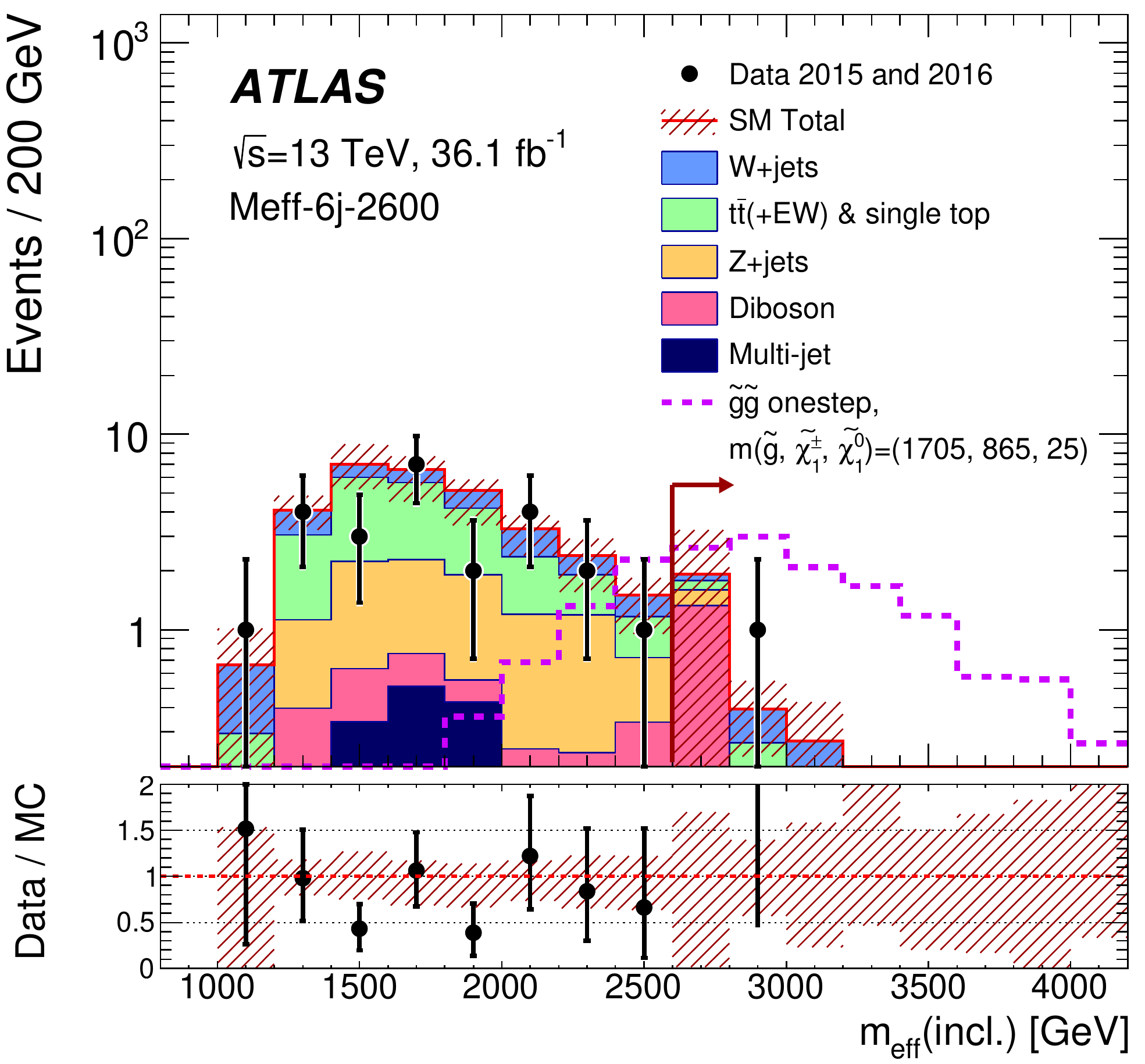}}
\subfigure[]{\includegraphics[width=0.4\textwidth]{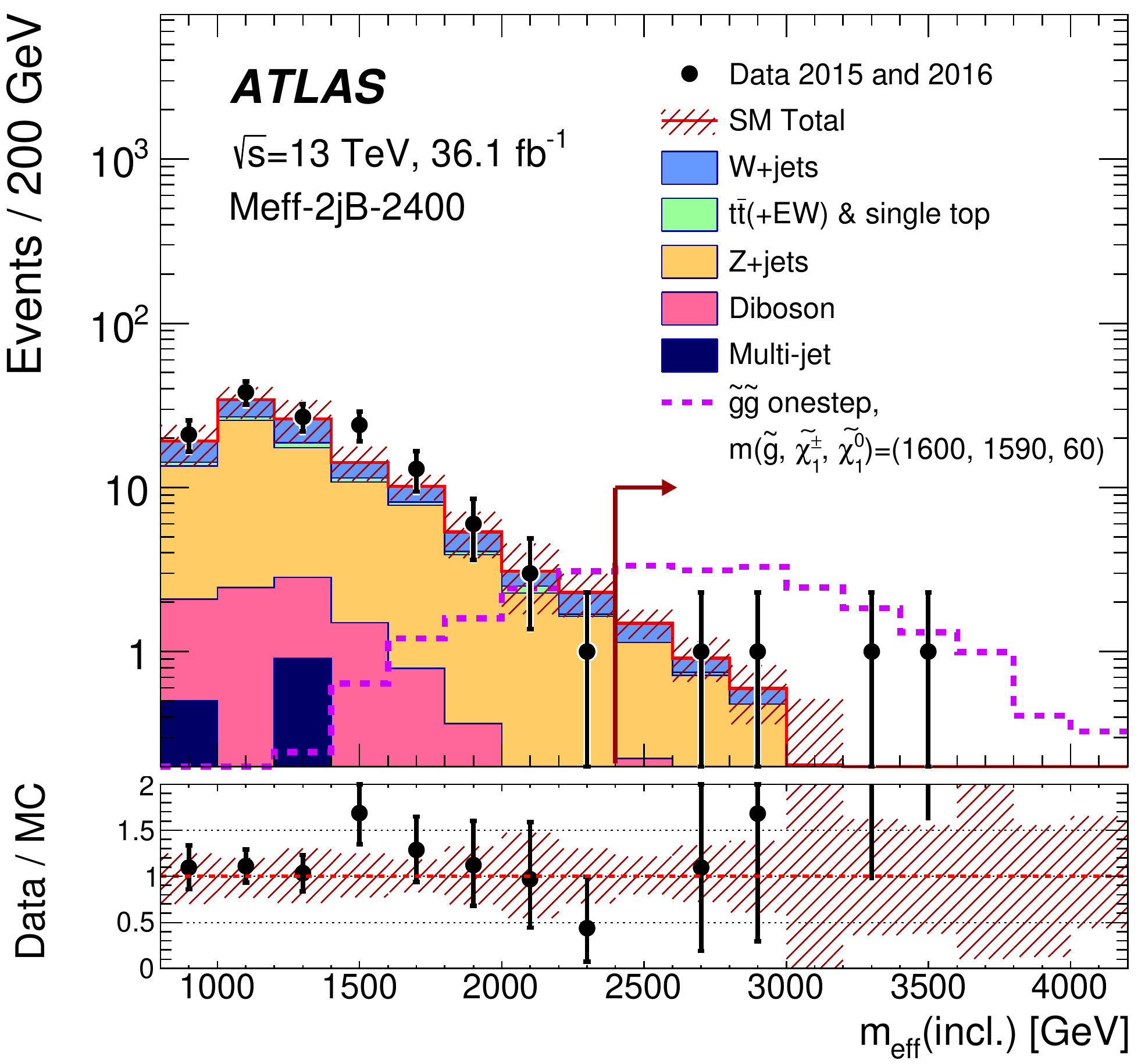}}
\end{center}
\vspace*{-0.04\textheight}\caption{\label{fig:srMeff}Observed $\meff({\textrm{incl.}})$ distributions for the (a) Meff-2j-2100, (b) Meff-2j-2800, (c) Meff-4j-1000, (d) Meff-4j-2200, (e) Meff-6j-2600 and (f) Meff-2jB-2400 signal regions, after applying all selection requirements except those on the plotted variable. The histograms show the MC background predictions prior to the fits described in the text, normalized using cross-section times integrated luminosity. The last bin includes the overflow. The hatched (red) error bands indicate the combined experimental and MC statistical uncertainties. The arrows indicate the values at which the requirements on $\meff({\textrm{incl.}})$ are applied. 
Expected distributions for benchmark signal model points, normalized using NLO+NLL cross-section (Section~\ref{sec:montecarlo}) times integrated luminosity, are also shown for comparison (masses in \GeV). 
}
\end{figure}

\begin{figure}[t]
\begin{center} 
\vspace*{-0.03\textheight}
\subfigure[]{\includegraphics[width=0.4\textwidth]{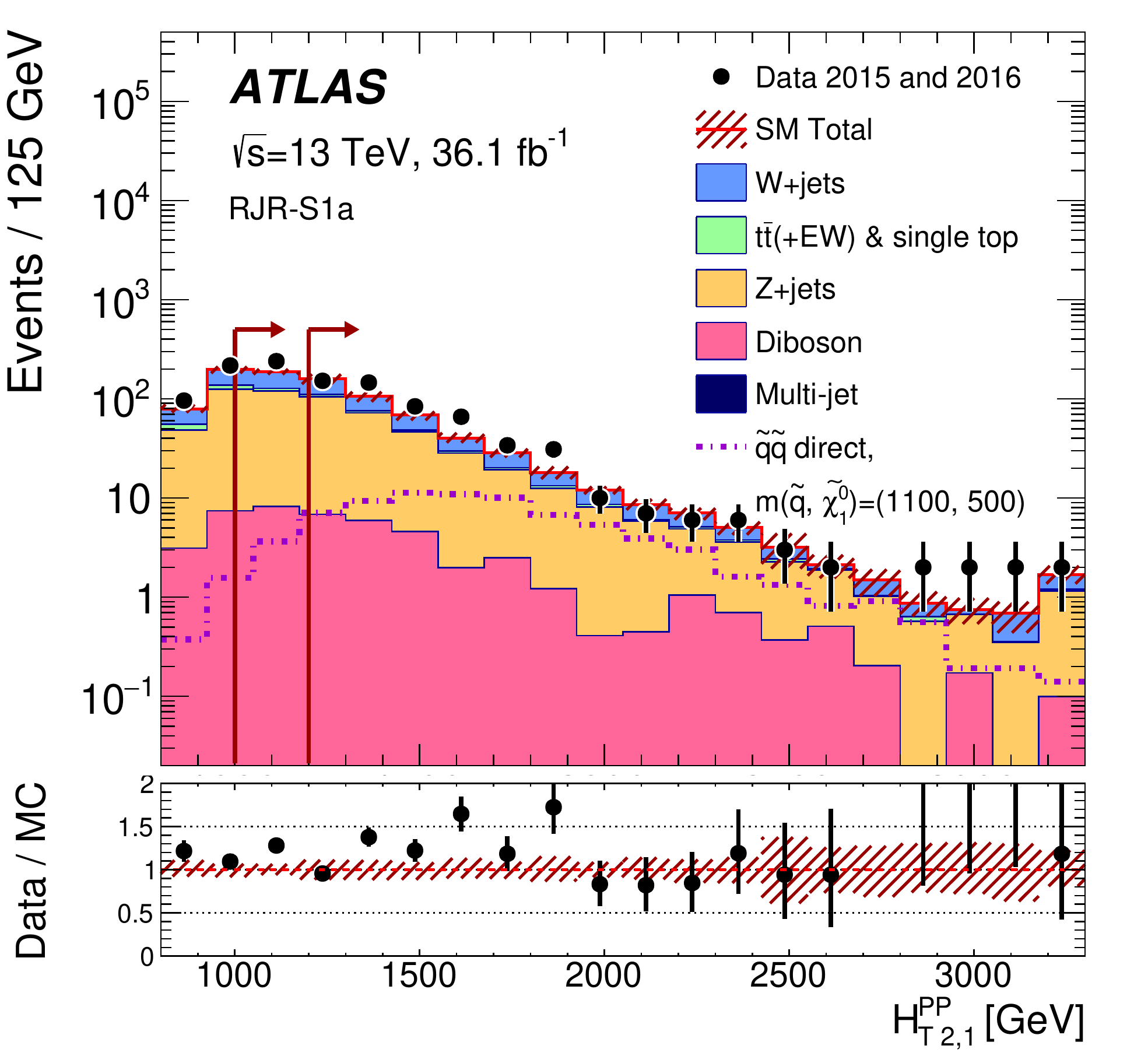}}
\subfigure[]{\includegraphics[width=0.4\textwidth]{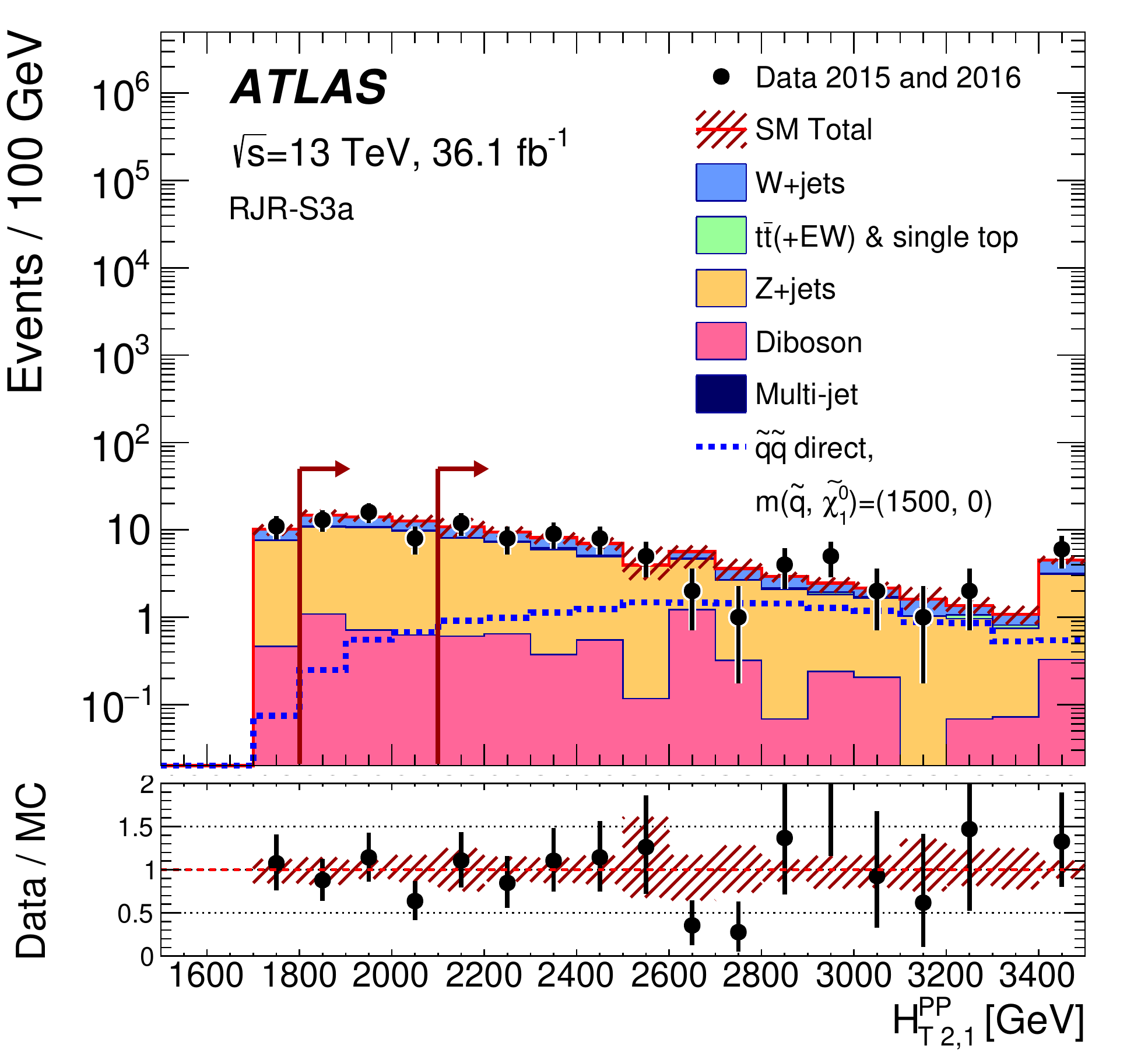}}\\
\vspace*{-0.005\textheight}\subfigure[]{\includegraphics[width=0.4\textwidth]{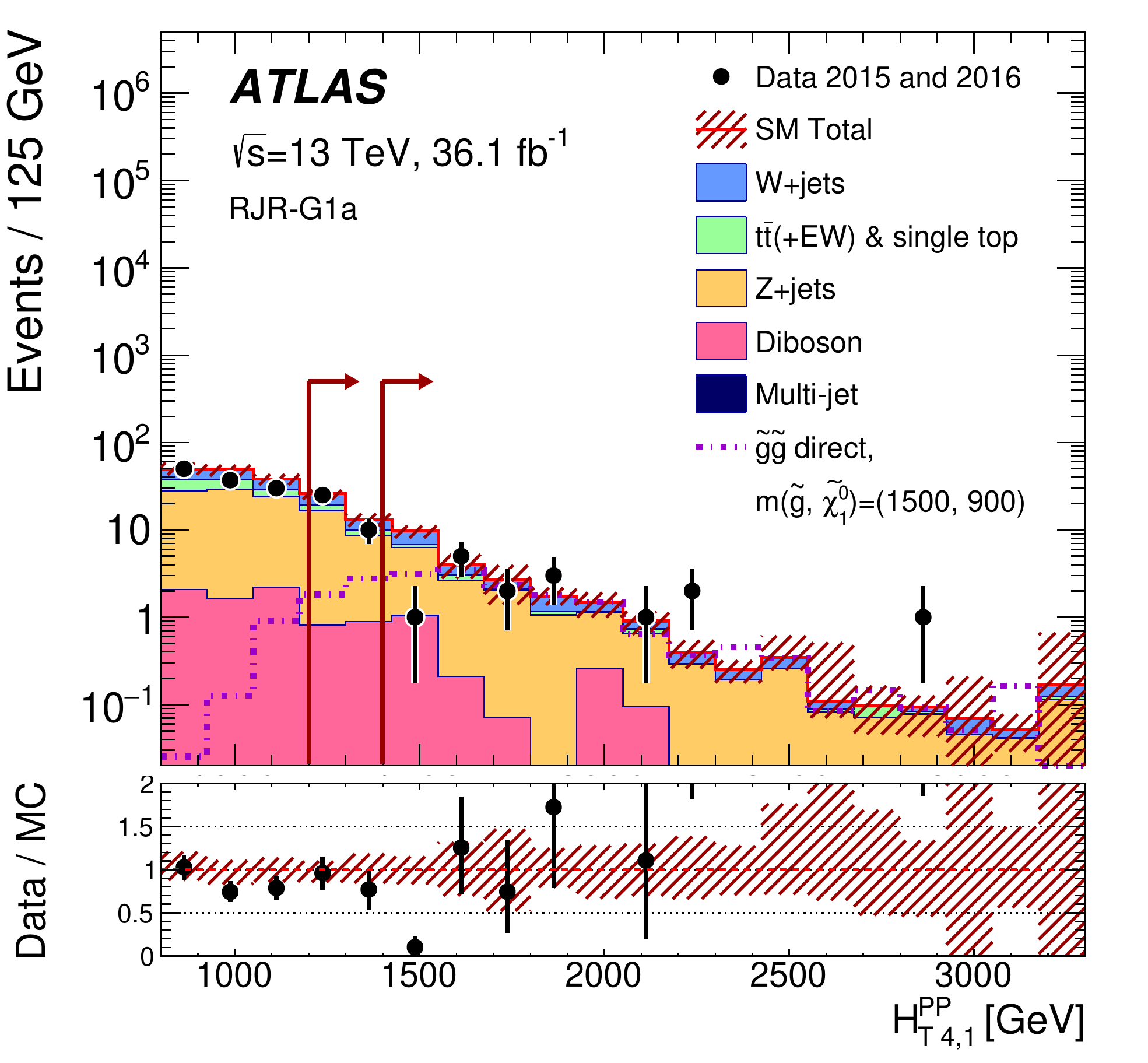}}
\vspace*{-0.005\textheight}\subfigure[]{\includegraphics[width=0.4\textwidth]{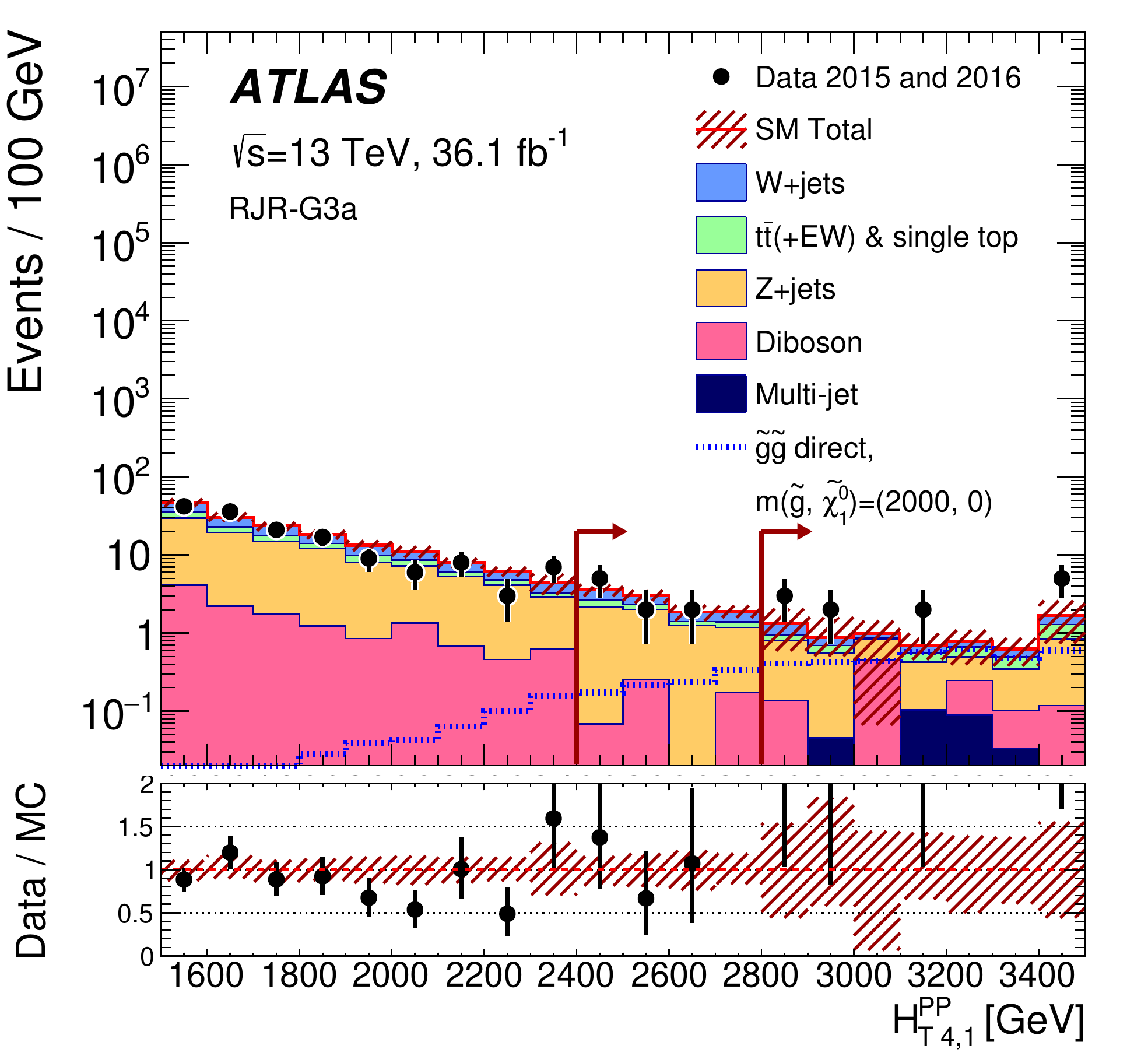}}\\
\vspace*{-0.005\textheight}\subfigure[]{\includegraphics[width=0.4\textwidth]{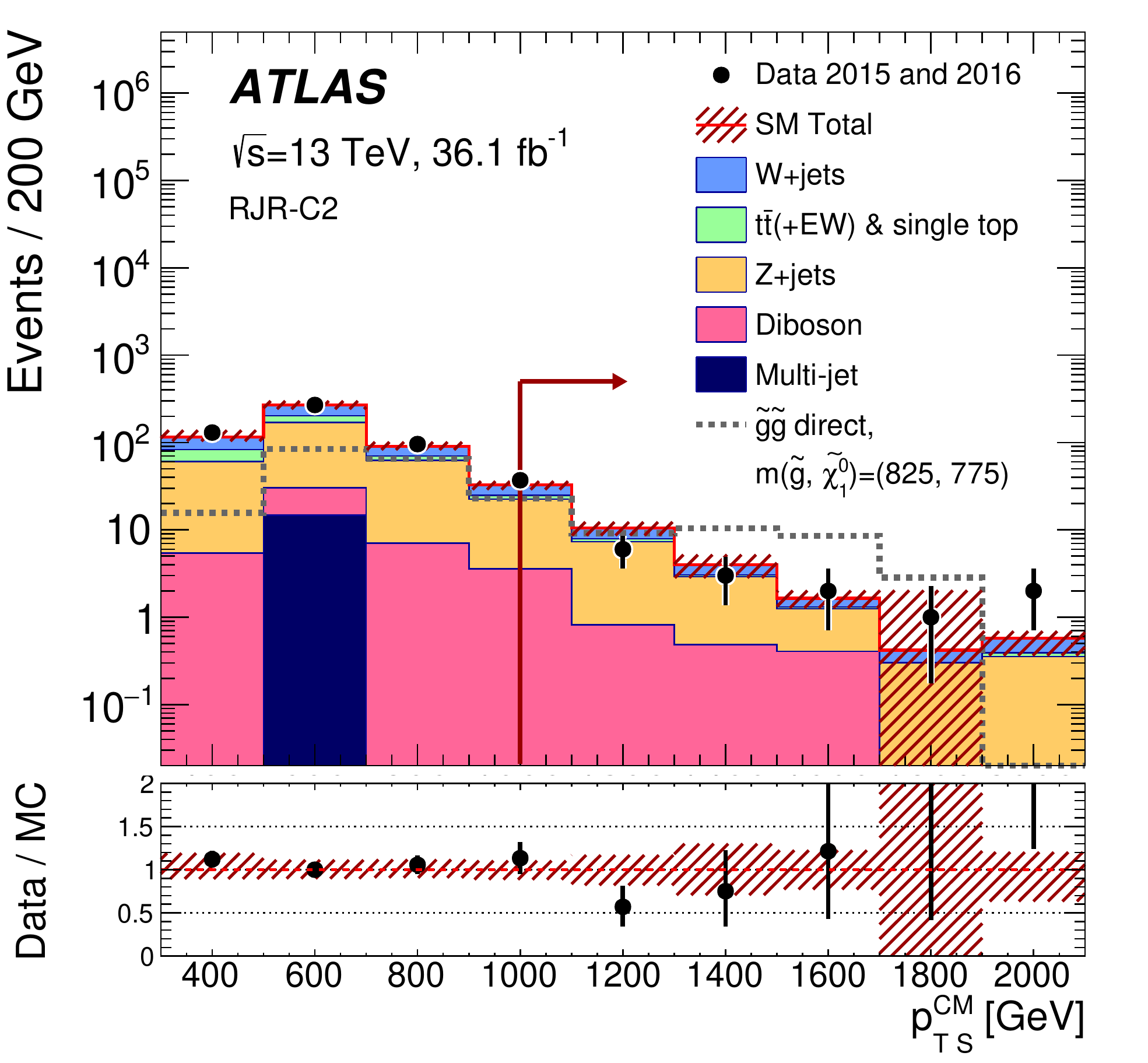}}
\vspace*{-0.005\textheight}\subfigure[]{\includegraphics[width=0.4\textwidth]{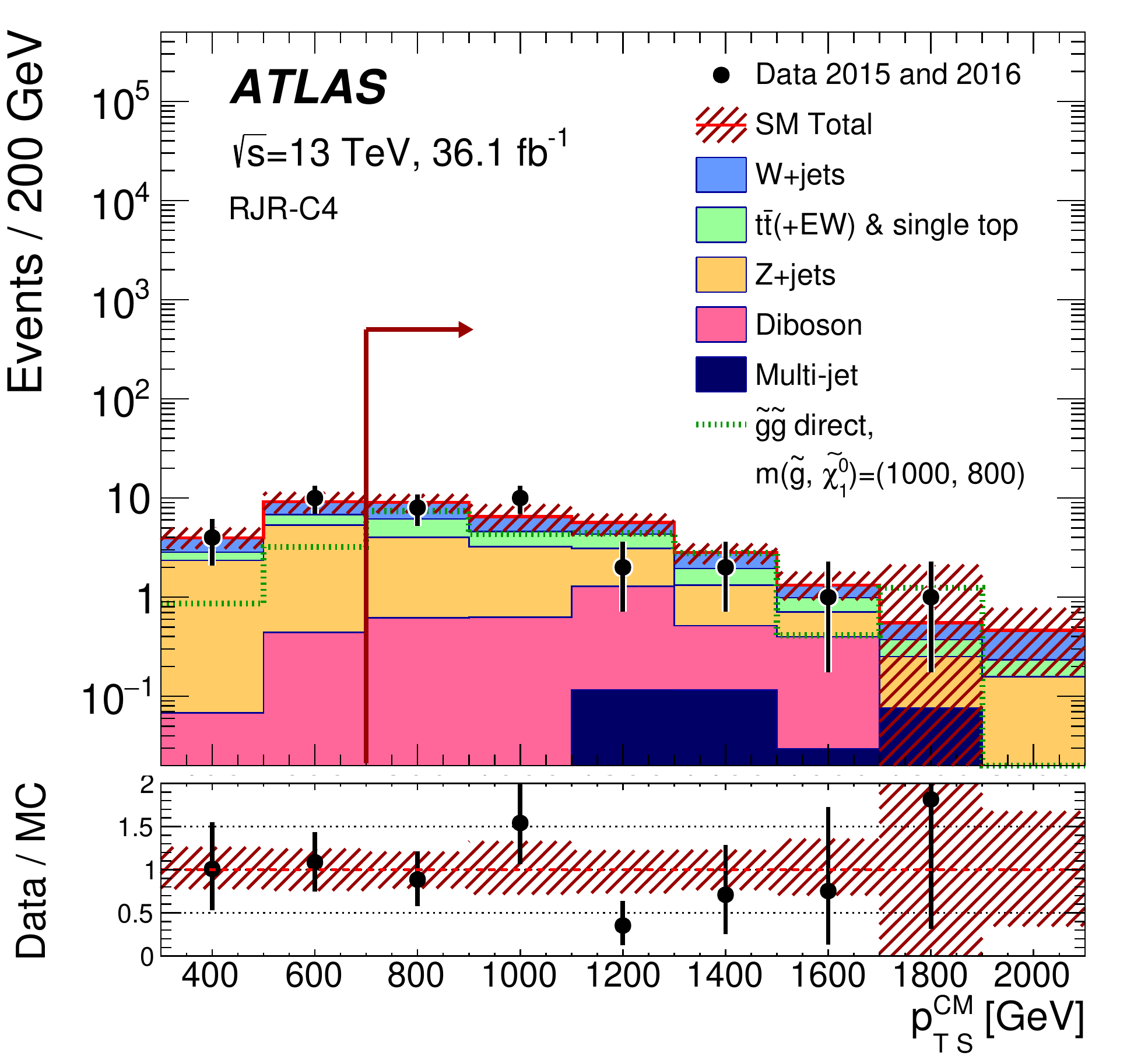}}
\end{center}
\vspace*{-0.04\textheight}\caption{\label{fig:srRJR}Observed $H_{\textrm T~2,1}^{~PP}$ distributions for the (a) RJR-S1a and (b) RJR-S3a signal regions, $H_{\textrm T~4,1}^{~PP}$ distributions for the (c) RJR-G1a and (d)  RJR-G3a signal regions, and $p_{\mathrm{T}~S}^{\textrm{~CM}}$ distributions for the (e) RJR-C2 and (f) RJR-C4 signal regions, after applying all selection requirements except those on the plotted variable. The histograms show the MC background predictions prior to the fits described in the text, normalized using cross-section times integrated luminosity. The last bin includes the overflow. The hatched (red) error bands indicate the combined experimental and MC statistical uncertainties. The arrows indicate the values at which the requirements on the plotted variable are applied. When two arrows are shown, these correspond to the looser SR variation `a' and the tighter variation `b'.
Expected distributions for benchmark signal model points, normalized using NLO+NLL cross-section (Section~\ref{sec:montecarlo}) times integrated luminosity, are also shown for comparison (masses in \GeV). 
}
\end{figure}

\clearpage

\begin{table}
\scriptsize
\begin{center}
\vspace*{-0.035\textwidth}
\begin{tabular}{|lrrrrrrrr|}
\hline
Signal Region  [\textbf{Meff-}] & \textbf{2j-1200 } & \textbf{2j-1600 } & \textbf{2j-2000 } & \textbf{2j-2400 } & \textbf{2j-2800 } & \textbf{2j-3600 } & \textbf{2j-B1600 } & \textbf{2j-B2400 } \\
\hline
\multicolumn{9}{|c|}{MC expected events} \\ \hline
Diboson &  $28$               &  $14.8$               &  $5.6$               &  $3.4$               &  $1.2$               &  $0.21$               &  $1.9$               &  $0.41$               \\
$Z/\gamma^{*}$+jets &  $345$               &  $140$               &  $54$               &  $24.2$               &  $10.2$               &  $2.3$               &  $16.6$               &  $2.5$               \\
$W$+jets &  $141$               &  $47$               &  $18$               &  $8.2$               &  $3.4$               &  $1.11$               &  $5.2$               &  $0.7$               \\
$\ttbar$(+EW) + single top &  $21.0$               &  $5.8$               &  $2.48$               &  $1.13$               &  $0.32$               &  $0.04$               &  $0.80$               &  $0.03$               \\
\hline
\multicolumn{9}{|c|}{Fitted background events} \\ \hline
Diboson & $28 \pm 4$ & $14.8 \pm 2.3$ & $5.5 \pm 1.2$ & $3.4 \pm 0.7$ & $1.2 \pm 0.2$ & $0.21 \pm 0.07$ & $1.9 \pm 0.5$ & $0.41 \pm 0.07$ \\
$Z/\gamma^{*}$+jets  & $336 \pm 19$ & $143 \pm 11$ & $64 \pm 8$ & $28.0 \pm 3.3$ & $12.2 \pm 1.5$ & $2.9 \pm 0.8$ & $14.6 \pm 1.9$ & $2.8 \pm 0.6$ \\
$W$+jets & $141 \pm 24$ & $68 \pm 16$ & $20 \pm 4$ & $9.6 \pm 2.6$ & $3.7 \pm 1.2$ & $0.37 \pm 0.32$ & $5.5 \pm 3.1$ & $0.7 \pm 0.7$ \\
$\ttbar$(+EW) + single top & $15 \pm 4$ & $2.9 \pm 1.6$ & $1.36 \pm 1.0$ & $0.5 \pm 0.5$ & $0.18 \pm 0.15$ &  $0.04_{-0.04}^{+0.05}$               & $0.5 \pm 0.5$ &  $0.02_{-0.02}^{+0.67}$               \\
Multi-jet & $6 \pm 6$ & $0.3 \pm 0.3$ & $0.07 \pm 0.07$ &  $0.02 \pm 0.02$               &  $<0.004$      & -- &  $0.03 \pm
0.03$               &  $<0.002$               \\
\hline
Total MC &  $538$               &  $208$               &  $80$               &  $37$               &  $15.1$               &  $3.6$               &  $24$               &  $3.6$               \\
\hline
Total bkg & $526 \pm 31$ & $228 \pm 19$ & $90 \pm 10$ & $42 \pm 4$ & $17.3 \pm 2.0$ & $3.6 \pm 0.9$ & $22 \pm 4$ & $3.9 \pm 1.2$ \\
\hline
Observed &  $611$                     &  $216$                     &  $73$                     &  $34$                     &  $19$                     &  $5$                     &  $26$                     &  $4$                     \\
\hline
$\langle\epsilon{\mathrm{\sigma}}\rangle_{\textrm{obs}}^{95}$ [fb] &  $4.14$  &  $1.03$  &  $0.47$  &  $0.32$  &  $0.33$  &  $0.20$  &  $0.46$  &  $0.17$  \\
$S_{\textrm{obs}}^{95}$ &   $149$  &   $37$  &   $17.0$  &   $11.4$  &   $11.9$  &   $7.2$  &   $16.7$  &   $6.1$  \\
$S_{\textrm{exp}}^{95}$ &  $ { 81 }^{ +31 }_{ -25 }$  &  $ { 44 }^{ +18 }_{ -12 }$  &  $ { 25.2 }^{ +9.0 }_{ -7.8 }$  &  $ { 15.3 }^{ +5.7 }_{ -3.9 }$  &  $ { 10.6 }^{ +3.9 }_{ -2.8 }$  &  $ { 5.5 }^{ +2.5 }_{ -1.4 }$  &  $ { 13.1 }^{ +5.5 }_{ -3.4 }$  &  $ { 5.7 }^{ +2.3 }_{ -1.2 }$  \\
$p_{0}$ ($\textrm Z$) &  $ 0.02$~$(2.03)$  &  $0.50$~$(0.00)$  &  $0.50$~$(0.00)$  &  $0.50$~$(0.00)$  &  $ 0.31$~$(0.50)$  &  $ 0.21$~$(0.81)$  &  $ 0.28$~$(0.57)$  &  $ 0.32$~$(0.47)$  \\
\hline
\end{tabular}

\vspace{0.2cm}
\begin{tabular}{|lrrrrrrrr|}
\hline
Signal Region  [\textbf{Meff-}] & \textbf{2j-2100 } & \textbf{3j-1300 } & \textbf{4j-1000 } & \textbf{4j-1400 } & \textbf{4j-1800 } & \textbf{4j-2200 } & \textbf{4j-2600 } & \textbf{4j-3000 } \\
\hline
\multicolumn{9}{|c|}{MC expected events} \\ \hline
Diboson &  $12$               &  $37$               &  $6.4$               &  $18.1$               &  $6.0$               &  $2.4$               &  $1.8$               &  $0.24$               \\
$Z/\gamma^{*}$+jets  &  $116$               &  $268$               &  $60$               &  $100$               &  $33$               &  $12.0$               &  $4.1$               &  $1.4$               \\
$W$+jets &  $34$               &  $107$               &  $29$               &  $52$               &  $15$               &  $4.5$               &  $1.66$               &  $0.6$               \\
$\ttbar$(+EW) + single top &  $5.0$               &  $36$               &  $43$               &  $42$               &  $7.7$               &  $1.6$               &  $0.64$               &  $0.21$               \\
\hline
\multicolumn{9}{|c|}{Fitted background events} \\ \hline
Diboson & $12 \pm 5$ & $37 \pm 6$ & $6.4 \pm 1.1$ & $18.1 \pm 3.0$ & $6.0 \pm 1.1$ & $2.4 \pm 0.7$ & $1.8 \pm 0.8$ & $0.24 \pm 0.07$ \\
$Z/\gamma^{*}$+jets  & $102 \pm 8$ & $221 \pm 20$ & $52 \pm 7$ & $85 \pm 10$ & $25 \pm 4$ & $9.9 \pm 2.0$ & $2.3 \pm 0.9$ & $1.2 \pm 0.5$ \\
$W$+jets & $35 \pm 10$ & $106 \pm 19$ & $22 \pm 7$ & $42 \pm 10$ & $12 \pm 6$ & $3.3 \pm 1.1$ & $1.57 \pm 1.0$ & $0.39 \pm 0.3$ \\
$\ttbar$(+EW) + single top & $2.6 \pm 1.4$ & $25 \pm 9$ & $43 \pm 8$ & $35 \pm 10$ & $5.0 \pm 3.3$ & $0.8 \pm 0.4$ &  $0.13_{-0.13}^{+0.17}$               & $0.12 \pm 0.11$ \\
Multi-jet & $0.11 \pm 0.11$ &  $1.4 \pm 1.4$   &$0.39 \pm 0.39$               & $0.5 \pm 0.5$ &  $0.10 \pm 0.10$               &$0.02 \pm 0.02$               &  $0.03 \pm 0.03$ &  $0.01 \pm 0.01$               \\
\hline
Total MC &  $167$               &  $449$               &  $138$               &  $212$               &  $61$               &  $20.5$               &  $8.2$               &  $2.4$               \\
\hline
Total bkg & $153 \pm 14$ & $390 \pm 29$ & $124 \pm 12$ & $182 \pm 16$ & $49 \pm 7$ & $16.5 \pm 2.7$ & $5.8 \pm 2.0$ & $2.0 \pm 0.6$ \\
\hline
Observed &  $190$                     &  $429$                     &  $142$                     &  $199$                     &  $55$                     &  $24$                     &  $4$                     &  $2$                     \\
\hline
$\langle\epsilon\sigma\rangle_{\textrm{obs}}^{95}$ [fb] &  $1.98$  &  $2.84$  &  $1.40$  &  $1.76$  &  $0.79$  &  $0.49$  &  $0.16$  &  $0.12$  \\
$S_{\textrm{obs}}^{95}$ &   $72$  &   $103$  &   $50.6$  &   $64$  &   $28.3$  &   $17.6$  &   $5.8$  &   $4.5$  \\
$S_{\textrm{exp}}^{95}$ &  $ { 38 }^{ +16 }_{ -10 }$  &  $ { 72 }^{ +29 }_{ -21 }$  &  $ { 37.1 }^{ +12.5 }_{ -9.1 }$  &  $ { 45 }^{ +15 }_{ -12 }$  &  $ { 22.2 }^{ +6.9 }_{ -7.0 }$  &  $ { 11.3 }^{ +4.8 }_{ -2.7 }$  &  $ { 6.7 }^{ +2.7 }_{ -1.9 }$  &  $ { 4.6 }^{ +1.6 }_{ -1.2 }$  \\
$p_{0}$ ($\textrm Z$) &  $ 0.02$~$(2.03)$  &  $ 0.13$~$(1.12)$  &  $ 0.10$~$(1.26)$  &  $ 0.08$~$(1.39)$  &  $ 0.18$~$(0.90)$  &  $ 0.09$~$(1.34)$  &  $0.50$~$(0.00)$  &  $0.50$~$(0.00)$  \\
\hline
\end{tabular}

\vspace{0.2cm}
\begin{tabular}{|lrrrrrrrr|}
\hline
Signal Region  [\textbf{Meff-}] & \textbf{5j-1600 } & \textbf{5j-1700 } & \textbf{5j-2000 } & \textbf{5j-2600 } & \textbf{6j-1200 } & \textbf{6j-1800 } & \textbf{6j-2200 } & \textbf{6j-2600 } \\
\hline
\multicolumn{9}{|c|}{MC expected events} \\ \hline
Diboson &  $10.8$               &  $6.6$               &  $8.9$               &  $2.6$               &  $20.5$               &  $1.9$               &  $1.7$               &  $1.3$               \\
$Z/\gamma^{*}$+jets  &  $56$               &  $31$               &  $50$               &  $7.4$               &  $109$               &  $3.3$               &  $1.3$               &  $0.76$               \\
$W$+jets &  $42$               &  $15.5$               &  $18.6$               &  $2.57$               &  $81$               &  $2.2$               &  $0.67$               &  $0.44$               \\
$\ttbar$(+EW) + single top &  $45$               &  $12.0$               &  $9.9$               &  $0.8$               &  $144$               &  $4.3$               &  $0.63$               &  $0.39$               \\
\hline
\multicolumn{9}{|c|}{Fitted background events} \\ \hline
Diboson & $10.8 \pm 1.8$ & $6.6 \pm 1.1$ & $8.9 \pm 1.5$ & $2.6 \pm 0.7$ & $20.5 \pm 3.5$ & $1.9 \pm 0.7$ & $1.7 \pm 0.8$ & $1.3 \pm 0.9$ \\
$Z/\gamma^{*}$+jets  & $42 \pm 5$ & $21 \pm 4$ & $37 \pm 6$ & $6.0 \pm 1.7$ & $61 \pm 11$ & $1.1 \pm 0.7$ & $0.9 \pm 0.5$ & $0.38 \pm 0.29$ \\
$W$+jets & $26 \pm 7$ & $8.0 \pm 2.6$ & $13.3 \pm 3.3$ &  $0.41_{-0.41}^{+0.45}$               & $46 \pm 22$ &  $0.8_{-0.8}^{+1.1}$               &  $0.10_{-0.10}^{+0.16}$               &  $0.16_{-0.16}^{+0.24}$               \\
$\ttbar$(+EW) + single top & $40 \pm 9$ & $7.1 \pm 2.8$ & $6.5 \pm 2.6$ & $0.4 \pm 0.4$ & $145 \pm 25$ & $1.2 \pm 1.0$ & $0.37 \pm 0.27$ &  $0.24_{-0.24}^{+0.41}$               \\
Multi-jet & $9 \pm 9$ &  $0.08_{-0.08}^{+0.09}$               & $0.09 \pm 0.09$               &  $0.01 \pm 0.01$ &  $1.29_{-1.29}^{+1.30}$               &  $0.12 \pm 0.12$&  $0.02_{-0.02}^{+0.03}$               &  $0.06 \pm 0.06 $               \\
\hline
Total MC &  $158$               &  $65$               &  $88$               &  $13.3$               &  $355$               &  $11.7$               &  $4.3$               &  $2.9$               \\
\hline
Total bkg & $128 \pm 14$ & $43 \pm 5$ & $65 \pm 7$ & $9.4 \pm 2.1$ & $274 \pm 32$ & $5.1 \pm 1.8$ & $3.1 \pm 1.3$ & $2.2 \pm 1.4$ \\
\hline
Observed &  $135$                     &  $49$                     &  $59$                     &  $10$                     &  $276$                     &  $9$                     &  $3$                     &  $1$                     \\
\hline
$\langle\epsilon\sigma\rangle_{\textrm{obs}}^{95}$ [fb] &  $1.26$  &  $0.64$  &  $0.49$  &  $0.24$  &  $2.19$  &  $0.33$  &  $0.15$  &  $0.11$  \\
$S_{\textrm{obs}}^{95}$ &   $45.4$  &   $23.0$  &   $17.8$  &   $8.8$  &   $79$  &   $11.9$  &   $5.4$  &   $3.8$  \\
$S_{\textrm{exp}}^{95}$ &  $ { 39 }^{ +14 }_{ -8 }$  &  $ { 18.2 }^{ +6.7 }_{ -5.6 }$  &  $ { 20.7 }^{ +8.6 }_{ -5.3 }$  &  $ { 8.5 }^{ +3.0 }_{ -2.1 }$  &  $ { 70 }^{ +19 }_{ -20 }$  &  $ { 8.2 }^{ +3.6 }_{ -1.7 }$  &  $ { 5.4 }^{ +1.8 }_{ -1.2 }$  &  $ { 4.3 }^{ +1.8 }_{ -0.6 }$  \\
$p_{0}$ ($\textrm Z$) &  $ 0.32$~$(0.46)$  &  $ 0.21$~$(0.82)$  &  $0.50$~$(0.00)$  &  $ 0.46$~$(0.09)$  &  $0.50$~$(0.00)$  &  $ 0.11$~$(1.25)$  &  $0.50$~$(0.00)$  &  $0.50$~$(0.00)$  \\
\hline
\end{tabular}

\vspace*{-0.01\textheight}\caption[p0 and UL]{Numbers of events
  observed in the signal regions used in the Meff-based analysis
  compared with background predictions obtained from the fits
  described in the text. The $p$-values ($p_{0}$) are the
  probabilities to obtain a value equal to or larger than that
  observed in the data. For an observed number of events lower than expected, the $p$-value is truncated at 0.5. In addition to $p$-values, the number of equivalent Gaussian standard deviations (Z) is given in parentheses.
Also shown are 95\% CL upper limits on the visible cross-section ($\langle\epsilon\sigma\rangle_{\textrm{obs}}^{95}$), 
the visible number of signal events ($S_{\textrm{obs}}^{95}$) and the number of signal events ($S_{\textrm{exp}}^{95}$) 
given the expected number of background events (and $\pm 1\sigma$ excursions of the expected number).
\label{tab:p0_UL}}
\end{center}
\end{table}

\begin{table}
\scriptsize
\begin{center}
\vspace*{-0.035\textwidth}
\begin{tabular}{|lrrrrrrr|}
\hline
Signal Region & \textbf{RJR-S1a } & \textbf{RJR-S1b } & \textbf{RJR-S2a } & \textbf{RJR-S2b } & \textbf{RJR-S3a } & \textbf{RJR-S3b } & \textbf{RJR-S4 } \\
\hline
\multicolumn{8}{|c|}{MC expected events} \\ \hline
Diboson &  $37$               &  $17$               &  $23$               &  $10.3$               &  $7.2$               &  $3.5$               &  $2.0$               \\
$Z/\gamma^{*}$+jets &  $495$               &  $189$               &  $222$               &  $102$               &  $70$               &  $30.5$               &  $17.9$               \\
$W$+jets &  $220$               &  $77$               &  $84$               &  $36$               &  $22.6$               &  $9.2$               &  $5.3$               \\
$\ttbar$(+EW) + single top &  $32$               &  $9.2$               &  $10.9$               &  $4.7$               &  $2.6$               &  $1.17$               &  $0.68$               \\
\hline
\multicolumn{8}{|c|}{Fitted background events} \\ \hline
Diboson & $37 \pm 8$ & $17 \pm 4$ & $23 \pm 5$ & $10.3 \pm 2.6$ & $7.2 \pm 1.5$ & $3.5 \pm 1.1$ & $2.0 \pm 0.5$ \\
$Z/\gamma^{*}$+jets & $450 \pm 40$ & $170 \pm 14$ & $211 \pm 17$ & $97 \pm 8$ & $67 \pm 5$ & $29.0 \pm 2.4$ & $17.0 \pm 1.5$ \\
$W$+jets & $208 \pm 27$ & $73 \pm 9$ & $83 \pm 12$ & $35 \pm 5$ & $22.3 \pm 3.0$ & $9.0 \pm 1.3$ & $5.2 \pm 0.9$ \\
$\ttbar$(+EW) + single top & $27 \pm 26$ & $7.4 \pm 2.0$ & $7.6 \pm 3.2$ & $3.3 \pm 1.2$ & $1.9 \pm 0.5$ & $0.82 \pm 0.34$ &  $0.49_{-0.49}^{+0.51}$               \\
Multi-jet & $18 \pm 17$ & $1.3 \pm 1.3$ & $0.6 \pm 0.6$ & $0.31 \pm 0.31$ & $0.27 \pm 0.27$ & $0.03 \pm 0.03$ & $0.03 \pm 0.03$ \\
\hline
Total MC &  $1830$               &  $370$               &  $378$               &  $172$               &  $120$               &  $45.9$               &  $27.7$               \\
\hline
Total bkg & $740 \pm 50$ & $268 \pm 18$ & $326 \pm 22$ & $146 \pm 10$ & $98 \pm 6$ & $42.4 \pm 3.0$ & $24.7 \pm 2.1$ \\
\hline
Observed &  $880$                     &  $325$                     &  $365$                     &  $170$                     &  $102$                     &  $46$                     &  $23$                     \\
\hline

$\langle\epsilon\sigma\rangle_{\textrm{obs}}^{95}$ [fb] &  $6.45$      & $2.76$      & $1.89$      & $1.38$      & $0.69$      & $0.51$      & $0.30$  \\
$S_{\textrm{obs}}^{95}$                                       & $233$        & $99.5$      & $68.3$      & $49.9$      & $24.7$      & $18.3$       & $10.7$  \\ 
$S_{\textrm{exp}}^{95}$                                       & $ { 120 }^{ +44 }_{ -34 }$ & $ { 50 }^{ +18 }_{ -13 }$ & $ { 50 }^{ +14 }_{ -10 }$ & $ { 32 }^{ +14 }_{ -8 }$ & $ { 24 }^{ +11 }_{ -6 }$ & $ { 15.5 }^{ +5.9 }_{ -3.4 }$ & $ { 11.6 }^{ +4.5 }_{ -3.9 }$ \\
$p_{0}$ ($\textrm Z$)                                        &  $0.01$~$(2.52)$ & $ 0.01$~$(2.34)$ & $ 0.14$~$(1.07)$ & $ 0.10$~$(1.30)$ & $ 0.50 $~$(0.00)$ & $ 0.50$~$(0.00)$ & $ 0.50$~$(0.00)$  \\
\hline 
\end{tabular}

\vspace{0.2cm}
\begin{tabular}{|lrrrrrrr|}
\hline
Signal Region & \textbf{RJR-G1a } & \textbf{RJR-G1b } & \textbf{RJR-G2a } & \textbf{RJR-G2b } & \textbf{RJR-G3a } & \textbf{RJR-G3b } & \textbf{RJR-G4 } \\
\hline
\multicolumn{8}{|c|}{MC expected events} \\ \hline
Diboson &  $3.1$               &  $1.6$               &  $2.8$               &  $1.34$               &  $0.80$               &  $0.37$               &  $0.24$               \\
$Z/\gamma^{*}$+jets &  $28.7$               &  $13.1$               &  $28.1$               &  $9.4$               &  $8.8$               &  $3.0$               &  $2.09$               \\
$W$+jets &  $14.0$               &  $6.4$               &  $14.6$               &  $5.0$               &  $4.7$               &  $1.7$               &  $1.0$               \\
$\ttbar$(+EW) + single top &  $6.0$               &  $2.0$               &  $6.5$               &  $2.0$               &  $3.1$               &  $1.5$               &  $1.1$               \\
\hline
\multicolumn{8}{|c|}{Fitted background events} \\ \hline
Diboson & $3.1 \pm 0.7$ & $1.6 \pm 0.5$ & $2.8 \pm 0.8$ & $1.34 \pm 0.33$ & $0.80 \pm 0.27$ & $0.36 \pm 0.29$ & $0.24 \pm 0.11$ \\
$Z/\gamma^{*}$+jets & $24.8 \pm 2.7$ & $11.3 \pm 1.4$ & $25.4 \pm 2.9$ & $8.4 \pm 1.2$ & $7.9 \pm 1.1$ & $2.7 \pm 0.7$ & $1.89 \pm 0.35$ \\
$W$+jets & $12.0 \pm 1.7$ & $5.5 \pm 0.9$ & $12.3 \pm 2.1$ & $4.2 \pm 0.8$ & $3.9 \pm 0.7$ & $1.5 \pm 0.6$ & $0.85 \pm 0.29$ \\
$\ttbar$(+EW) + single top & $4.8 \pm 0.9$ & $1.6 \pm 1.4$ & $5.2 \pm 1.9$ & $1.6 \pm 0.6$ & $2.4 \pm 0.9$ & $1.2 \pm 1.0$ & $0.9 \pm 0.8$ \\
Multi-jet & $0.25 \pm 0.25$ & $0.13 \pm 0.13$ & $0.5 \pm 0.5$ & $0.2 \pm 0.2$ & $0.5 \pm 0.5$ & $0.26 \pm 0.25$ &  $0.18_{-0.18}^{+0.18}$               \\
\hline
Total MC &  $66.8$               &  $30.9$               &  $80.4$               &  $28.9$               &  $44.4$               &  $21.1$               &  $14.4$               \\
\hline
Total bkg & $45 \pm 4$ & $20.1 \pm 2.3$ & $46 \pm 4$ & $15.8 \pm 1.8$ & $15.6 \pm 1.7$ & $6.0 \pm 1.4$ & $4.1 \pm 0.9$ \\
\hline
Observed &  $42$                     &  $16$                     &  $52$                     &  $15$                     &  $21$                     &  $12$                     &  $6$                     \\
\hline

$\langle\epsilon\sigma\rangle_{\textrm{obs}}^{95}$ [fb] &  $0.44$     & $0.25$      & $0.63$      & $0.26$      & $0.42$      & $0.38$      & $0.22$  \\
$S_{\textrm{obs}}^{95}$                                       & $15.9$      & $8.9$       & $22.7$      & $9.4$      & $15.2$      & $13.9$      & $7.8$  \\ 
$S_{\textrm{exp}}^{95}$                                       & $ { 16.6 }^{ +6.7 }_{ -5.0 }$ & $ { 11.0 }^{ +4.1 }_{ -2.7 }$ & $ { 16.7 }^{ +6.8 }_{ -4.8 }$ & $ { 9.9 }^{ +4.1 }_{ -2.5 }$ & $ { 10.7 }^{ +3.4 }_{ -3.1 }$ & $ { 10.3 }^{ +2.7 }_{ -2.1 }$ & $ { 6.3 }^{ +1.9 }_{ -2.1 }$ \\
$p_{0}$ ($\textrm Z$)                                        & $ 0.50$~$(0.00)$ & $ 0.50$~$(0.00)$ & $ 0.19$~$(0.89)$ & $ 0.50$~$(0.00)$ & $ 0.11$~$(1.21)$ & $ 0.07$~$(1.50)$ & $ 0.24$~$(0.72)$  \\
\hline 
\end{tabular}

\vspace{0.2cm}
\begin{tabular}{|lrrrrr|}
\hline
Signal Region & \textbf{RJR-C1 } & \textbf{RJR-C2 } & \textbf{RJR-C3 } & \textbf{RJR-C4 } & \textbf{RJR-C5 } \\
\hline
\multicolumn{6}{|c|}{MC expected events} \\ \hline
Diboson &  $4.5$               &  $3.4$               &  $1.6$               &  $2.7$               &  $0.8$               \\
$Z/\gamma^{*}$+jets &  $24.8$               &  $20.7$               &  $7.8$               &  $10.3$               &  $2.3$               \\
$W$+jets &  $9.8$               &  $7.4$               &  $8.3$               &  $8.0$               &  $2.4$               \\
$\ttbar$(+EW) + single top &  $1.32$               &  $1.6$               &  $5.5$               &  $6.9$               &  $3.39$               \\
\hline
\multicolumn{6}{|c|}{Fitted background events} \\ \hline
Diboson & $4.5 \pm 1.0$ & $3.4 \pm 0.8$ & $1.6 \pm 0.5$ & $2.7 \pm 0.7$ & $0.8 \pm 0.5$ \\
$Z/\gamma^{*}$+jets & $22.6 \pm 2.3$ & $18.9 \pm 2.0$ & $6.5 \pm 1.2$ & $8.6 \pm 1.2$ & $2.1 \pm 0.6$ \\
$W$+jets & $9.9 \pm 1.9$ & $7.5 \pm 1.4$ & $8.9 \pm 1.4$ & $8.6 \pm 1.4$ & $2.7 \pm 2.1$ \\
$\ttbar$(+EW) + single top &  $0.86_{-0.86}^{+1.00}$               & $1.0 \pm 0.7$ & $3.2 \pm 1.5$ & $4.0 \pm 2.4$ &  $0.89_{-0.89}^{+2.17}$               \\
Multi-jet & $0.06 \pm 0.06$ & $0.33 \pm 0.33$ & $0.5 \pm 0.5$ & $0.8 \pm 0.8$ &  $0.25_{-0.25}^{+0.26}$               \\
\hline
Total MC &  $43.9$               &  $53.3$               &  $54.8$               &  $84.0$               &  $28.0$               \\
\hline
Total bkg & $37.9 \pm 3.5$ & $31.2 \pm 2.9$ & $20.7 \pm 2.6$ & $24.8 \pm 3.3$ & $6.7 \pm 1.3$ \\
\hline
Observed &  $36$                     &  $29$                     &  $12$                     &  $24$                     &  $10$                     \\
\hline

$\langle\epsilon\sigma\rangle_{\textrm{obs}}^{95}$ [fb] & $0.38$       & $0.35$      & $0.18$      & $0.42$      & $0.30$   \\
$S_{\textrm{obs}}^{95}$                                       &  $13.8$      & $12.7$      & $6.4$       & $15.2$      & $10.7$ \\ 
$S_{\textrm{exp}}^{95}$                                       & $ { 15.3 }^{ +5.7 }_{ -4.7 }$ & $ { 14.0 }^{ +5.0 }_{ -4.2 }$ & $ { 11.2 }^{ +4.4 }_{ -3.5 }$ & $ { 15.2 }^{ +4.5 }_{ -3.5 }$ & $ { 7.8 }^{ +2.7 }_{ -2.0 }$ \\
$p_{0}$ ($\textrm Z$)                                        &  $ 0.50$~$(0.00)$ & $ 0.50$~$(0.00)$ & $ 0.50$~$(0.00)$ & $ 0.50$~$(0.00)$ & $ 0.14$~$(1.06)$  \\
\hline 
\end{tabular}

\vspace*{-0.01\textheight}\caption[p0 and UL]{Numbers of events observed in the signal regions used in the RJR-based analysis compared with background predictions obtained from the fits described in the text. The $p$-values ($p_{0}$) are the probabilities to obtain a value equal to or larger than that observed in the data. For an observed number of events lower than expected, the $p$-value is truncated at 0.5. In addition to $p$-values, the number of equivalent Gaussian standard deviations (Z) is given in parentheses.
Also shown are 95\% CL upper limits on the visible cross-section ($\langle\epsilon\sigma\rangle_{\textrm{obs}}^{95}$), 
the visible number of signal events ($S_{\textrm{obs}}^{95}$) and the number of signal events ($S_{\textrm{exp}}^{95}$) 
given the expected number of background events (and $\pm 1\sigma$ excursions of the expected number).
\label{tab:p0_UL_RJ}}
\end{center}
\end{table}

\clearpage

\begin{figure}[t]
\begin{center}
\subfigure[]{\includegraphics[width=0.7\textwidth]{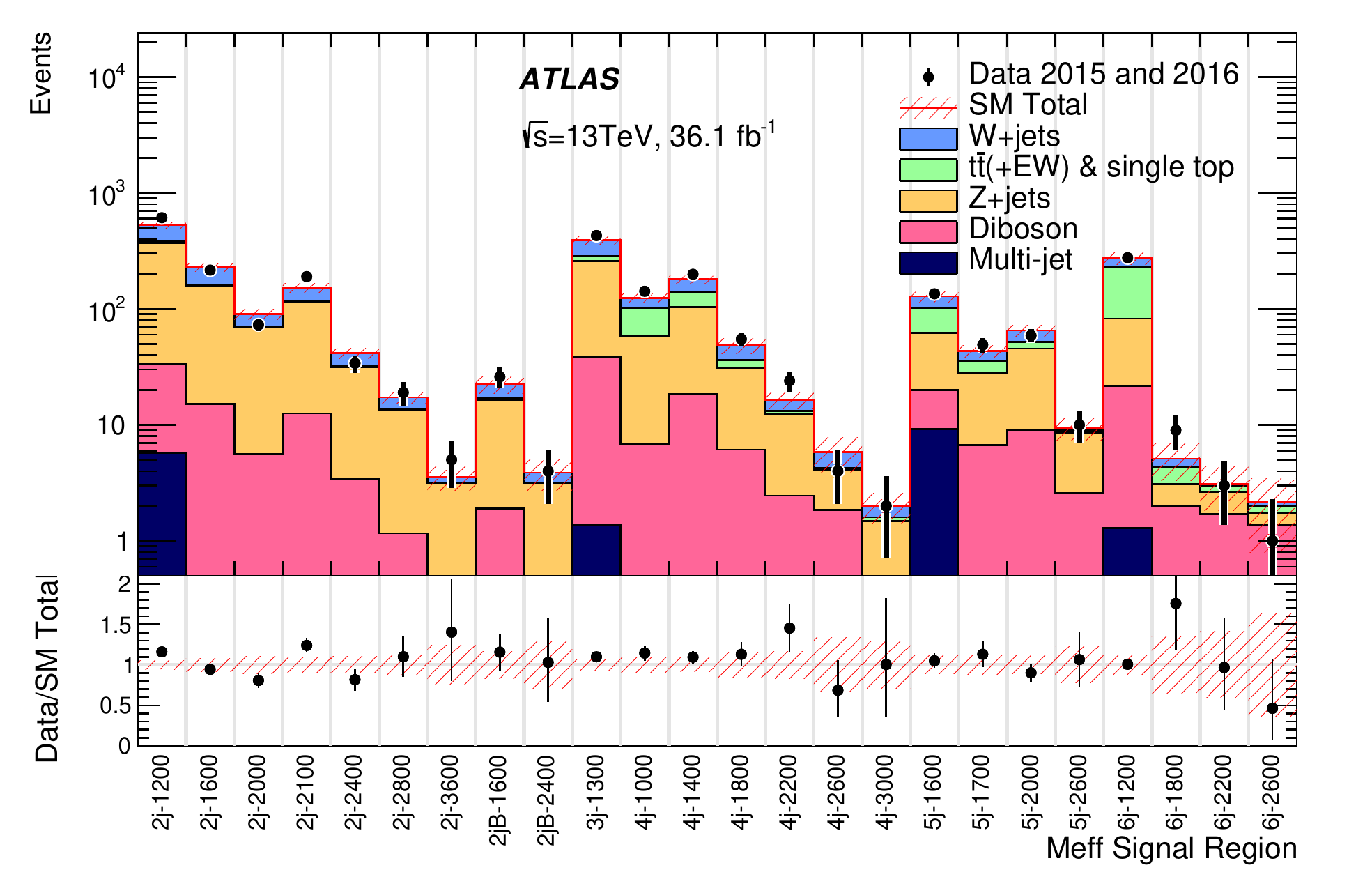}} 
\subfigure[]{\includegraphics[width=0.7\textwidth]{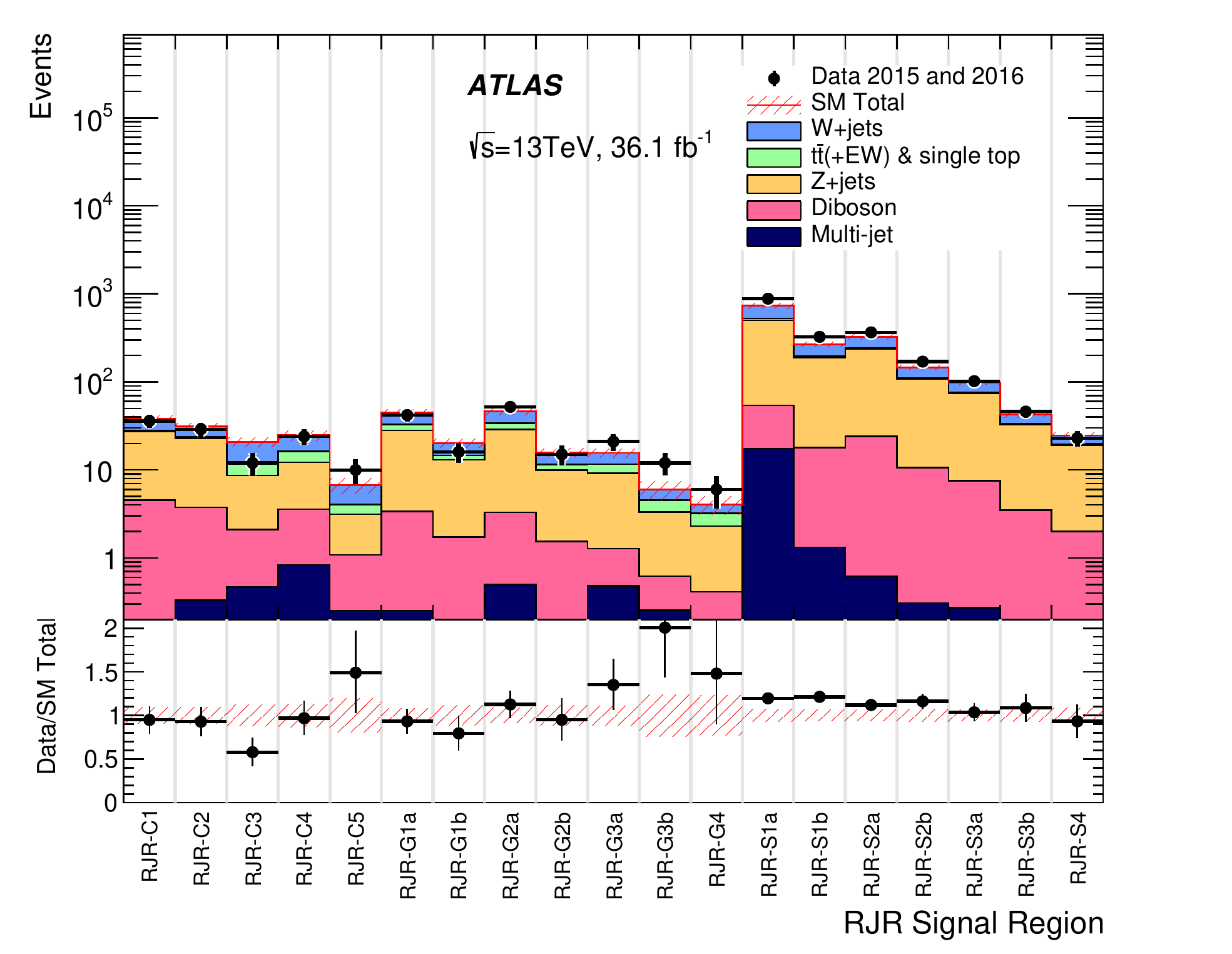}} 
\end{center}
\caption{\label{fig:PlotSR} Comparison of the observed and expected
  event yields as a function of signal region in the (a) Meff-based
  and (b) RJR-based searches. The background predictions are those
  obtained from the background-only fits, presented in
  Tables~\ref{tab:p0_UL} and ~\ref{tab:p0_UL_RJ}. The bottom graph
  shows the ratio of observed data yields to the total predicted
  background. The hatched (red) error bands indicate the combined
  experimental and MC statistical uncertainties. 
} 
\end{figure}

\clearpage

In the absence of a statistically significant excess, limits are set
on contributions to the SRs from BSM physics. Upper limits at 95\% CL
on the number of BSM signal events in each SR and the corresponding
visible BSM cross-section are derived from the model-independent fits
described in Section~\ref{sec:strategy} using the CL$_{\textrm S}$ prescription. Limits are evaluated using MC pseudo-experiments. The results are presented in Tables~\ref{tab:p0_UL} and \ref{tab:p0_UL_RJ}. 

The model-dependent fits in all the SRs are used to set limits on
specific classes of SUSY models {\color{black}using asymptotic formulae~\cite{statforumlimits}}. The two searches presented in this paper are combined such that the final observed and expected 95\% CL exclusion limits are obtained from the signal regions with the best expected CL$_{\textrm S}$ value. {\color{black}Fine structures
in the limit lines arise due to transitions between best SR's which then
also have an impact on the interpolations between grid points.}

In Figure~\ref{fig:directLimit}, limits are shown for two classes of
simplified models in which only direct production of first- and
second-generation mass-degenerate squark or gluino pairs are
considered. Limits are obtained by using the signal region with the
best expected sensitivity at each point. 
In these simplified-model scenarios, the upper limit of the excluded
first- and second-generation squark mass region is $1.55~\TeV$
assuming massless $\ninoone$, as obtained from the signal region
RJR-S4. {\color{black}The observed exclusion limit is worse than the expected limit in the
region with squark ($\ninoone$) mass of $1 ~\TeV$ ($500~\GeV$) due to
a 2 $\sigma$ excess in SR Meff-2j-1200.} The corresponding limit on the gluino mass is $2.03~\TeV$, if the
$\ninoone$ is massless, as obtained from the signal region
Meff-4j-3000. The best sensitivity in the region of parameter space
where the mass difference between the squark (gluino) and the lightest
neutralino is small, is obtained from the dedicated RJR-C signal
regions. In these regions with very compressed spectra and where the
mass difference is less than $50 ~\GeV$, squark (gluino) masses up to
$650~\GeV$ ($1~\TeV$) are excluded. {\color{black} In
Figure~\ref{fig:directLimit}(b), the compressed-mass region with a
gluino mass below 700$~\GeV$ is fully excluded by this analysis; small deviations in
the exclusion contour in this region, suggesting non-excluded areas,
are due to interpolation effects.} {\color{black} The observed exclusion limit is worse than the expected limit in the
region with gluino ($\ninoone$) mass of 1800 (700) $\GeV$ due to a moderate
excess (1.3 $\sigma$) in SR Meff-4j-2200.}

\begin{figure}[t]
\begin{center}
\subfigure[]{\includegraphics[width=0.49\textwidth]{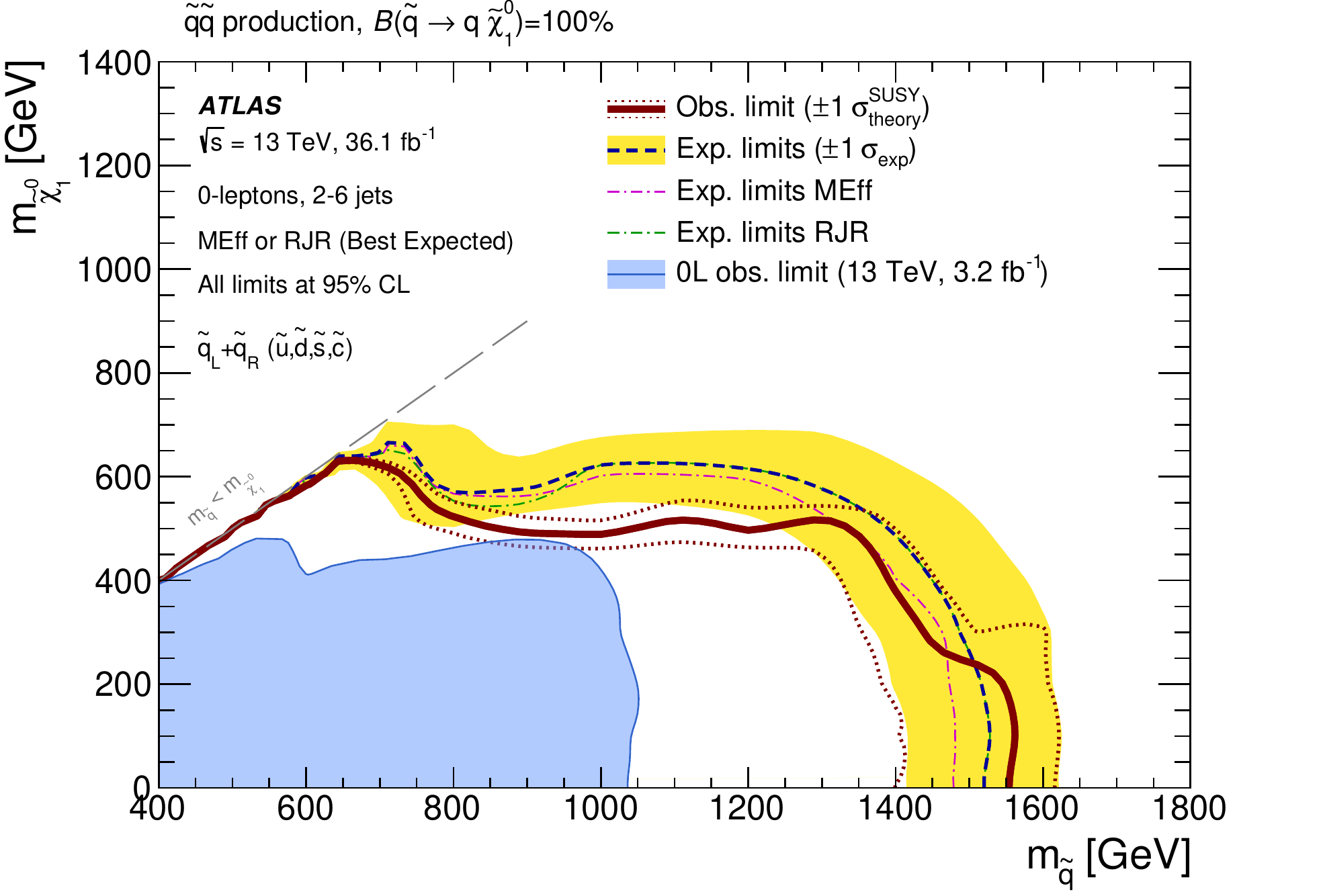}}
\subfigure[]{\includegraphics[width=0.49\textwidth]{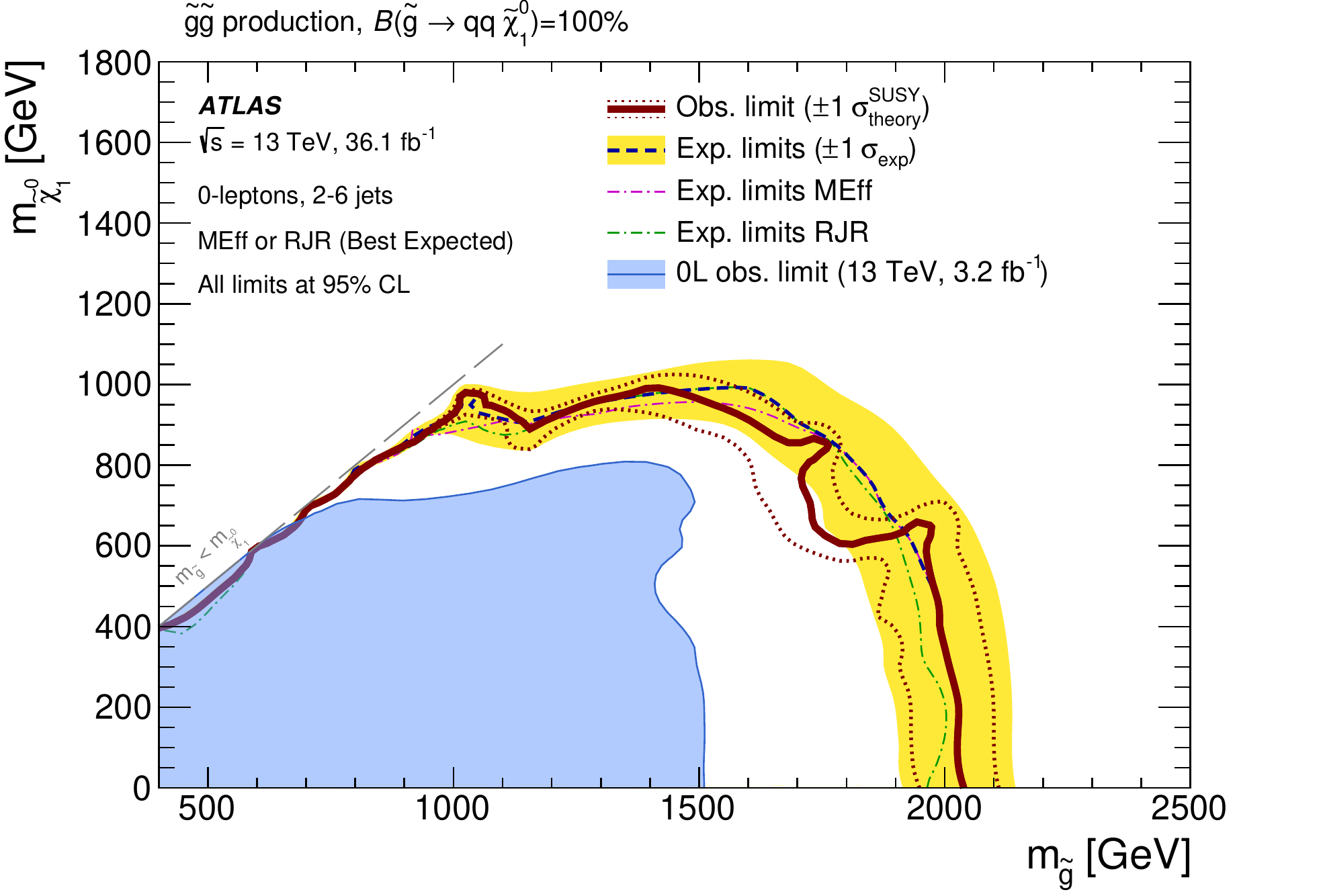}}
 \caption{Exclusion limits for direct production of (a) first- and second-generation squark pairs with decoupled gluinos and (b) gluino pairs with decoupled squarks. Gluinos (first- and second-generation squarks) are required to decay to two quarks (one quark) and a neutralino LSP.  Exclusion limits are obtained by using the signal region with the best expected sensitivity at each point. Expected limits from the Meff- and RJR-based searches separately are also shown for comparison. 
The blue dashed lines show the expected limits at 95\% CL, with the light (yellow) bands indicating the $1\sigma$ excursions due to experimental and background-only  theoretical uncertainties.
Observed limits are indicated by medium dark (maroon) curves where the solid contour represents the nominal limit, and the dotted lines are obtained by varying the signal cross-section by the renormalization and factorization scale and PDF uncertainties.
Results are compared with the observed limits obtained by the previous ATLAS searches with jets, missing transverse momentum, and no leptons~\cite{0LPaper_13TeV}. 
\label{fig:directLimit}}
\end{center}
\end{figure}

In Figure~\ref{fig:limitSMonestep}, limits are shown for pair-produced first- and second-generation squarks or gluinos each decaying via an intermediate $\chinoonepm$ to a quark (for squarks) or two quarks (for gluinos), 
a $W$ boson and a $\ninoone$. Two sets of models of mass spectra are considered for each case. One is with a fixed $m_{\chinoonepm}=(m_{\squark}+m_{\ninoone})/2$ (or $(m_{\gluino}+m_{\ninoone})/2$),
the other is with a fixed $m_{\ninoone}=60~\GeV$. In the former models with squark pair production, $m_{\squark}$ up to $1.15~\TeV$ are excluded for a massless $\ninoone$, as is $m_{\gluino}$ up to $1.98~\TeV$
with gluino pair production. These limits are obtained from the signal region RJR-G2b and Meff-6j-2600, respectively. In the regions with very compressed spectra with mass difference between the gluino (or squark) and $\ninoone$
less than $50~\GeV$, RJR-C signal regions also exclude squark (gluino)
masses up to $600~\GeV$ ($1~\TeV$). In the latter models,
Meff-2jB-1600 and Meff-2jB-2400 extend the limits on squark (gluino)
masses up to $1.1~\TeV$ ($1.85~\TeV$) in the regions with small mass
difference between the squark (gluino) and $\chinoonepm$.

\begin{figure}[tb]
\begin{center}
\subfigure[]{\includegraphics[width=0.455\textwidth]{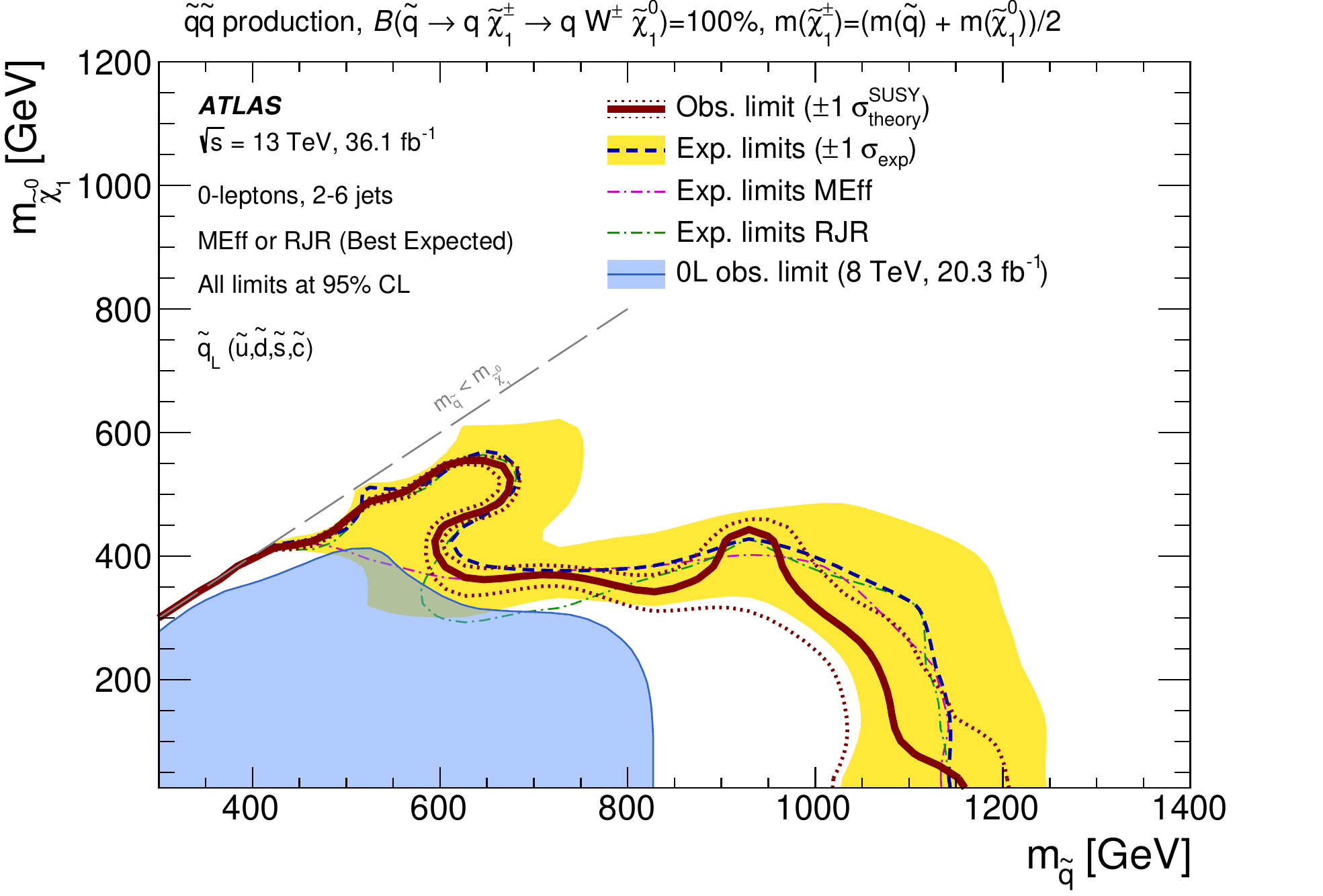}}
\subfigure[]{\includegraphics[width=0.455\textwidth]{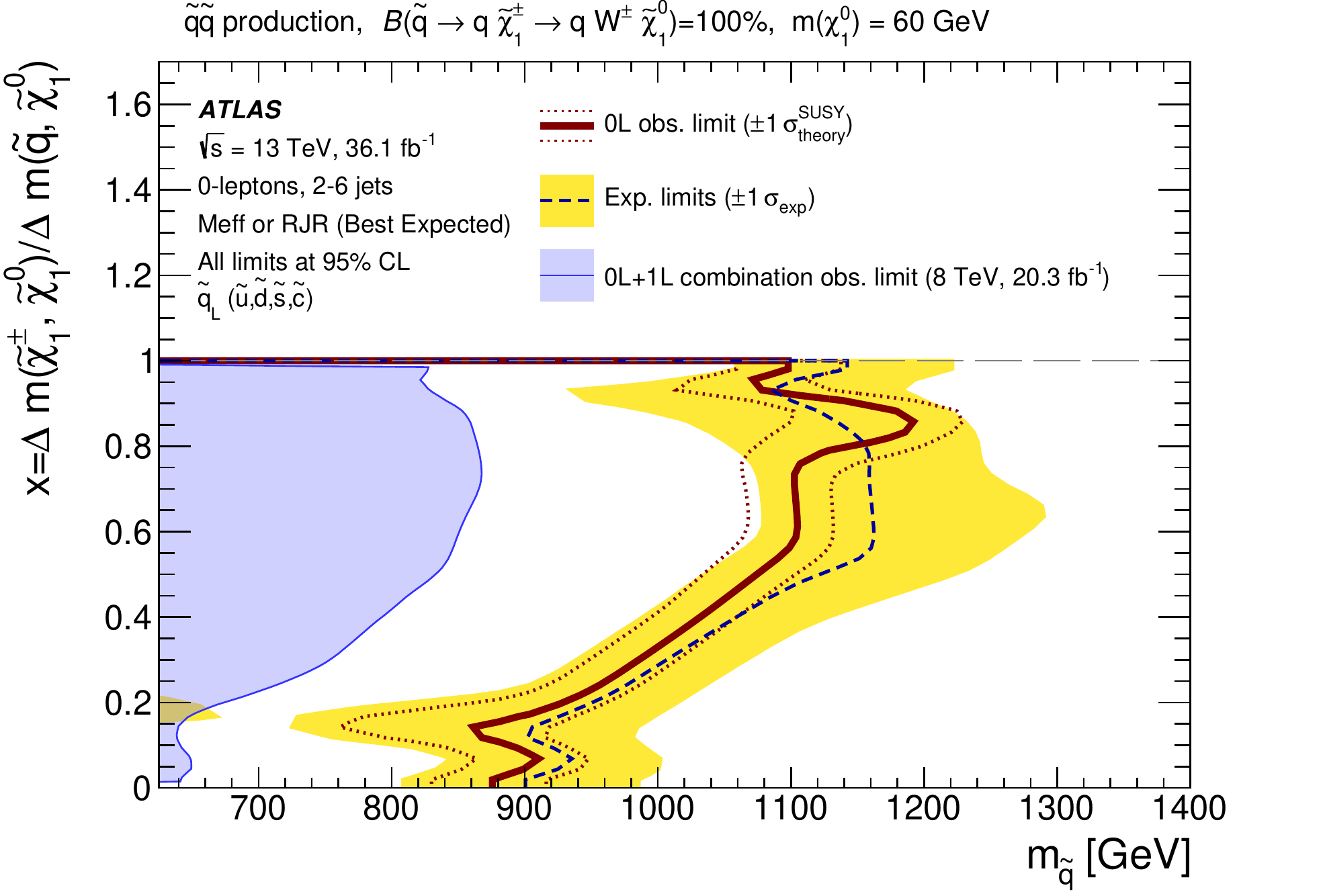}}
\subfigure[]{\includegraphics[width=0.455\textwidth]{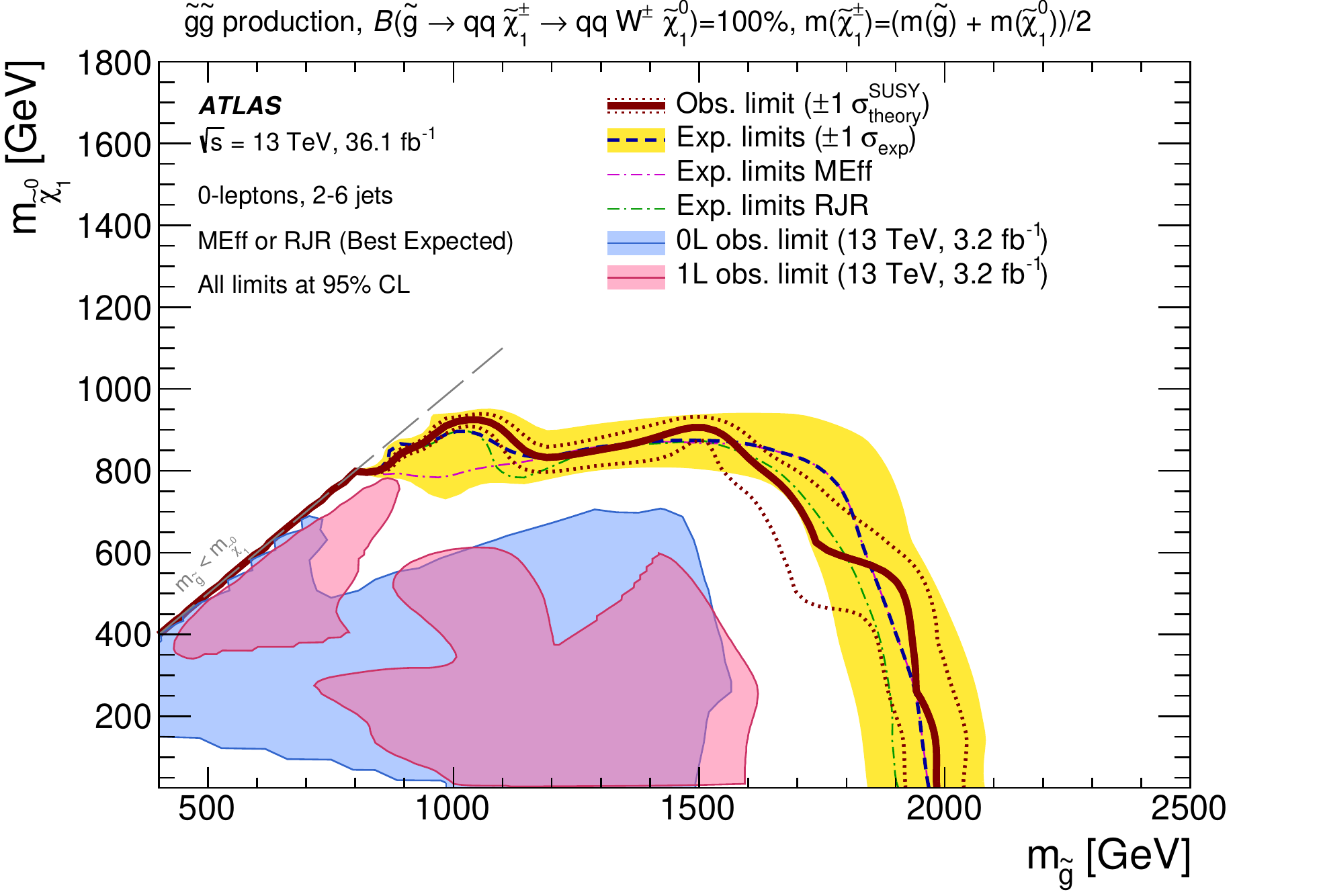}}
\subfigure[]{\includegraphics[width=0.455\textwidth]{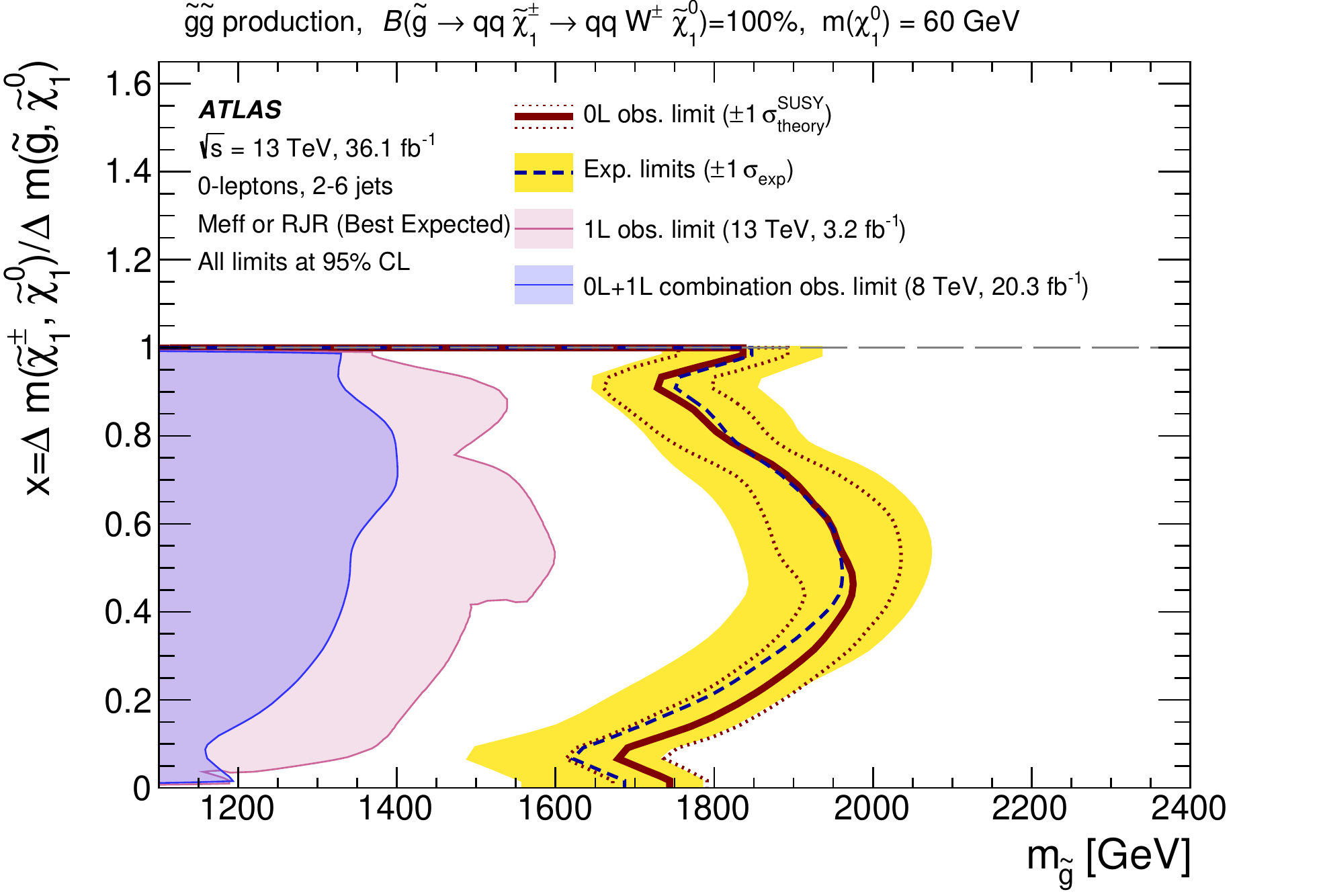}}
\caption{
Exclusion limits for direct production of (a,b) first- and second-generation left-handed squark pairs with decoupled gluinos and (c,d) gluino pairs with decoupled squarks. Gluinos (first- and second-generation squarks) are required to decay to two quarks (one quark) and an intermediate $\chinoonepm$, decaying to a $W$ boson and a $\ninoone$. Models with (a,c)  a fixed $m_{\chinoonepm}=(m_{\gluino}+m_{\ninoone})/2$ (or ($m_{\squark}+m_{\ninoone})/2$) and varying values of $m_{\gluino}$ (or $m_{\squark}$) and $m_{\ninoone}$, and (b,d) a fixed $m_{\ninoone}=60~\GeV$ and varying values of $m_{\gluino}$ (or $m_{\squark}$) and $m_{\chinoonepm}$ are considered. 
Exclusion limits are obtained by using the signal region with the best expected sensitivity at each point. Expected limits from the Meff- and RJR-based searches separately are also shown for comparison in (a,c). 
The blue dashed lines show the expected limits at 95\% CL, with the light (yellow) bands indicating the $1\sigma$ excursions due to experimental and background-only  theoretical uncertainties.
Observed limits are indicated by medium dark (maroon) curves where the solid contour represents the nominal limit, and the dotted lines are obtained by varying the signal cross-section by the renormalization and factorization scale and PDF uncertainties.
Results are compared with the observed limits obtained by the previous ATLAS searches with one or no leptons, jets and missing transverse momentum~\cite{summaryPaper,0LPaper_13TeV,1leptonPaper-arxiv}. }
\label{fig:limitSMonestep}
\end{center}
\end{figure}

\clearpage

In Figure~\ref{fig:limitSMonestepN2}, limits are shown for gluino pair production decaying via an intermediate $\ninotwo$ to two quarks, a $Z$ boson and a $\ninoone$. 
The mass of the $\ninoone$ is set to $1~\GeV$. In these models, gluino
masses below $2.0~\TeV$ are excluded for $\ninotwo$ masses of
$\sim1~\TeV$, as obtained from the signal region Meff-6j-2600.

\begin{figure}[t]
\begin{center}
\includegraphics[width=0.49\textwidth]{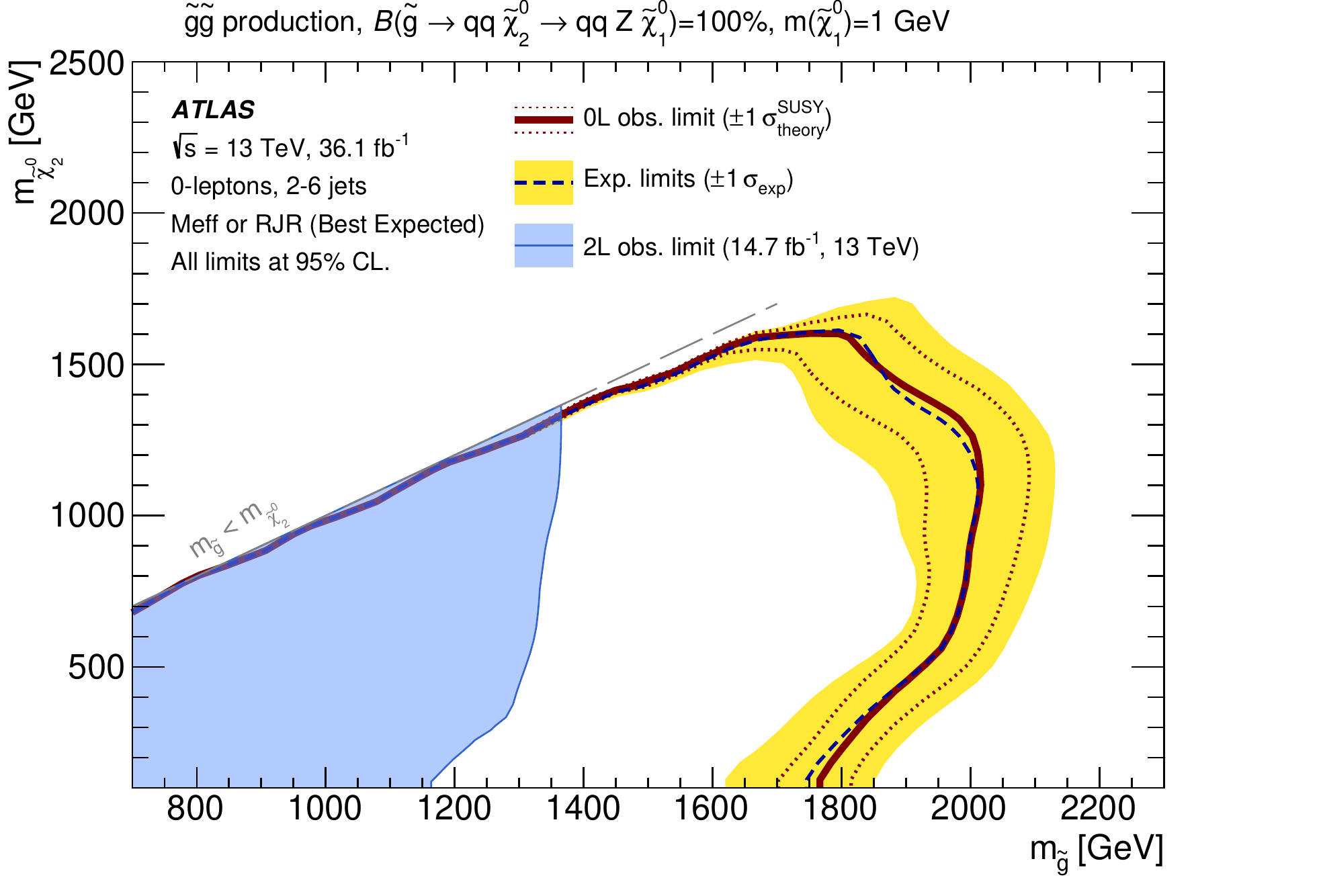}
\caption{Exclusion limits for pair-produced gluinos each decaying via an intermediate  $\ninotwo$ to two quarks, a $Z$ boson and a $\ninoone$ for models with a fixed $m_{\ninoone} = 1 ~\GeV$ and varying values of $m_{\gluino}$ and $m_{\ninotwo}$. Exclusion limits are obtained by using the signal region with the best expected sensitivity at each point. The blue dashed lines show the expected limits at 95\% CL, with the light (yellow) bands indicating the $1\sigma$ excursions due to experimental and background-only theoretical uncertainties. Observed limits are indicated by medium dark (maroon) curves where the solid contour represents the nominal limit, and the dotted lines are obtained by varying the signal cross-section by the renormalization and factorization scale and PDF uncertainties.  
Results are compared with the observed limits obtained by the previous ATLAS search in events containing a leptonically decaying $Z$ boson, jets and missing transverse momentum~\cite{ATLAS_ZMET_Run2_paper}. 
}
\label{fig:limitSMonestepN2}
\end{center}
\end{figure}

In Figure~\ref{fig:limitSMonestepN2C}, results are presented in the models with mixed decays of intermediate $\chinoonepm$ and $\ninotwo$
for squark pair and gluino pair production. The highest limits on the squark mass are $1.34~\TeV$ and on the gluino mass are $2.02~\TeV$, which are
similar to the models with 100\% branching fraction for $\chinoonepm$
($\ninotwo$) to a $W$ ($Z$) boson and $\ninoone$. {\color{black} In
Figure~\ref{fig:limitSMonestepN2C}(b), the limits are extended by the SR Meff-2jet in the region with small mass differences between the
gluino and $\ninotwo$.}

\begin{figure}[t]
\begin{center}
\subfigure[]{\includegraphics[width=0.49\textwidth]{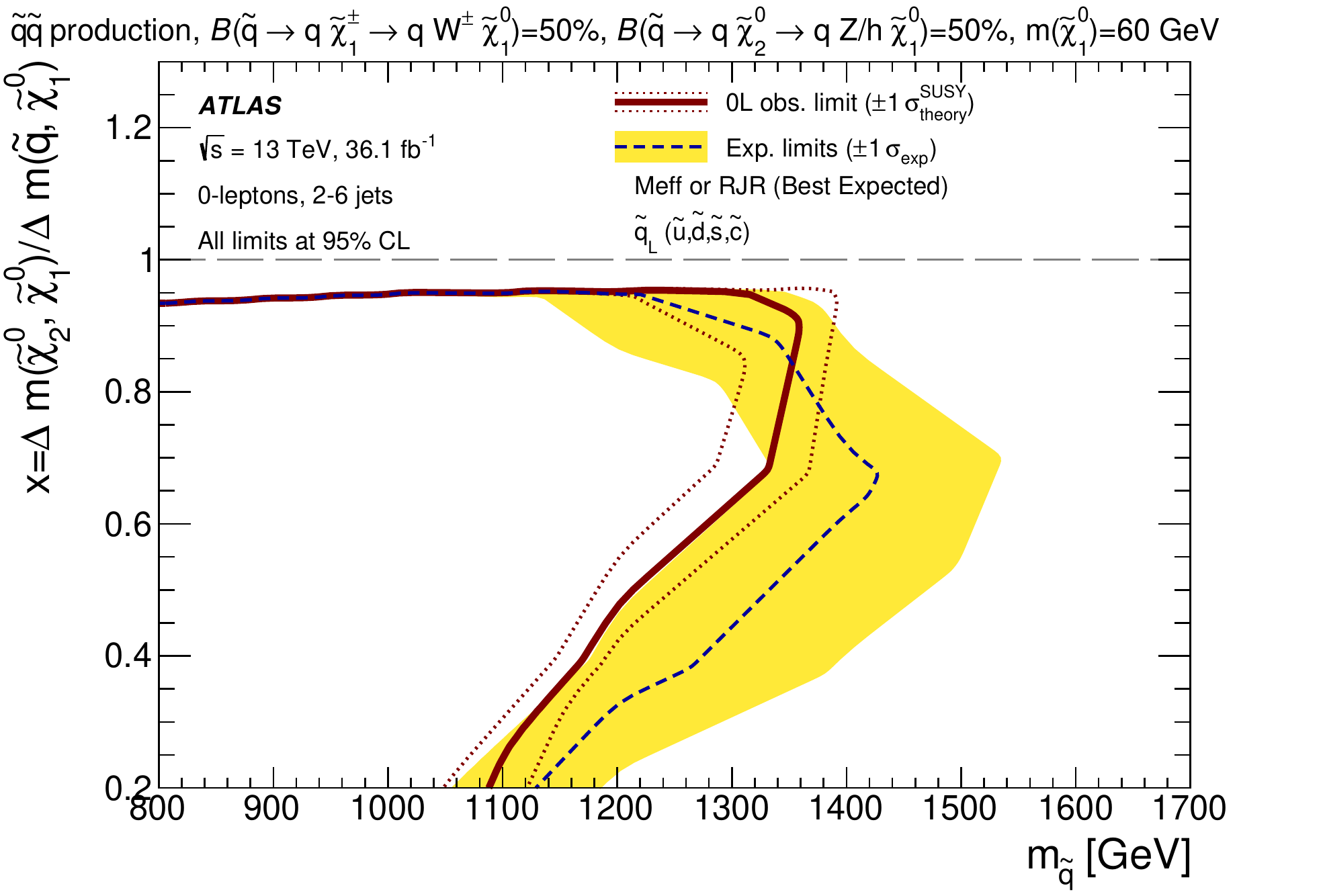}}
\subfigure[]{\includegraphics[width=0.49\textwidth]{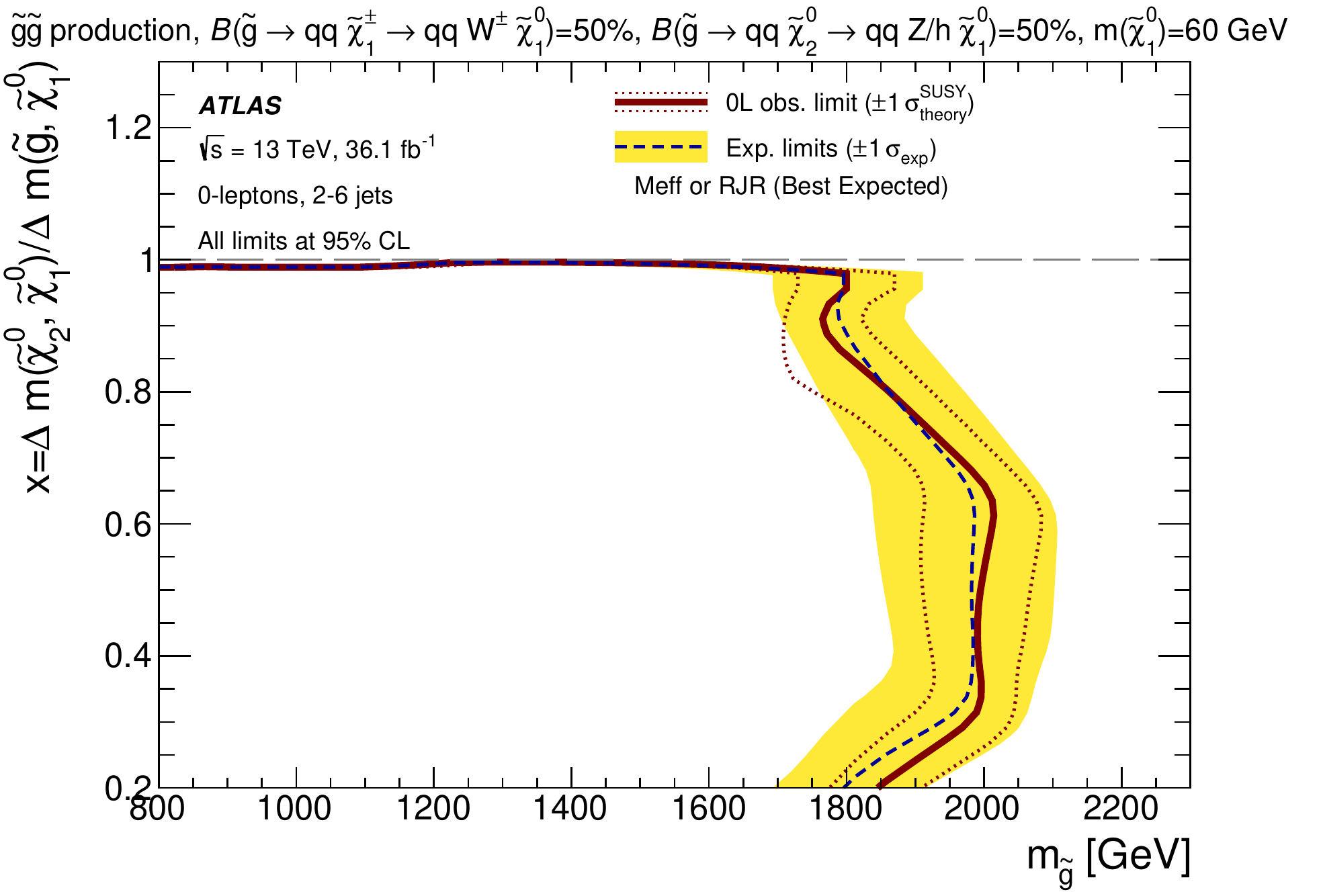}}
\caption{Exclusion limits for direct production of (a) first- and
  second-generation left-handed squark pairs with decoupled gluinos
  and (b) gluino pairs with decoupled squarks. Gluinos (first- and
  second-generation squarks) are required to decay to two quarks (one
  quark) and a intermediate $\chinoonepm$ or $\ninotwo$ with a 50\%
  branching fraction, respectively, with $\chinoonepm$ decays to a $W$
  boson and a $\ninoone$, and $\ninotwo$ decays to a $Z$ or a $h$
  boson and $\ninoone$. Models with fixed $m_{\ninoone}=60~\GeV$
are considered while varying $m_{\gluino}$ (or $m_{\squark}$) and $m_{\ninoone}$.
Exclusion limits are obtained by using the signal region with the best expected sensitivity at each point. The blue dashed lines show the expected limits at 95\% CL, with the light (yellow) bands indicating the $1\sigma$ excursions due to experimental and background-only theoretical uncertainties. Observed limits are indicated by medium dark (maroon) curves where the solid contour represents the nominal limit, and the dotted lines are obtained by varying the signal cross-section by the renormalization and factorization scale and PDF uncertainties.  
}
\label{fig:limitSMonestepN2C}
\end{center}
\end{figure}

In Figure~\ref{fig:limitPMSSM}, results are interpreted in simplified pMSSM models assuming only first- and second-generation squarks, gluino and $\ninoone$.
The $\ninoone$ is assumed to be purely bino. Models with a fixed $m_{\ninoone}=0, 695, 995~\GeV$ are considered while varying $m_{\gluino}$ and $m_{\squark}$.
In the limit of high squark mass, gluino masses up to $2~\TeV$ are
excluded for massless $\ninoone$, which is consistent with the simplified models of gluino pair production
with decoupled squarks. With a gluino mass of $6~\TeV$, squark masses up to $2.2~\TeV$ are excluded for a massless $\ninoone$, much higher than in
the simplified models of squark pair production with decoupled gluinos. 
This is due to the large cross-section of squark pair production via gluino exchange diagrams.

\begin{figure}[t]
\begin{center}
\subfigure[]{\includegraphics[width=0.49\textwidth]{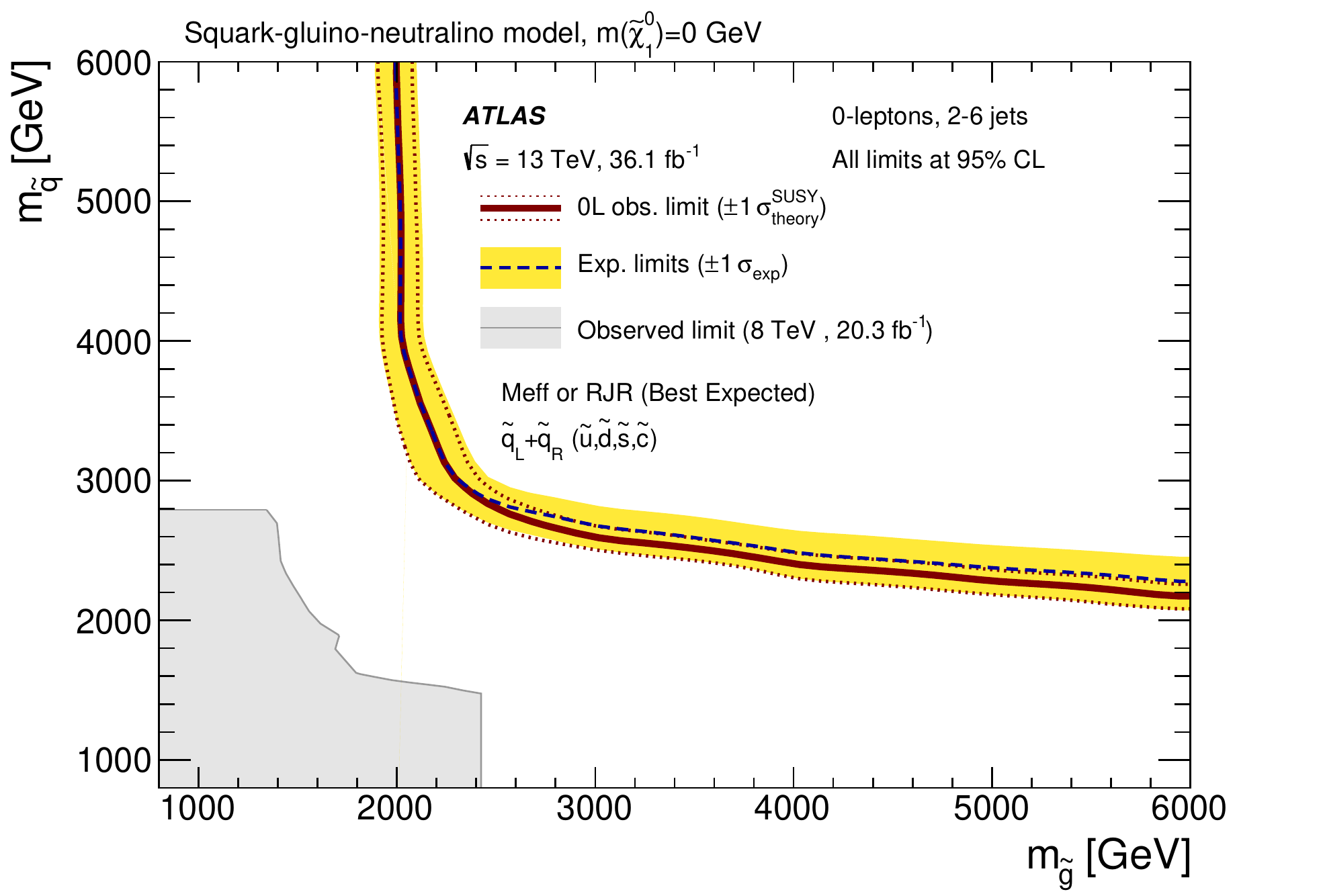}}
\subfigure[]{\includegraphics[width=0.49\textwidth]{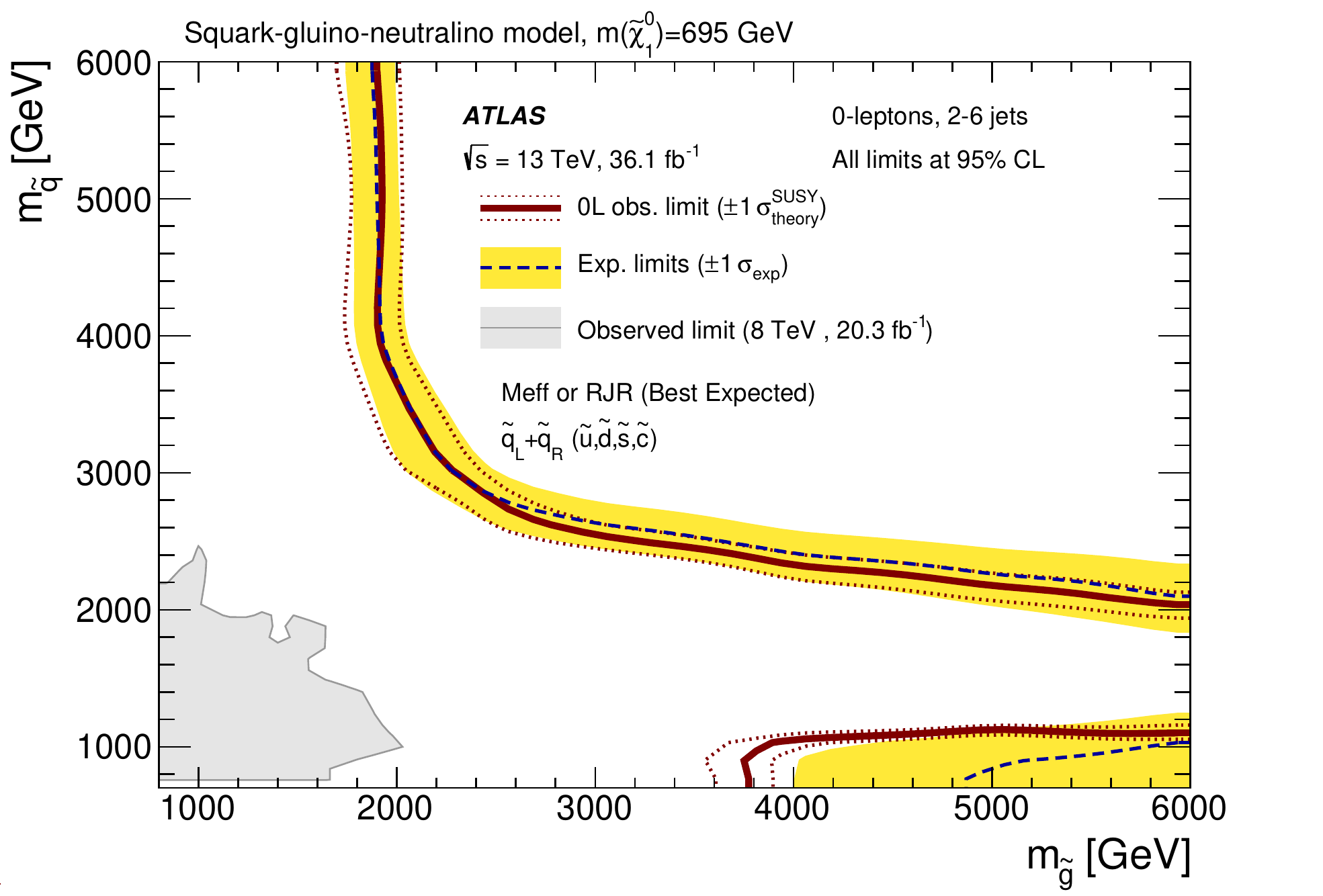}}
\subfigure[]{\includegraphics[width=0.49\textwidth]{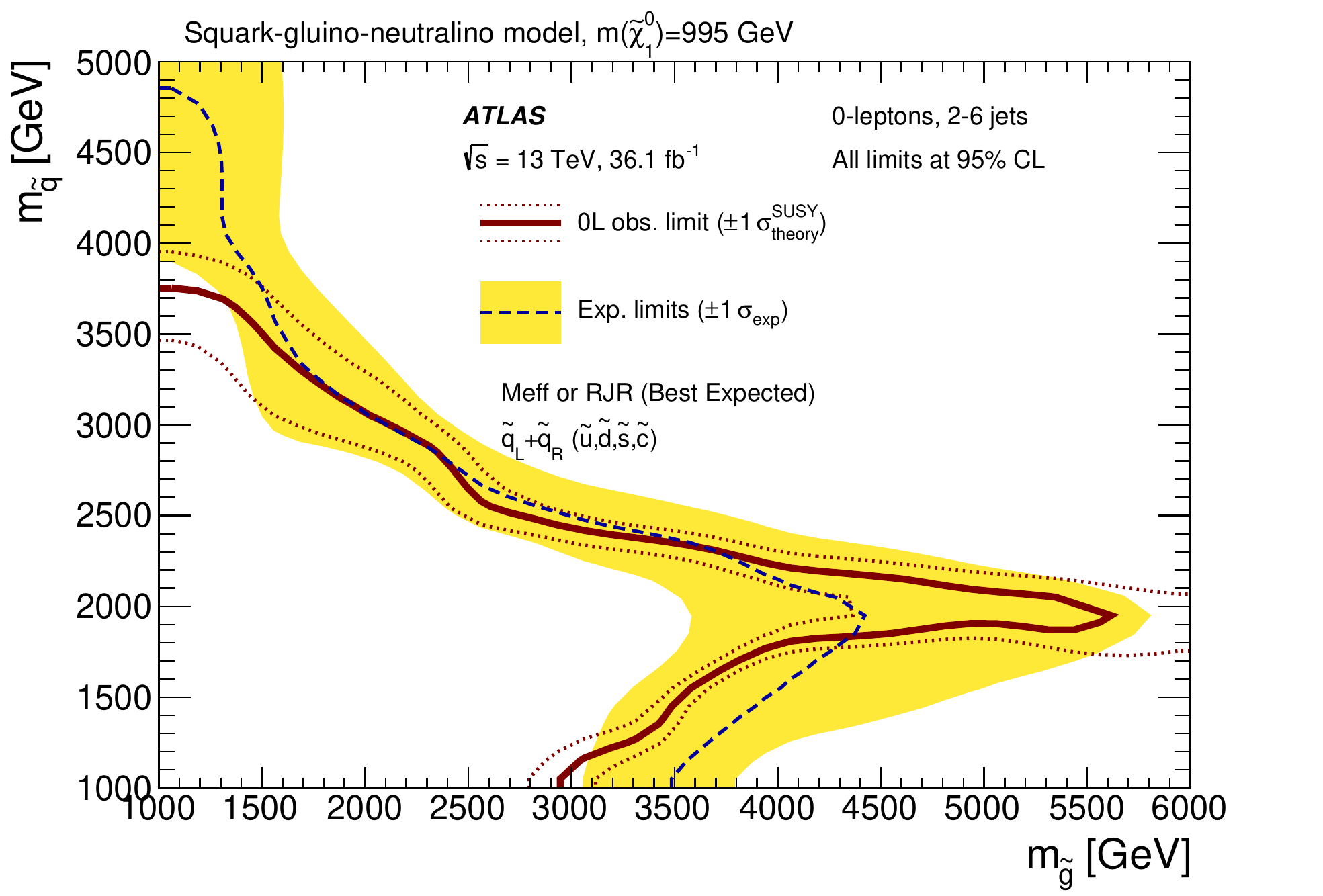}}
\caption{Exclusion limits for inclusive squark--gluino production in pMSSM models with (a) $m_{\tilde{\chi}^{0}_{1}}=0~\GeV$, (b) $m_{\ninoone}=695~\GeV$ and (c) $m_{\tilde{\chi}^{0}_{1}}=995~\GeV$  varying values of $m_{\gluino}$ and $m_{\squark}$ and assuming purely bino $\ninoone$. Exclusion limits are obtained by using the signal region with the best expected sensitivity at each point. The blue dashed lines show the expected limits at 95\% CL, with the light (yellow) bands indicating the $1\sigma$ excursions due to experimental and background-only theoretical uncertainties. Observed limits are indicated by medium dark (maroon) curves where the solid contour represents the nominal limit, and the dotted lines are obtained by varying the signal cross-section by the renormalization and factorization scale and PDF uncertainties.  
Results (a,b) are compared with the observed limits obtained by the previous ATLAS searches with no leptons, jets and missing transverse momentum~\cite{0-leptonPaper}. }
\label{fig:limitPMSSM}
\end{center}
\end{figure}

A comparison of the Meff-based and RJR-based results highlights some notable features.
The RJR-Cx signal regions provide additional sensitivity in the most compressed mass regions beyond their Meff-based counterparts, extending exclusion limits up to $200 ~\GeV$ in \ninoone mass for the smallest mass splitting, as is the case in Figure~\ref{fig:limitSMonestep}(a) for first- and second-generation squarks decaying via an intermediate $\chinoonepm$. In general, the RJR-Cx regions are only mildly sensitive to the specific decays of squarks and gluinos, resulting in similar sensitivity as a function of \squark/\gluino and \ninoone masses between signal models with direct decays in Figure~\ref{fig:directLimit} and those with intermediate sparticle decays as in Figure~\ref{fig:limitSMonestep}. 

Despite being largely orthogonal, the RJR-based and Meff-based SRs targeting squark and gluino direct decay signals tend to result in similar sensitivity, with the RJR-based regions generally performing better for intermediate mass splittings. This is the result of tighter restrictions placed on dimensionless variables in the RJR-based regions, resulting in generally lower background yields.

For models with additional jets in the final state expected from intermediate sparticle decays, the Meff-5j-x and Meff-6j-x provide significant additional sensitivity with respect to lower multiplicity SRs, extending exclusion limits by close to $100 ~\GeV$ in gluino mass when intermediate $\chinoonepm$ decays are considered. These more stringent jet multiplicity requirements compensate for the modest $\met/\meff(N_{\textrm j})$ values characteristic of these models.

With requirements aimed at tagging hadronic decays of $W/Z$ bosons, the Meff-2jB-x SRs provide higher sensitivity to models with intermediate $\chinoonepm$ and $\ninotwo$ decays when these sparticles are almost degenerate in mass with their parent squarks and gluinos, corresponding to Figures~\ref{fig:limitSMonestep}(b),~\ref{fig:limitSMonestep}(d),~\ref{fig:limitSMonestepN2}, and~\ref{fig:limitSMonestepN2C}. In these cases, the sensitivity of the Meff-2jB-x regions far surpasses those of the RJR-based and other Meff-based SRs.

\FloatBarrier

\section{Conclusion}
\label{sec:conclusion}

This paper presents results of two selection strategies to search for squarks and gluinos in final states containing
high-\pT{} jets, large missing transverse momentum but no electrons or
muons, based on a 36.1~\ifb\ dataset of $\sqrt{s}=13 ~\TeV$
proton--proton collisions recorded by the ATLAS experiment at the LHC
in 2015 and 2016. No significant deviation from the background expectation is found. 

Results are interpreted in terms of simplified models or pMSSM models
with only first- and second-generation squarks, or gluinos, together with a neutralino
LSP, with the masses of all the other SUSY particles set such that the
particles are effectively decoupled.   
For a massless lightest neutralino, gluino masses below 2.03~\TeV\ are excluded at the 95\% confidence level in a simplified model with only gluinos and the lightest neutralino. 
For a simplified model involving the strong production of squarks of the first and second generations, with decays to a massless lightest neutralino, squark masses below 1.55~\TeV\ are excluded, assuming mass-degenerate squarks of the first two generations. 
No exclusion is obtained for simplified models of squark (gluino) pair
production with lightest neutralino masses above 630~\GeV\ (970~\GeV).
In simplified models with pair-produced squarks and gluinos, each decaying via an intermediate $\chinoonepm$ to one quark or two quarks, a $W$ boson and a $\ninoone$, 
squark masses below $1.15~\TeV$ and gluino masses below 1.98~\TeV\ are excluded for massless $\ninoone$. 
In pMSSM models assuming squarks, gluinos and $\ninoone$, gluino masses below $2.0~\TeV$ are excluded for a squark mass of $6~\TeV$
or squark masses below $2.2~\TeV$ are excluded for a gluino mass of $6~\TeV$ for massless $\ninoone$.

These results substantially extend the region of supersymmetric parameter space previously excluded by ATLAS searches.

\section*{Acknowledgments}
We thank CERN for the very successful operation of the LHC, as well as the
support staff from our institutions without whom ATLAS could not be
operated efficiently.

We acknowledge the support of ANPCyT, Argentina; YerPhI, Armenia; ARC, Australia; BMWFW and FWF, Austria; ANAS, Azerbaijan; SSTC, Belarus; CNPq and FAPESP, Brazil; NSERC, NRC and CFI, Canada; CERN; CONICYT, Chile; CAS, MOST and NSFC, China; COLCIENCIAS, Colombia; MSMT CR, MPO CR and VSC CR, Czech Republic; DNRF and DNSRC, Denmark; IN2P3-CNRS, CEA-DRF/IRFU, France; SRNSF, Georgia; BMBF, HGF, and MPG, Germany; GSRT, Greece; RGC, Hong Kong SAR, China; ISF, I-CORE and Benoziyo Center, Israel; INFN, Italy; MEXT and JSPS, Japan; CNRST, Morocco; NWO, Netherlands; RCN, Norway; MNiSW and NCN, Poland; FCT, Portugal; MNE/IFA, Romania; MES of Russia and NRC KI, Russian Federation; JINR; MESTD, Serbia; MSSR, Slovakia; ARRS and MIZ\v{S}, Slovenia; DST/NRF, South Africa; MINECO, Spain; SRC and Wallenberg Foundation, Sweden; SERI, SNSF and Cantons of Bern and Geneva, Switzerland; MOST, Taiwan; TAEK, Turkey; STFC, United Kingdom; DOE and NSF, United States of America. In addition, individual groups and members have received support from BCKDF, the Canada Council, CANARIE, CRC, Compute Canada, FQRNT, and the Ontario Innovation Trust, Canada; EPLANET, ERC, ERDF, FP7, Horizon 2020 and Marie Sk{\l}odowska-Curie Actions, European Union; Investissements d'Avenir Labex and Idex, ANR, R{\'e}gion Auvergne and Fondation Partager le Savoir, France; DFG and AvH Foundation, Germany; Herakleitos, Thales and Aristeia programmes co-financed by EU-ESF and the Greek NSRF; BSF, GIF and Minerva, Israel; BRF, Norway; CERCA Programme Generalitat de Catalunya, Generalitat Valenciana, Spain; the Royal Society and Leverhulme Trust, United Kingdom.

The crucial computing support from all WLCG partners is acknowledged gratefully, in particular from CERN, the ATLAS Tier-1 facilities at TRIUMF (Canada), NDGF (Denmark, Norway, Sweden), CC-IN2P3 (France), KIT/GridKA (Germany), INFN-CNAF (Italy), NL-T1 (Netherlands), PIC (Spain), ASGC (Taiwan), RAL (UK) and BNL (USA), the Tier-2 facilities worldwide and large non-WLCG resource providers. Major contributors of computing resources are listed in Ref.~\cite{ATL-GEN-PUB-2016-002}.

\printbibliography
\clearpage
\begin{flushleft}
{\Large The ATLAS Collaboration}

\bigskip

M.~Aaboud$^\textrm{\scriptsize 137d}$,
G.~Aad$^\textrm{\scriptsize 88}$,
B.~Abbott$^\textrm{\scriptsize 115}$,
O.~Abdinov$^\textrm{\scriptsize 12}$$^{,*}$,
B.~Abeloos$^\textrm{\scriptsize 119}$,
S.H.~Abidi$^\textrm{\scriptsize 161}$,
O.S.~AbouZeid$^\textrm{\scriptsize 139}$,
N.L.~Abraham$^\textrm{\scriptsize 151}$,
H.~Abramowicz$^\textrm{\scriptsize 155}$,
H.~Abreu$^\textrm{\scriptsize 154}$
% ...
\bigskip
\\
% ...
$^{12}$ Institute of Physics, Azerbaijan Academy of Sciences, Baku, Azerbaijan\\
% ...
$^{88}$ CPPM, Aix-Marseille Universit{\'e} and CNRS/IN2P3, Marseille, France\\
% ...
$^{115}$ Homer L. Dodge Department of Physics and Astronomy, University of Oklahoma, Norman OK, United States of America\\
% ...
$^{119}$ LAL, Univ. Paris-Sud, CNRS/IN2P3, Universit{\'e} Paris-Saclay, Orsay, France\\
% ...
$^{137}$ $^{(a)}$ Facult{\'e} des Sciences Ain Chock, R{\'e}seau Universitaire de Physique des Hautes Energies - Universit{\'e} Hassan II, Casablanca; $^{(b)}$ Centre National de l'Energie des Sciences Techniques Nucleaires, Rabat; $^{(c)}$ Facult{\'e} des Sciences Semlalia, Universit{\'e} Cadi Ayyad, LPHEA-Marrakech; $^{(d)}$ Facult{\'e} des Sciences, Universit{\'e} Mohamed Premier and LPTPM, Oujda; $^{(e)}$ Facult{\'e} des sciences, Universit{\'e} Mohammed V, Rabat, Morocco\\
% ...
$^{139}$ Santa Cruz Institute for Particle Physics, University of California Santa Cruz, Santa Cruz CA, United States of America\\
% ...
$^{151}$ Department of Physics and Astronomy, University of Sussex, Brighton, United Kingdom\\
% ...
$^{154}$ Department of Physics, Technion: Israel Institute of Technology, Haifa, Israel\\
$^{155}$ Raymond and Beverly Sackler School of Physics and Astronomy, Tel Aviv University, Tel Aviv, Israel\\
% ...
$^{161}$ Department of Physics, University of Toronto, Toronto ON, Canada\\
% ...
$^{a}$ Also at Department of Physics, King's College London, London, United Kingdom\\
$^{b}$ Also at Institute of Physics, Azerbaijan Academy of Sciences, Baku, Azerbaijan\\
$^{c}$ Also at Novosibirsk State University, Novosibirsk, Russia\\
$^{d}$ Also at TRIUMF, Vancouver BC, Canada\\
$^{e}$ Also at Department of Physics {\&} Astronomy, University of Louisville, Louisville, KY, United States of America\\
% ...
$^{*}$ Deceased
\end{flushleft}

 


\end{document}